\documentclass{ws-ijmpab}
\def\draftnote{~}
\usepackage{float}
\usepackage{epsfig}
\usepackage{amsmath}
\usepackage{amssymb}

\makeatletter
\@addtoreset{footnote}{section}
\makeatother

\newcommand{\ep}{\varepsilon}

\newcommand{\vect}[1]{\vec{#1}}

\newcommand{\eqs}[1]{\begin{equation} \begin{split} #1\end{split} \end{equation} }

\newcommand{\ks}[1]{#1 \!\!\! \slash } 
\newcommand{\kks}[1]{#1 \!\!\!\! \slash }

\newcommand{\tr}{{\rm Tr}}
\newcommand{\ie}{{\it i.e.}}
\newcommand{\eg}{{\it e.g.}}
\newcommand{\etal}{{\it et al.}}

\newcommand{\ce}[1]{Eq.~(\ref{#1})}

\newcommand{\cf}[1]{{Fig.~\ref{#1}}}
\newcommand{\ct}[1]{{Table.~\ref{#1}}}

\begin{document}

\markboth{Jean-Philippe Lansberg}{$J/\psi$, $\psi'$ and $\Upsilon$ Production at Hadron Colliders: a review}
\catchline{}{}{}{}{}

\title{
$J/\psi$, $\psi'$ AND $\Upsilon$ PRODUCTION AT HADRON COLLIDERS: \\A REVIEW
}

\author{JEAN-PHILIPPE LANSBERG\footnote{Jean-Philippe.Lansberg@cpht.polytechnique.fr}}

\address{Centre de Physique Th\'eorique, \'Ecole Polytechnique\footnote{Unit\'e mixte 7644 du CNRS}\\
F-91128 Palaiseau, France
}

\address{Physique Th\'eorique Fondamentale, Universit\'e de Li\`ege\\
17 All\'ee du 6 Ao\^ut, B\^at. B5, B-4000 Li\`ege-1, Belgique\\~\\
}

\maketitle

\begin{abstract}
We give an overview of the present status of knowledge of the production of $J/\psi$, $\psi'$ and $\Upsilon$
in high-energy hadron collisions. We first present two early models,
namely the Colour-Singlet Model (CSM) and the Colour-Evaporation Model (CEM). The first is the
natural application of pQCD to quarkonium production and has been shown to fail dramatically to describe
experimental data, the second is its phenomenological counterpart and was introduced in the 
spirit of the quark-hadron duality in the late seventies. 
Then, we expose the most recent experimental measurements of $J/\psi$, $\psi'$ 
and $\Upsilon$ prompt and direct production at nonzero $p_T$ from two high-energy hadron colliders, 
the Tevatron and RHIC. In a third part, we review six contemporary models describing $J/\psi$, 
$\psi'$ and $\Upsilon$ production at nonzero $p_T$.

\end{abstract}
\ccode{PACS numbers: 14.40.Gx 13.85.Ni  }
\newpage



\section{History: from a revolution to an anomaly}

\subsection{$J/\psi$ and the November revolution}

The era of quarkonia has started at the simultaneous discovery\footnote{Nobel prize of 1976.} 
of the $J/\psi$ in November 1974 by Ting {\it et al.}\cite{Aubert:1974js} 
at BNL and by Richter {\it et al.}\cite{Augustin:1974xw} at SLAC.
Richter's experiment used the electron-positron 
storage ring SPEAR, whose center-of-momentum energy could be tuned at the desired value. With 
the Mark I detector, they discovered a sharp enhancement of the production
cross section in different channels:  $e^+ e^-$, $\mu^+ \mu^-$,
$\pi^+ \pi^-$,\dots On the other hand, Ting's experiment was based on the high-intensity proton beams of 
the Alternating Gradient Synchrotron (AGS) working at the energy of 30 GeV, 
which bombarded a fixed target with the consequence of producing showers of 
particles detectable by the appropriate apparatus.

In the following weeks, the Frascati group (Bacci~\etal\cite{Bacci:1974za}) 
confirmed the presence of this new particle 
whose mass was approximately 3.1 GeV. The confirmation was so fast that it was actually 
published in the same issue of Physical Review Letters, {\it ie.} vol. 33, no. 23, 
issued the second of December 1974. In the meantime, Richter's group discovered another 
resonant state with a slightly higher mass, 
which was called\footnote{We shall also make us of the name $\psi(2S)$.} $\psi'$.

It was also promptly established
that the quantum numbers of the $J/\psi$ were the same as those of the photon , 
\ie~$1^{--}$. Moreover, since the following ratio
\eqs{R= \frac{\hbox{cross section for $e^+ e^- \to$ hadrons}}{\hbox{cross section for 
$e^+ e^- \to \mu^+\mu^-$}}} was much larger on-resonance than off, it was then clear that the
$J/\psi$ did have direct hadronic decays. The same conclusion held for the $\psi'$ as well. 
The study of multiplicity in pion decays
indicated that $\psi$ decays were restricted by $G$-parity 
conservation, holding only for hadrons. Consequently, $J/\psi$ and $\psi'$ 
were rapidly considered as hadrons of isospin 0 and $G$-parity -1.

Particles with charge conjugation $C$ different from -1 were found later. 
Indeed, they were only produced by decay of 
$\psi''$ and the detection of the radiated photon during the (electromagnetic)
decay was then required. The first to achieve this task and discover a new state 
was the DASP collaboration\cite{Braunschweig:1975ac} based at DESY 
(Deutsches Elektronen-Synchrotron), Hamburg, 
working at an $e^+e^-$ storage ring called DORIS. This new particle, named\footnote{The name chosen 
here reveals that physicists already had at this time an interpretation of this state as bound-state
of quarks $c$. Indeed, the symbol $P$ follows from the classical spectroscopic notation for
a $\ell=1$ state, where $\ell$ stands for the angular momentum quantum number.} 
$P_c$, had a mass of approximately 3.5 GeV. At SPEAR, other resonances at 
3.415, 3.45 and 3.55 GeV were discovered, the 3.5 GeV state was confirmed. 
Later, these states were shown to be $C=+1$.

Coming back to the ratio $R$, it is instructive to analyse the implication in the framework of the 
quark model postulated by Gell-Mann and Zweig in 1963. In 1974 at the London conference, 
Richter presented\cite{Richter:1974vz} the experimental situation as in~\cf{fig:R-richter}.

\begin{figure}
\centerline{\includegraphics[width=5cm]{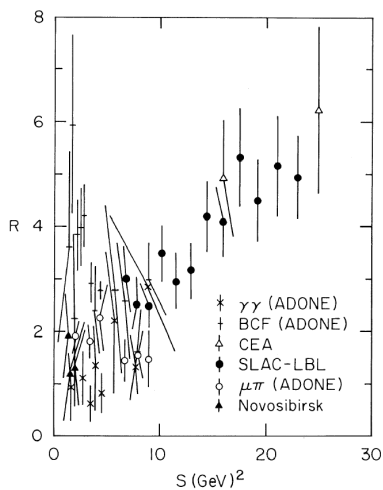}}
\caption{Experimental status of $R$ as of July 1974 (from Ref.~\protect\refcite{Richter:1974vz})
.}
\label{fig:R-richter}
\end{figure} 

In the framework of the Gell-Mann-Zweig quark model with 
three quarks,
a plateau was expected with a value of $2/3$ or $2$ if the quark were considered as 
``coloured'' -- a new concept introduced recently then --. The sign of a new quark -- as the 
charm quark first proposed by Bjorken and Glashow in 1964\cite{Bjorken:1964gz}-- would 
have been another plateau from a certain energy (roughly its mass) with a 
height depending on its charge. Retrospectively, one cannot blame anyone for not having seen
this ``plateau'' on this kind of plots. Quoting Richter, ``the situation that prevailed in 
the Summer of 1974'' was ``vast confusion''.

This charm quark  was also required by the mechanism of Glashow, Iliopoulos and Maiani 
(GIM)\cite{Glashow:1970gm}, in order to cancel the anomaly in weak decays. The charm quark 
was then expected to exist and to have an electric charge $2/3$. $R$ was therefore 
to be $10/3$ in the coloured quark model, still not obvious in~\cf{fig:R-richter}. This explains 
why the discovery of such sharp and easily interpreted resonances in November 1974 
was a {\it revolution} in particle physics.

It became quite rapidly obvious that the $J/\psi$ was the lowest-mass $c\bar c$ system 
with the same quantum numbers as photons -- explaining why  it was so much produced 
compared to some other members of its family. These  $c\bar c$  bound states were named ``charmonium'', 
firstly by Appelquist, De R\'ujula, Politzer and Glashow\cite{Appelquist:1974yr}
in analogy with positronium, whose bound-state level structure was similar.

At that time, the charm had nevertheless always been found hidden, that is in charm-anti-charm 
bound states. In order to study these explicitly charmed mesons, named $D$, the investigations 
were based on the assumption that 
the $D$ was to decay weakly and preferentially into strange quarks. The weak character of 
the decay motivated physicists to search for parity-violation signals. The 
first resonance attributed to $D^0$ meson was found in $K^+\pi^+$ decay by Goldhaber 
{\it et al.}\cite{Goldhaber:1976xn} in 1976. A little later, $D^+$ and $D^-$ were also 
discovered as well as an excited state, $D^*$, with a mass compatible with the 
decay $\psi'''\to D^0 D^*$. And, finally, the most conclusive evidence 
for the discovery of charmed meson was the observation of parity violation in $D$ 
decays\cite{Wiss:1976gd}. To complete the picture of the charm family, the first charmed baryon
was discovered during the same year\cite{Knapp:1976qw}. The quarks were not anymore just a way to 
interpret symmetry in masses and spins of particles, they had acquired an existence. 

\subsection{The bottomonium family}

In the meantime, in 1975, another brand new particle was unmasked at SLAC by Perl \etal\cite{Perl:1975bf}. This was the first particle of a third generation of quarks and leptons, the $\tau$, 
a lepton. Following the standard model, more and more trusted, two other quarks were 
expected to exist. Their discovery 
would then be the very proof of the theory that was developed since the sixties. 
Two names for the fifth quark were already chosen: ``beauty'' 
and ``bottom'', and in both cases represented by the letter $b$; the sixth quark was as well 
already christened with the letter $t$ for the suggested names
 ``true'' or ``top''.

The wait was not long: two years. After a false discovery\footnote{Nonetheless,  
this paper\cite{Hom:1976cv} first suggested the notation $\Upsilon$ for any ``onset of high-mass 
dilepton physics''.} of a resonance at 6.0 GeV, a new dimuon resonance similar 
to $J/\psi$ and called $\Upsilon$ was brought to light at Fermilab, thanks to the FNAL 
proton synchrotron accelerator, by 
Herb \etal\cite{Herb:1977ek}, with a mass of 9.0 GeV; as for charmonia, the first radial excited 
state ($\Upsilon(2S)$) was directly found\cite{Innes:1977ae} thereafter. Again, the discovery of a
new quarkonium was a decisive step forward towards the comprehension of particle physics.

Various confirmations of these discoveries were not long to come. The $3S$ state was then 
found\cite{Ueno:1978vr} at Fermilab 
as well as an evidence that the $4S$ state was lying above the threshold for the production of $B$ mesons. 
The latter was confirmed at the Cornell $e^+e^-$ storage ring  with the CLEO detector. The 
first evidence for $B$ meson and, thus, for unhidden $b$ quark, was also brought
 by the CLEO collaboration\cite{Bebek:1980zd} 
in 1980. One year later, ``the first evidence for baryons with naked beauty''({\it sic}) was reported by 
CERN physicists\cite{Basile:1981wr}.

Another decade was needed for the discovery of the sixth quark which was definitely christened
 ``top''. Indeed, in 1994, the CDF Collaboration found the first evidence for it at the Tevatron 
collider at Fermilab\cite{Abe:1994st}. The discovery was published in 1995 both by 
CDF\cite{Abe:1995hr} and D$\emptyset$\cite{Abachi:1995iq}.
 Unfortunately, due to its very short lifetime, this quark cannot bind with 
its antiquark to form the toponium. To conclude this historical prelude, we give 
the spectra (Figs. 2 \& 3) of the $c\bar c$ and $b\bar b$ systems as well as two tables (Tables 1 \& 2) summing up the
characteristics of the observed states as of today.

\begin{table}[H]
\centering\label{table:sum_charmonium}
\tbl{Properties of charmonia (cf. Ref.~\protect\refcite{pdg}).}
{
\begin{tabular}{|c||c|c|c|c|}
\hline Meson & $n ^{2S+1}L_J$ & $J^{PC}$& Mass (GeV)& $\Gamma_{\mu\mu}$ (keV)\\
\hline \hline 
$\eta_c$  &  $1\ ^1S_0$  &  $0^{-+}$ & 2.980 & N/A\\
$J/\psi$  &  $1\ ^3S_1$  &  $1^{--}$ & 3.097 & 5.40\\
$\chi_{c0}$, $\chi_{c1}$, $\chi_{c2}$  &  $1\ ^3P_{0,1,2}$  &  $0^{++}$,$1^{++}$,$2^{++}$ & 
3.415,3.511,3.556 & N/A\\
$h_c$  &  $1\ ^1P_0$  &  $1^{+-}$ & 3.523 & N/A\\
$\eta_c(2S)$  &  $2\ ^1S_0$  &  $0^{-+}$ & 3.594 & N/A\\
$\psi'$          & $2\ ^3S_1$  & $1^{--}$ & 3.686 & 2.12\\
\hline
\end{tabular}
}
\end{table}

\begin{figure}[h]
\centering\includegraphics[width=\textwidth]{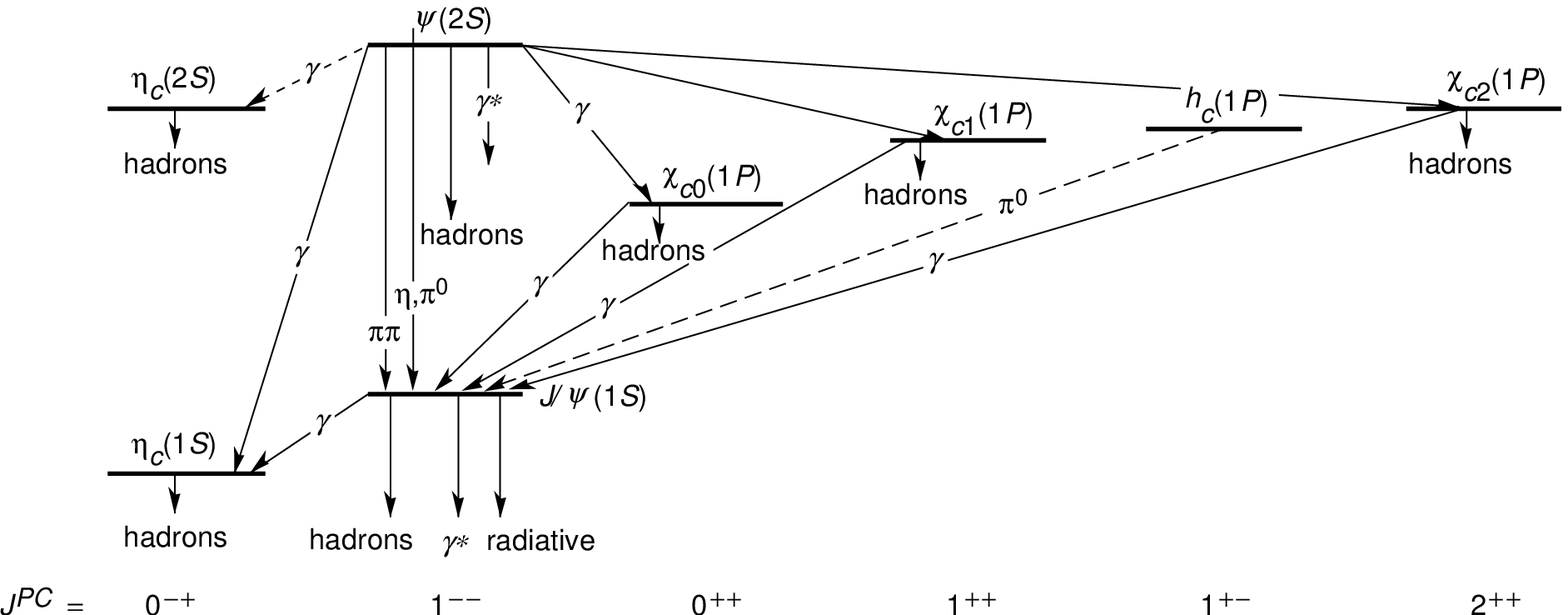}
\caption{Spectrum and transitions of the charmonium family (Reprinted figure from Ref.~\protect\refcite{pdg}
with permission of Elsevier. Copyright (2004).).}
\label{fig:charm_spectrum}
\end{figure}

\begin{figure}[h]
\centering\includegraphics[width=10cm]{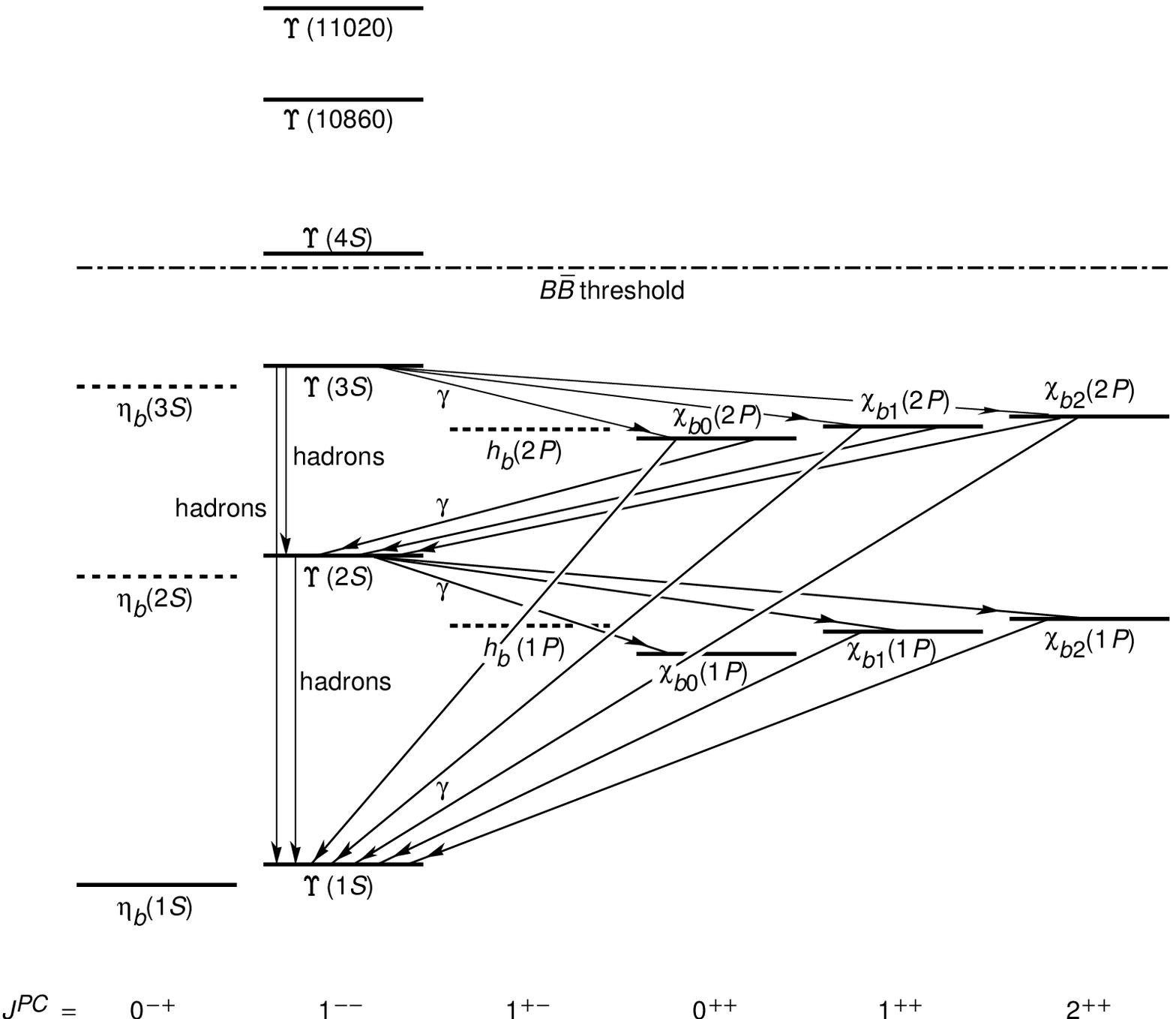}
\caption{Spectrum and transitions of the bottomonium family (Reprinted figure from Ref.~\protect\refcite{pdg}. 
with permission of Elsevier. Copyright (2004).).}
\label{fig:bottom_spectrum}
\end{figure}

\begin{table}[h]
\centering\label{table:sum_bottomonium}
\tbl{Properties of bottomonia (cf. Ref.~\protect\refcite{pdg}) .}{
\begin{tabular}{|c||c|c|c|c|}
\hline Meson & $n ^{2S+1}L_J$ & $J^{PC}$& Mass (GeV)& $\Gamma_{\mu\mu}$ (keV)\\
\hline \hline 
$\Upsilon(1S)$   & $1 ^3S_1$  & $1^{--}$ & 9.460 & 1.26\\
$\chi_{b0}$, $\chi_{b1}$ , $\chi_{b2} (1P)$  &  $1\ ^3P_{0,1,2}$  &  $0^{++}$,$1^{++}$,$2^{++}$ & 
9.860,9.893,9.913 & N/A\\
$\Upsilon(2S)$   &  $2\ ^3S_1$  & $1^{--}$ & 10.023 & 0.32\\
$\chi_{b0}$, $\chi_{b1}$, $\chi_{b2}$ (2P) &  $2\ ^3P_{0,1,2}$  &  $0^{++}$,$1^{++}$,$2^{++}$ & 
10.232,10.255,10.269 & N/A\\
$\Upsilon(3S)$   &  $3\ ^3S_1$ & $1^{--}$ & 10.355 & 0.48\\
\hline
\end{tabular}}
\end{table}

\subsection{Early predictions for quarkonium production}

\subsubsection{The Colour-Singlet Model}
This model is\footnote{before the inclusion of fragmentation contributions.} 
the most natural application of QCD to heavy-quarkonium production in the high-energy regime. 
It takes its inspiration in the factorisation theorem of QCD\cite{facto1,facto2,facto3,facto4}\footnote{proven for 
some definite cases, {\it e.g.} Drell-Yan process.}
where the hard part is calculated by the strict application pQCD and the soft part
is factorised in a universal wave function. This model is meant to describe the
production not only of $J/\psi$, $\psi(2S)$, $\Upsilon(1S)$, $\Upsilon(2S)$ and $\Upsilon(3S)$, \ie~the 
$^3S_1$ states, but also the singlet $S$ states $\eta_c$ and $\eta_b$ as well as the 
$P$ ($\chi$) and $D$ states.

Its greatest quality resides in its predictive power as the only input, apart from the PDF, 
 namely
the wave function, can be determined from data on decay processes or by application
of potential models. Nothing more is required. It was first applied to 
hadron colliders\cite{CSM_hadron1,CSM_hadron2,CSM_hadron3}, then to electron-proton 
colliders\cite{Berger:1980ni}. 
The cross sections for $^3S_1$ states were then calculated, as well as also for $\eta$ and $\chi$, 
for charmonium and for bottomonium. These calculations were compared to ISR and FNAL
data from $\sqrt{s}=27$ GeV to $\sqrt{s}=63$ GeV for which the data extended to 6 GeV 
for the transverse momentum. Updates\cite{Halzen:1984rq,Glover:1987az} of the model
to describe collisions at the CERN $p\bar p$ collider ($\sqrt{s}=630$ GeV) were then presented. 
At that energy, the possibility that the charmonium be produced from the decay of a beauty 
hadron was becoming competitive. Predictions for Tevatron energies were also made\cite{Glover:1987az}.

In order to introduce the reader to several concepts and quantities that will be useful
throughout this Review, let us proceed with a detailed description of this model.

\subsubsection{The Model as of early 90's}

It was then  based on several approximations or postulates:
\begin{itemize}
\item If we decompose the quarkonium production in two steps, first the creation of 
two {\it on-shell} heavy quarks ($Q$ and $\bar Q$) and then their binding to make 
the meson, one {\it postulates the factorisation} of these two processes.
\item As the scale of the first process is approximately $M^2+p_T^2$, one  
considers it as a {\it   perturbative} one. One supposes that its 
cross section be computable with Feynman-diagram methods.
\item As we consider only bound states of heavy quarks (charm and bottom quarks), 
their velocity in the meson must be small. One therefore supposes that the meson be created 
with its 2 constituent quarks {\it   at rest} in the meson frame. This 
is {\it the static approximation.} 
\item One finally assumes that the colour and the spin of the $Q\bar Q$ pair do not 
change during the binding. Besides, as physical states are colourless, one requires the pair 
 be produced in a {\it   colour-singlet state}. This explains the name Colour-Singlet Model (CSM).
\end{itemize}

In high-energy hadronic collisions, the leading contribution comes from a gluon 
fusion process; as the energy of the collider increases, the initial parton momentum fraction $x_i$ needed 
to produce the quarkonium decreases to reach the region in $x$ where the number of gluons 
becomes much larger than the number of quarks. One has then only six Feynman diagrams  
for the $^3S_1$ states production associated with a gluon\footnote{This is 
the dominant process when the transverse momentum of the meson
is non-vanishing.} (see \cf{fig:CSM_box}).

\begin{figure}[h]
\centerline{\includegraphics[width=8cm]{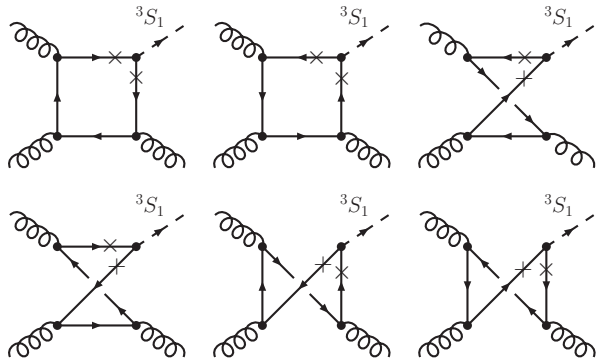}}
\caption{The 6 diagrams for $gg \to \!\!\ ^3S_1 g$ at LO within the CSM.}
\label{fig:CSM_box}
\end{figure} 

One usually starts with  ${\cal M}(p)$, the perturbative amplitude to produce 
the heavy-quark pair on-shell with relative momentum $p$ and in a configuration similar 
to the one of the meson. To realise the latter constraint, 
one introduces a projection operator\footnote{In fact, this amounts to associate a $\gamma^5$ 
matrix to pseudoscalars, $\gamma^\mu$ to vectors, etc.}; the amplitude ${\cal M}(p)$ is then 
simply calculated with the usual Feynman rules.

The amplitude to produce the meson is thence given by 
\begin{equation}\label{eq:amp_CSM_1}
{\cal A}=\int \Phi(\vect{p}) {\cal M}(p)  \delta(2p^0) dp, 
\end{equation}
where $\Phi(\vect{p})$ is the usual Schr\"odinger wave-function.

Fortunately, one does not have to carry out the integration thanks to the {\it the static approximation}
which amounts to considering the first non-vanishing term of $\cal A$ when the perturbative
part $\cal M$ is expanded in $p$. For $S$-wave, this gives

\begin{equation}\label{eq:psim2}
\int \Phi(\vect{p}) {\cal M}(p) \delta(2p^0)dp \simeq \left.{\cal M}\right|_{p=0} 
\left.\Psi\right|_{\vect{x}=0},
\end{equation}
where $\Psi$ is the wave-function in coordinate space,  and $\left.\Psi\right|_{\vect{x}=0}$ 
or $\Psi(0)$ is its value at the origin. For $P$-waves, $\Psi(0)$ is zero, and the second term
in the Taylor expansion must be considered; this makes appear $\Psi'(0)$.  $\Psi(0)$ (or  $\Psi'(0)$)  is the 
non-perturbative input, which is also present in the leptonic decay width from 
which it can be extracted.

If the perturbative part ${\cal M}(p)$ is calculated at the leading order in $\alpha_s$, 
this will be referred to as the Leading Order CSM (LO CSM)\cite{CSM_hadron1,CSM_hadron2,CSM_hadron3,Berger:1980ni}.

\subsubsection{The Colour Evaporation Model}

This model was initially introduced in 1977\cite{Fritzsch:1977ay,Halzen:1977rs} and was revived 
in 1996 by Halzen~\etal\cite{CEM1,CEM2}.Contrarily 
to the CSM, the heavy-quark pair produced by the perturbative interaction is not assumed to 
be in a colour-singlet state. One simply considers that the colour and the spin of the 
asymptotic $Q\bar Q$ state is randomised by numerous soft interactions occurring after its production,
and that, as a consequence, it is not correlated with the quantum numbers of the pair right after 
its production.

A first outcome of this statement is that the production of a $^3S_1$ state by one gluon
is possible, whereas in the CSM it was  forbidden solely by colour conservation. In addition,
the probability that the $Q\bar Q$ pair eventually be in a colour-singlet state is therefore 
$\frac{1}{9}$, which gives the total cross section to produce a quarkonium:
\begin{equation}\label{eq:CEM1}
\sigma_{onium}=\frac{1}{9}\int_{2m_Q}^{2m_{\bar qQ}} dm \frac{d\sigma_{Q\bar Q}}{dm}.
\end{equation}
This amounts to integrating the cross section of production of $Q\bar Q$ from the threshold $2m_Q$ up
to the threshold to produce two charm or beauty mesons (designated here by $\bar qQ$) 
and one divides by 9 to get the probability of having a colour-singlet state.

The procedure to get the cross section for a definite state, for instance a $J/\psi$, is tantamount
to ``distributing'' the cross sections among all states:
\begin{equation}\label{eq:CEM2}
\sigma_{J/\psi}=\rho_{J/\psi}\sigma_{onium}.
\end{equation}
The natural value for $\rho_{J/\psi}$, as well as for the other states in that approximation, is the 
inverse of the number of quarkonium lying between $2m_c$ and $2m_D$. This can be refined by 
factors arising from spin multiplicity arguments
or by the consideration of the mass difference between the produced  and the final
states. These are included in the Soft-Colour-Interactions approach (SCI) (see section~\ref{subsec:SCI}).

By construction, this model is unable to give information about the polarisation of
the quarkonium produced, which is a key test for the other models\cite{yr}. Furthermore,
nothing but fits can determine the values to input for $\rho$. Considering production 
ratios for charmonium states within the simplest
approach for spin, we should have for instance $\sigma[\eta_c]:\sigma[J/\psi] = 
1:3$ and $\sigma[\chi_{c0}]:\sigma[\chi_{c1}]:\sigma[\chi_{c2}] = 
1:3:5$, whereas deviations from the predicted ratio for $\chi_{c1}$ 
and $\chi_{c2}$ have been observed. Moreover, it is unable to describe the observed variation 
-- from one process to another -- of the production ratios for charmonium states. For example, 
the ratio of the cross sections for $\chi_c$ and $J/\psi$ differs significantly in 
photoproduction and hadroproduction, whereas for the CEM these number are strictly constant. 

All these arguments make us think that despite its simplicity and its 
phenomenological reasonable grounds, this model is less reliable than the CSM. It is also
instructive to point out that the invocation of reinteractions after the $Q\bar Q$ pair 
production contradicts factorisation, which is albeit required when
the PDF are used. However, as we shall see now, the CSM has undergone in the nineties 
a lashing denial from the data.

\subsection{$\psi'$ anomaly}

In 1984, Halzen \etal\cite{Halzen:1984rq} noticed that charmonium production from 
a weak $B$ decay  could
be significant at high energy -- their prediction were made for $\sqrt{s}=540$ GeV -- 
and could even dominate 
at high enough $p_T$. This can be easily understood having in mind that the $B$ meson is produced
by fragmentation of a $b$ quark -- the latter dresses with a 
light quark --. And to produce only one $b$ quark at high  $p_T$ is not so burdensome; 
the price to pay is only to put one quark propagator off-shell, instead of two for $gg \to \psi g$. 

This idea was confirmed by the calculations of Glover \etal\cite{Glover:1987az}, 
which were used as a benchmark by the UA1 Collaboration\cite{Albajar:1990hf}. After the introduction
of a $K$ factor of 2, the measurements agreed with the predictions; however the $p_T$ slope 
 was not compatible with $b$-quark fragmentation simulations.

From 1992, the CDF collaboration undertook an analysis 
of $\psi$\footnote{In the following, $\psi$ stands for both $J/\psi$ and $\psi'$.} production. 
They managed to extract unambiguously the {\it prompt} component of the signal (not from $B$ 
decay) using a Silicon Vertex Detector (SVX)\cite{Abe:1992ww}\footnote{The details of the 
analysis are given in the following section.}.

The preliminary\footnote{The final results --confirming the preliminary ones-- were in fact 
published in 1997 \cite{CDF7997a}.} results\cite{:1994dx} showed an 
unexpectedly large {\it prompt} component. 
For the $\psi'$, the {\it prompt} cross section was orders of magnitude above the predictions of
the LO CSM (compare the data to the dashed curve on \cf{fig:dsdpt_braaten-94} (right)) . 
This problem was then referred to as the $\psi'$ anomaly.  For  
$J/\psi$, the discrepancy was smaller (\cf{fig:dsdpt_braaten-94} (left)), but it was conceivably blurred 
by the fact that a significant part of the production was believed to come from $\chi_c$ 
radiative feed-down, but no experimental results were there to confirm this hypothesis.

\begin{figure}[h]
\centering
\mbox{{\includegraphics[clip=true,width=0.5\textwidth]{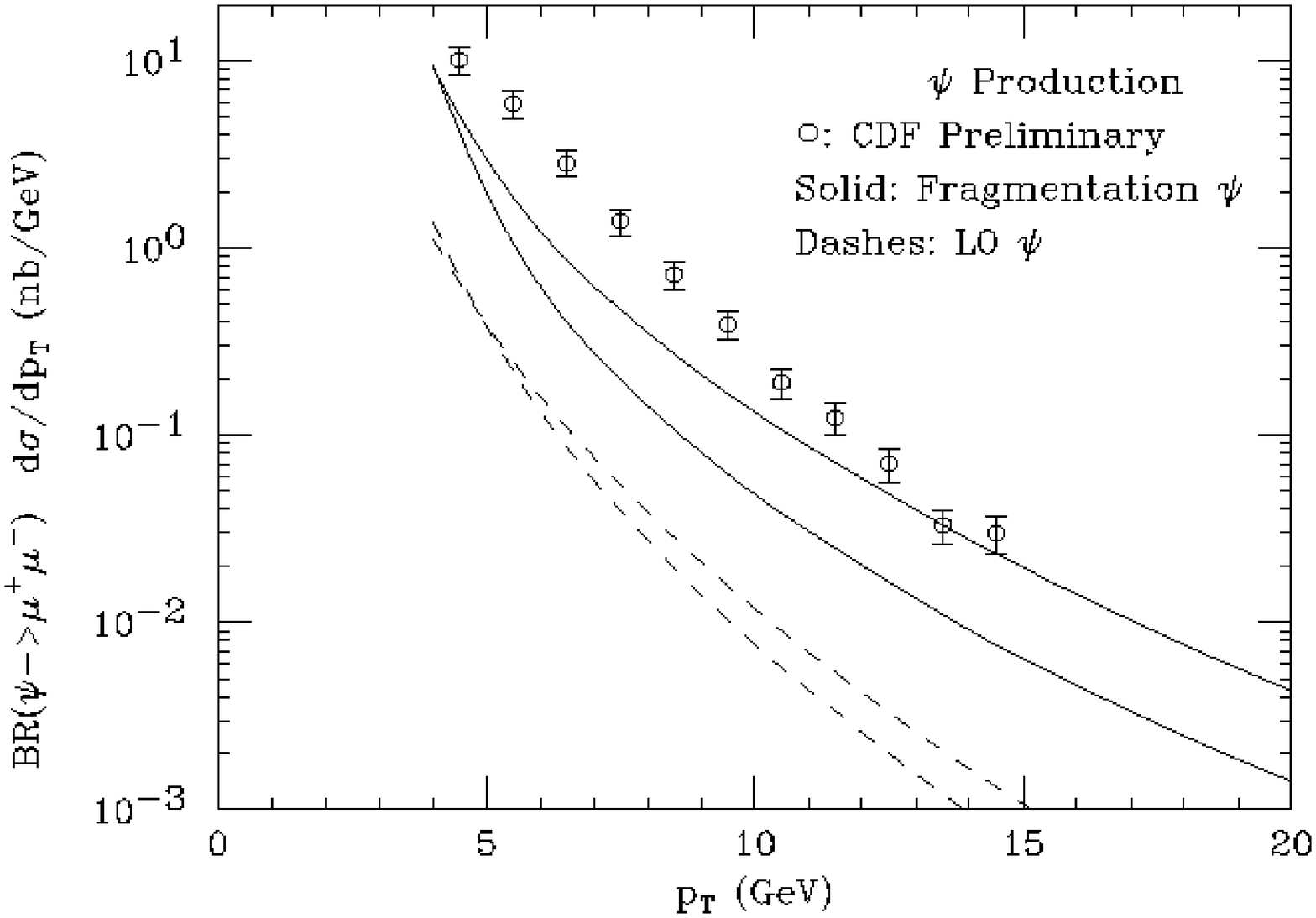}}
      {\includegraphics[clip=true,width=0.5\textwidth]{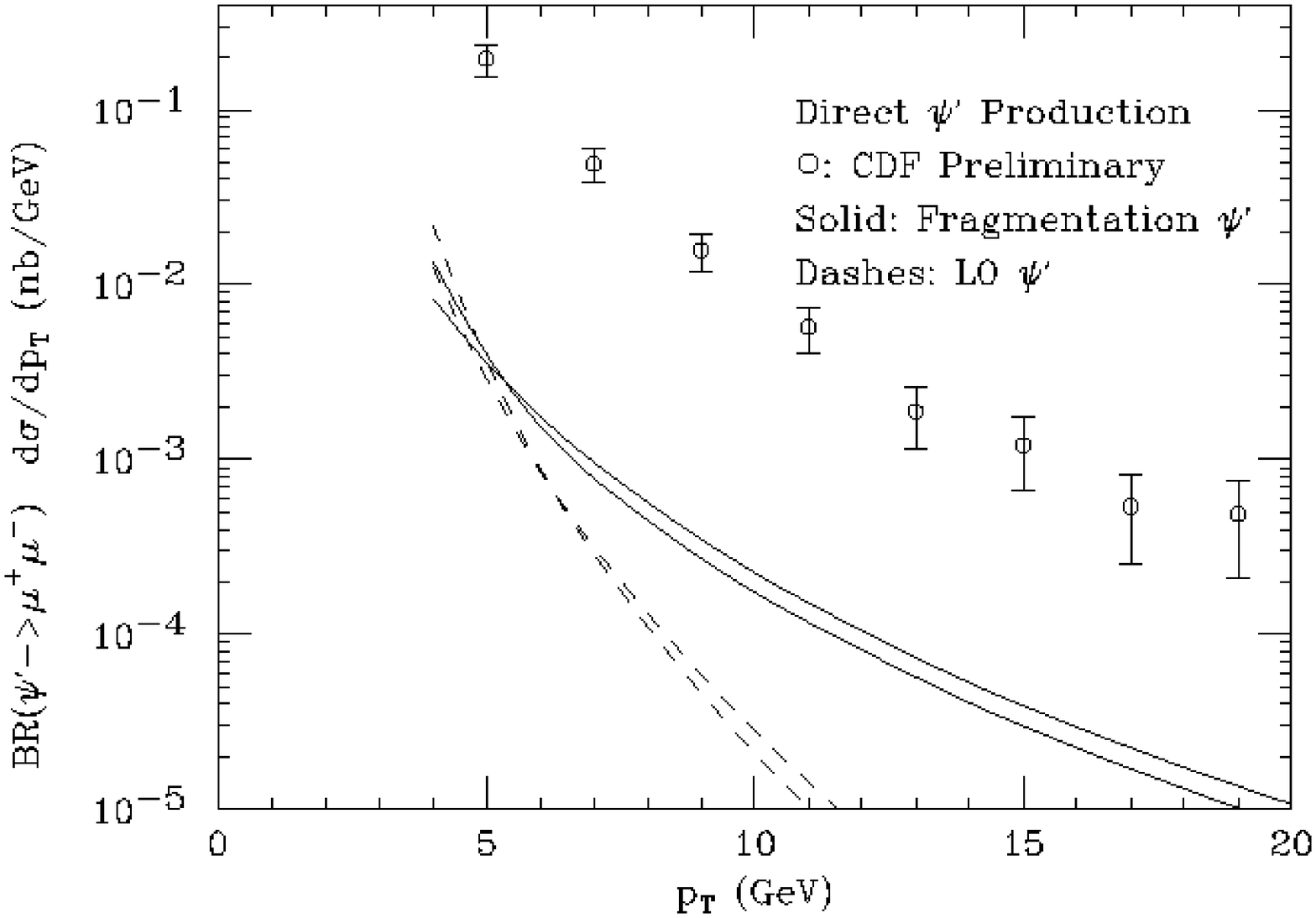}}}
\caption{ Comparison between preliminary measurements of Ref.~\protect\refcite{:1994dx} from CDF and the cross sections
 obtained by Braaten \etal~for LO CSM (dashed curves) 
of  $J/\psi$ (left) and $\psi'$ (right), as well as for CSM fragmentation contribution (solid curves). In 
each case, the two curves depict the extremum values obtained by varying parameters such as $m_c$ and 
the different scales: $\mu_R$, $\mu_F$, $\mu_{frag}$ 
(Reprinted figures from Ref.~\protect\refcite{Braaten:1994xb}
with permission of Elsevier. Copyright (1994).)}
\label{fig:dsdpt_braaten-94}
\end{figure}

\subsubsection{Fragmentation within the CSM: the anomaly confirmed}

The prediction from the LO CSM for {\it prompt} $\psi$ being significantly below 
these preliminary measurements
by CDF, Braaten and Yuan\cite{Braaten:1993rw} pointed out in 1993 that gluon fragmentation 
processes, even though of higher order in $\alpha_s$, were to prevail over the LO CSM for $S$-wave 
mesons at large $p_T$, in same spirit as the production from $B$ was dominant as it came from 
a fragmentation process. Following this paper, with Cheung and Fleming\cite{Braaten:1993mp}, 
they considered the fragmentation of $c$ quark into a $\psi$ in $Z_0$ decay. From this
calculation, they extracted the corresponding fragmentation function. In another 
paper\cite{Braaten:1994kd}, 
they considered gluon fragmentation into $P$-wave mesons. All the tools were then at hand 
for a full prediction of the prompt component of the $J/\psi$ and $\psi'$ at the Tevatron.
This was realised simultaneously by Cacciari and Greco \cite{Cacciari:1994dr} and by
Braaten \etal\cite{Braaten:1994xb}. Let us now review briefly the approach followed.

To all orders in $\alpha_s$, we have the following fragmentation cross section for a quarkonium
$\cal Q$:
\eqs{
\sigma_{{\cal Q}}(P) \simeq \sum_i\int^1_0 dz d \sigma_i(\frac{P}{z},\mu_{frag}) D_{i\to{\cal Q}}(z,\mu_{frag}).
}
The fragmentation scale, $\mu_{frag}$,  is as usual chosen to avoid large logarithms 
of $p_T/\mu_{frag}$ in 
$\sigma_i(\frac{P}{z},\mu_{frag})$, that is $\mu_{frag}\simeq p_T$. The summation of the corresponding
large logarithms of $\mu_{frag}/m_Q$ appearing in the fragmentation function is realised via
 an evolution equation\cite{Curci:1980uw,Collins:1981uw}.

\begin{figure}[H]
\centering
\includegraphics[width=8.5cm,clip=true]{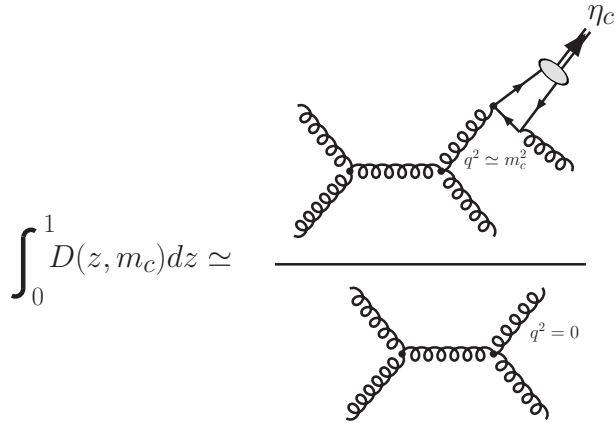}
\caption{Illustration of how to obtain $\int^1_0 D(z,m_c)$.}
\label{fig:etac_fragm_ratio_CSM}
\end{figure}

The interesting point raised  by Braaten and Yuan is that the fragmentation functions can be 
calculated perturbatively in $\alpha_s$ at the scale $\mu_{frag}=2m_Q$. For instance, in the case of gluon
fragmentation into an  $\eta_c$, the trick is to note that $\int^1_0 dz D_z(z,m_c)$ is the 
ratio to the rates for the well-known $gg\to gg$ process and
$gg\to \eta_c gg$ (\cf{fig:etac_fragm_ratio_CSM}). After some manipulations, 
the fragmentation function can be obtained from this ratio
by identifying the integrand in the $z$ integral. This gives :
\eqs{
D_{g\to\eta_c}(z,2m_c)= \frac{1}{6} \alpha^2_s(2m_c) \frac{|\psi(0)|^2}{m_c^3} [3z-2z^2+2(1-z)\ln(1-z)].
}

The other fragmentation functions of a given parton $i$ into a given quarkonium $\cal Q$, 
$D_{i\to{\cal Q}}(z,\mu_{frag})$, were obtained in the same spirit. For 
the Tevatron, the differential cross section versus $p_T$ of 
various CSM fragmentation processes are plotted 
in \cf{fig:dsdpt_divers} (left).

\begin{figure}[h]
\centering
\mbox{{\includegraphics[width=0.5\textwidth]{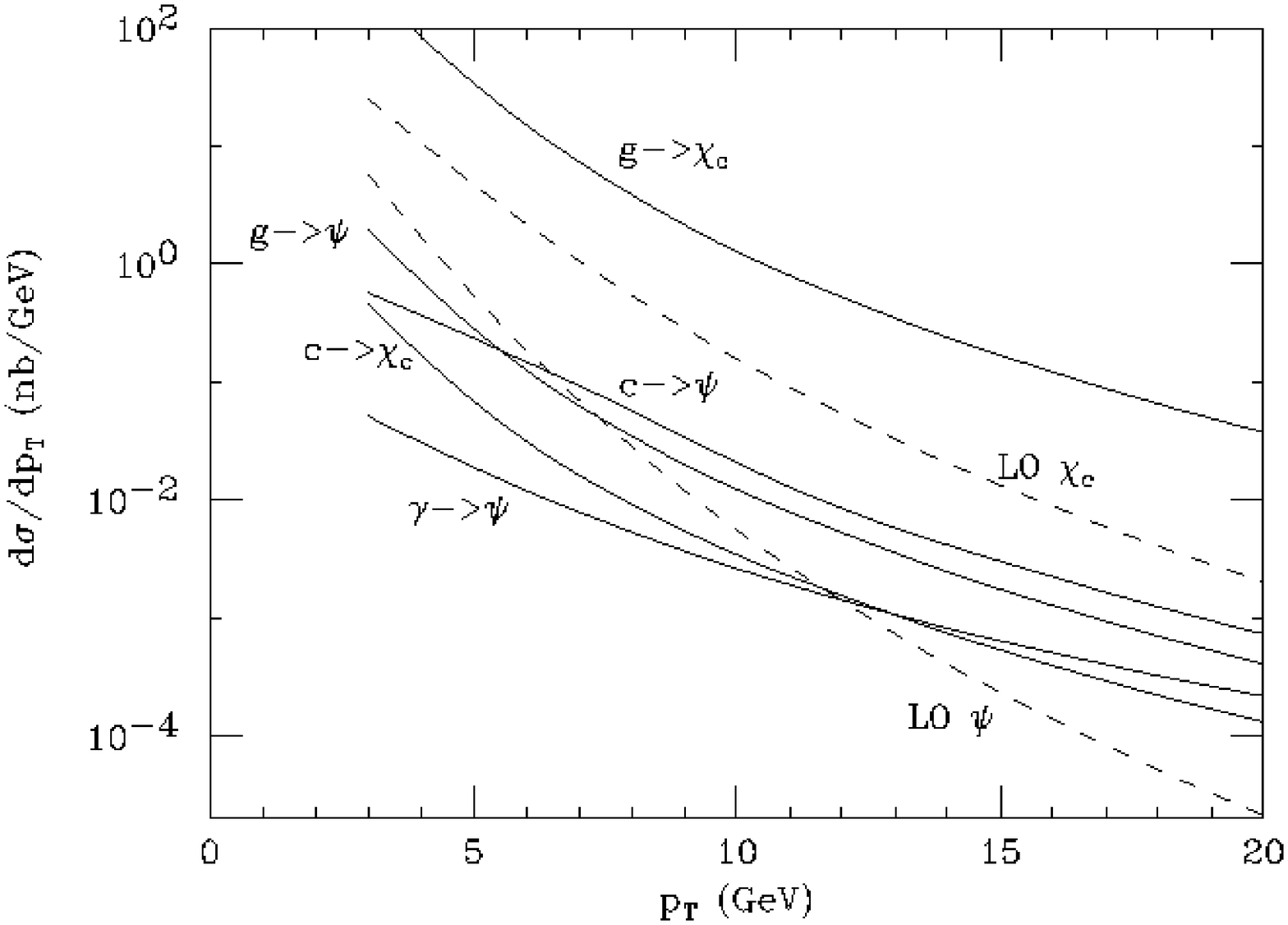}}
      {\includegraphics[width=0.5\textwidth]{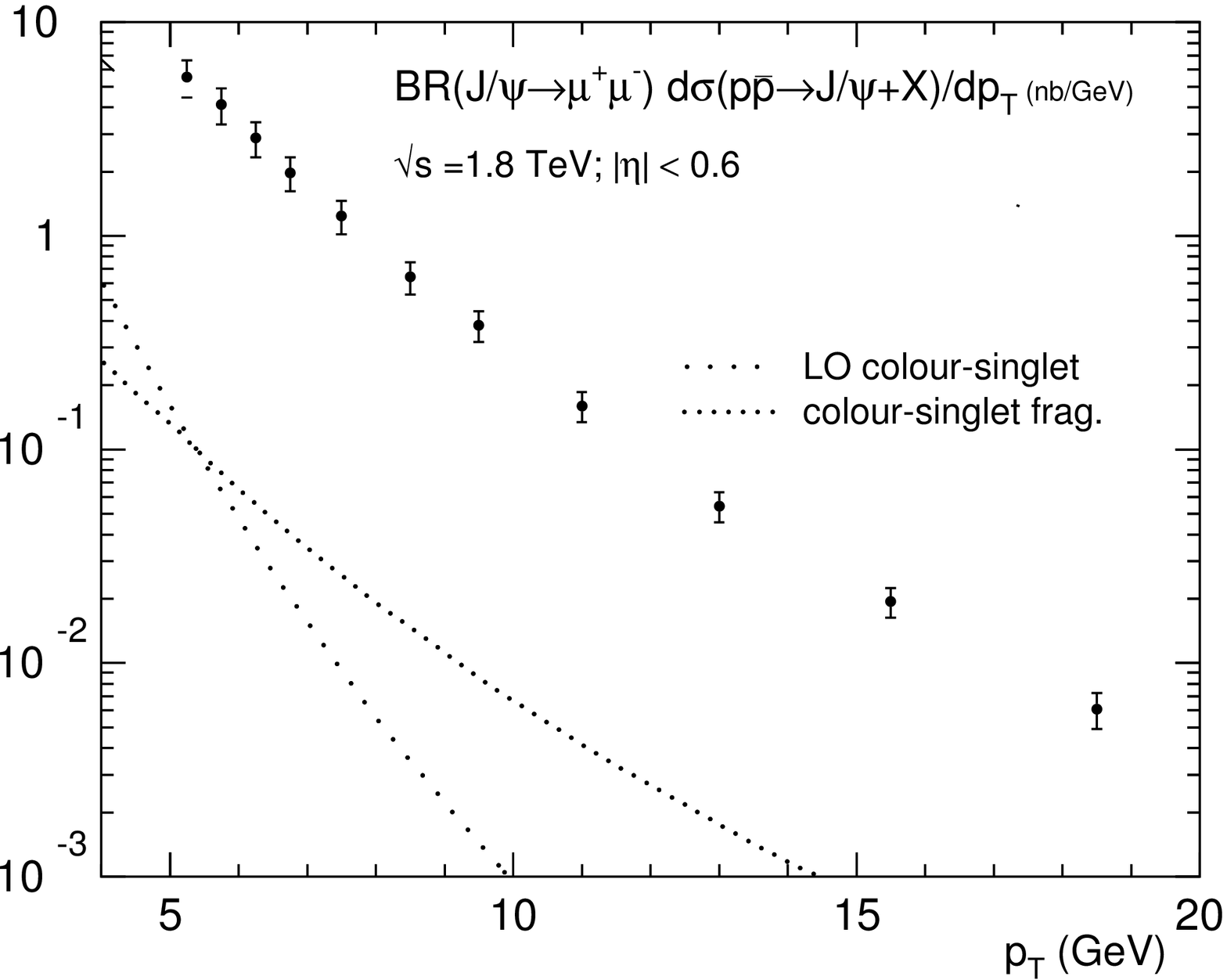}}}
\caption{ (left) Differential cross section versus $p_T$ of various CSM fragmentation processes 
for $J/\psi$ to be compared with the LO contributions 
(Reprinted figure from Ref.~\protect\refcite{Braaten:1994xb} with permission of Elsevier. Copyright (1994).). (right) 
Differential cross section 
versus $p_T$ of the CSM (fragmentation and LO) production 
to be compared with the direct production of $J/\psi$ from CDF
(Reprinted figure from Ref.~\protect\refcite{Kramer:2001hh} with permission of Elsevier. Copyright (2001).)}
\label{fig:dsdpt_divers}
\end{figure}

 The {\it prompt} component of the $J/\psi$ and the direct component of the $\psi'$ could 
in turn be obtained and compared with the preliminary data of CDF (see the solid curves
in plots in \cf{fig:dsdpt_braaten-94} above). For the $J/\psi$, the previous disagreement 
was reduced and could be accounted for by the theoretical and experimental 
uncertainties; on the other hand, for the $\psi'$, the disagreement
continued to be dramatic. The situation would be clarified by the extraction
of the direct component for $J/\psi$, for which theoretical uncertainties are reduced 
and are similar to those for the $\psi'$.

The CDF collaboration undertook the disentanglement of the direct $J/\psi$ signal\cite{CDF7997b}. 
They searched for $J/\psi$ associated with the photon emitted during this radiative decay: 
the result was a direct cross section 30 times above expectations from LO CSM plus fragmentation. 
This was the confirmation that the CSM was not the suitable model for heavy-quarkonium production
in high-energy hadron collisions.

It is a common misconception of the CSM to believe that the well-known factor 30 of discrepancy
between data and theory for direct production of $J/\psi$ arises when the data are compared 
with the predictions for the LO CSM following Baier, R\"uckl, Chang, \dots,\cite{CSM_hadron1,CSM_hadron2,CSM_hadron3} 
tuned to the right energy. As you can see on \cf{fig:dsdpt_divers} (right), the factor would be 
rather of two orders of magnitude at large $p_T$ 
for $J/\psi$. The same conclusion holds also for $\psi'$ (see \cf{fig:dsdpt_braaten-94} (right)).

It is also worth pointing that, in the CSM, the direct component of $\psi$ ($^3S_1$ charmonia) produced 
by fragmentation is mainly from $c$-quark fragmentation (see \cf{fig:dsdpt_divers} (left))
as soon as $P_T$ reaches 5 GeV and the $\alpha_s$ penalty of the gluon fragmentation is not 
compensated anymore by the $c$-quark mass. It was further pointed out in 2003 by Qiao~\cite{Qiao:2003pu} 
 that sea-quark initiated contributions could dominate in the fragmentation region (large $p_T$).

\section[Review of contemporary measurements from the Tevatron and RHIC]{Review of contemporary measurements of direct production of $J/\psi$, $\psi'$ and $\Upsilon$ 
from the Tevatron and RHIC}

\subsection{Foreword}

Limiting ourselves to high-energy collisions, the most recent published results for quarkonium
hadroproduction come from two accelerators:

\begin{enumerate}
\item The Tevatron at Fermilab which -- as stated by its name -- runs at TeV energy with 
proton-antiproton collisions. For Run I (``Run IA'' 1993-94 and ``Run IB'' 1995-96), the 
energy in c.m. was 1.8 TeV. For Run II, it has been increased to 1.96 TeV. The experimental 
analyses for this Run are still being carried out.
\item RHIC at BNL running at 200 GeV for the $J/\psi$ study with proton-proton collisions.
\end{enumerate}

It is worth pointing out here that high-energy  $p$-$\bar p$ and $p$-$p$ collisions 
give similar results for the same kinematics, due to the small contribution of 
valence quarks.

\subsection{Different types of production: prompt, non-prompt and direct}\label{sec:genmeson}

As we have already explained, the detection of quarkonia proceeds via the identification
of their leptonic-decay products. We give in~\ct{table:dimuon} the relative decay widths into
muons.

\begin{table}[H]
\centering\tbl{Table of branching ratios in dimuons (Ref.~\protect\refcite{pdg}).}{
\begin{tabular}{|l|l|}
\hline Meson & $\Gamma(\mu^+\mu^-)/\Gamma(total)$ \\
\hline \hline 
$J/\psi $        &  $0.0588 \pm 0.0010$  \\
$\psi(2S) $      &  $0.0073 \pm 0.0008$  \\
$\Upsilon(1S)$   &  $0.0248 \pm 0.0006$  \\
$\Upsilon(2S)$   &  $0.0131 \pm 0.0021$  \\
$\Upsilon(3S)$   &  $0.0181 \pm 0.0017$  \\
\hline
\end{tabular}}
\label{table:dimuon}
\end{table}

Briefly, the problem of direct $J/\psi$ production separation comprises three steps:
\begin{itemize}
\item muon detection;
\item elimination of $J/\psi$ produced by hadrons containing $b$ quarks ;
\item elimination of $\chi_c$ radiative-decay production;
\end{itemize}

Table 4 summarises the different processes to be discussed and the quantities linked.
\begin{table}[H]
\tbl{Different processes involved in $J/\psi$ production accompanied by quantities
used in the following discussion.}{\footnotesize
\begin{tabular}{|c|l|c|c|c|}
\hline $1^{st}$ step & $2^{nd}$ step &$3^{rd}$ step & Type & \begin{tabular}{c}Associated
\\ quantity \end{tabular}\\
\hline \hline 
& $c\bar c \to J/\psi$ &$-$&  Direct prod.& $ F(b \!\!\!/, \chi\!\!\!/ ,\psi\!\!\!/' )^{J/\psi}$\\$p\bar p \to c\bar c + X$ & $c\bar c \to \chi_c$ &  $\chi_c \to J/\psi+ \gamma$ & 
\begin{tabular}{c} \hbox{Prompt prod. by}\\\hbox{decay of}$\chi_c$ \end{tabular}
 & $ F(b \!\!\!/)_\chi^{J/\psi}$ \\ & $c\bar c \to \psi'$ &  $\psi' \to J/\psi+ X$ & 
\begin{tabular}{c} 
\hbox{Prompt prod. by}\\\hbox{decay of} $\psi'$ \end{tabular}
 & $ F(b \!\!\!/)_{\psi'}^{J/\psi}$ \\
\hline 
$\begin{array}{c}
p\bar p \to \bar bc +X\\  
\bar bc\to\bar c  c + \ell^-+\bar\nu_\ell
\end{array}$  
&$\begin{array}{ll}
\\ c\bar c \to J/\psi \\
c\bar c \to \chi_c
\end{array}$
&  $\begin{array}{cl} \\ - \\
\chi_c \to J/\psi+ \gamma
\end{array}$
& $\begin{array}{ll}
\\-\\-\end{array}$
& $\begin{array}{c}\\ f_b (or F_b) \\
\end{array}$
\\
etc \dots& & & &\\
\hline
\end{tabular}}
\label{table:jpsi00}
\end{table}
Let us explain the different fractions that appear in the table: 
\begin{itemize}
\item $ F(b \!\!\!/, \chi\!\!\!/,\psi\!\!\!/ (2S) )^{J/\psi}$ is the {\it prompt}  fraction of $J/\psi$ 
that do not come $\chi$, neither from $\psi(2S)$, \ie~the direct fraction.
\item $F(b \!\!\!/)_{\psi(2S)}^{J/\psi}$ is the {\it prompt} fraction of $J/\psi$ that
come from $\psi(2S)$. 
\item $F(b \!\!\!/)_\chi^{J/\psi}$ is the {\it prompt} fraction of $J/\psi$ that
come from $\chi$.
\item $F_b$ (or $f_b$) is the {\it non-prompt} fraction or equally the fraction that come for $b$ quarks.
\end{itemize}

Concerning $\psi(2S)$,  due to the absence of stable higher excited states likely to decay into it, we have
the summary shown in Table 5 for the different processes to be discussed and the quantities linked:
\begin{table}[H]
\tbl{Different processes involved in $\psi'$ production accompanied by quantities
used in the following discussion.}{\footnotesize
\begin{tabular}{|c|l|c|c|c|}
\hline $1^{st}$ step & $2^{nd}$ step &$3^{rd}$  step & Type & \begin{tabular}{c}Associated
\\ quantity \end{tabular}\\
\hline \hline 
 &  & &  
& \\
 $p\bar p \to c\bar c + X$  & $c\bar c \to \psi(2S)$ &  $-$ & 
Direct/Prompt prod.
 & $ F(b \!\!\!/)^{\psi(2S)}$\\    &  &   & 
 &  \\

\hline 
$\begin{array}{c}
p\bar p \to \bar bc +X\\  
\bar bc\to\bar c  c + \ell^-+\bar\nu_\ell
\end{array}$  
&$ c\bar c \to \psi(2S)$
&   $-$ 
& $-$
& $f_b (or F_b)$
\\
etc \dots& & & &\\
\hline
\end{tabular}
}
\label{table:psi200}
\end{table}
Let us explain the different fractions that appear in the corresponding table: 
\begin{itemize}
\item $F(b \!\!\!/)^{\psi(2S)}$ is the {\it prompt}  fraction of $\psi(2S)$, \ie~the direct production.
\item $F_b$ (or $f_b$) is the {\it non-prompt} fraction or equally the fraction that come for $b$ quarks.
\end{itemize}

 As can be seen in \ct{table:dimuon},
the leptonic branching ratio of $\Upsilon$ are also relatively high. The detection and 
the analysis of the bottomonia are therefore carried out in the same fashion. 
For the extraction of the direct production, the $b$-quark feed-down is 
obviously not relevant, only the decays from stable higher resonances
of the family are to be considered.

All the quantities useful for the bottomonium discussion are summarised in Table 6: 
\begin{table}[H]
\tbl{Different processes involved in $\Upsilon(nS)$ ($n=1,2,3$) production accompanied by quantities
used in the following discussion.}{\footnotesize
\begin{tabular}{|c|l|c|c|c|}
\hline $1^{st}$ step & $2^{nd}$ step &$3^{rd}$ step & Type & \begin{tabular}{c}Associated
\\ quantity \end{tabular}\\
\hline \hline 
$p\bar p \to b\bar b + X$ & $b\bar b \to \Upsilon(nS)$ &$-$          &  
Direct prod.& $F_{direct}^{\Upsilon(nS)}$\\
&  $b\bar b \to \chi_b$  & $\chi_b \to  \Upsilon(nS)+ \gamma$  & 
\begin{tabular}{c} \hbox{Prod. by }  \\ \hbox{decay of } $\chi_b$ \end{tabular}
 & $F_{\chi_b}^{\Upsilon(nS)}$ \\    & $b\bar b \to \Upsilon(n'S)$ &  $\Upsilon(nS) \to \Upsilon(n'S)+ X$  & 
\begin{tabular}{c} 
\hbox{Prod. by}  \\\hbox{decay of } \\$\Upsilon(n'S)$ \end{tabular}
 &  $F_{\Upsilon(n'S)}^{\Upsilon(nS)}$ \\
\hline
\end{tabular}
}
\label{table:upsi00}
\end{table}
Let us explain the different fractions that appear in the latter table: 
\begin{itemize}
\item  $F_{direct}^{\Upsilon(nS)}$ is the direct  fraction of $\Upsilon(nS)$.
\item  $F_{\chi_b}^{\Upsilon(nS)}$ is the fraction of $\Upsilon(nS)$ that come from $\chi_b$. 
\item $F_{\Upsilon(n'S)}^{\Upsilon(nS)}$ is the  fraction of $\Upsilon(nS)$ that come from a higher $\Upsilon(n'S)$.
\end{itemize}

\subsection{CDF analysis for $\psi$ production cross sections}

The sample of $p\bar p$ collisions amounts to $17.8 \pm 0.6$ pb$^{-1}$ 
at $\sqrt{s}=1.8$ TeV~\cite{CDF7997a}. For $J/\psi$, the considered sample consists however of
 $15.4 \pm 0.6$ pb$^{-1}$  of integrated luminosity\footnote{This difference
is due to the subtraction of data taken during a period of reduced
level 3 tracking efficiency. These data have however been taken into account
for $\psi(2S)$ after a correction derived from the $J/\psi$ sample.}.

The number of $\psi$ candidates is therefore determined by fitting
the mass distribution of the muons in the c.m. frame after subtraction of the  
noise. The mass distribution is fit to the signal shape fixed
by simulation and to a linear background. The fit also yields the mass of
the particle and a background estimate.

From the branching ratio and after corrections due to the experimental efficiency, 
the number of produced particles is easily obtained from the number of the candidates.

For the present study of CDF, the fits are reasonably good for each $p_T$-bin,
the $\chi^2$ per degree of freedom ranging from 0.5 to 1.5. The measured width
of the mass peak was from 17 MeV to 35 MeV for $p_T$ from 5 to 20 GeV.

Approximatively, 22100 $J/\psi$ candidates and 800 $\psi'$ candidates above
a background of 1000 events are observed. The $J/\psi$ efficiency is $97.0 \pm 0.2 \%$
and the  $\psi'$ one is $92.3 \pm 0.2 \%$\cite{CDF7997a}.

The integrated cross sections are measured to be 
\begin{equation}
\sigma(J/\psi)\cdot{\cal B}(J/\psi \to \mu^+\mu^-)= 17.4 \pm 0.1 (stat.)^{+2.6}
_{-2.8}(syst.) \ { \rm nb}
\end{equation}

\begin{equation}
\sigma(\psi(2S))\cdot{\cal B}(\psi(2S) \to \mu^+\mu^-)= 0.57 \pm 0.04 (stat.)^
{+0.08}_{-0.09}(syst.) \ { \rm nb}.
\end{equation}

\ 

\noindent where $\sigma(\psi)\equiv \sigma(p\bar p\to \psi X, p_T(\psi)>5 \hbox{ GeV},
|\eta(\psi)|<0.6)$.

\subsubsection{Disentangling prompt charmonia}\label{subsec:psipr}

As already seen, prompt $\psi$'s are the ones which do not come from the decay of $B$ mesons.
Their production pattern has the distinctive feature, compared to non-prompt ones, that
there exists a measurable distance between the $B$ production vertex and its decay into
charmonium.

To proceed, CDF uses the SVX, whose resolution is $40 \mu$m, whereas 
$B$ lifetime is $c\tau_B\approx 450\mu$m. Muons are constrained to come from the 
same point which is called the secondary vertex, as opposed to the 
primary vertex, that is the collision point
of protons. Then the projection of the distance between these two vertices on the
$\psi$ momentum, $L_{xy}$, can be evaluated. It is converted into a proper time equivalent quantity
using the formula, $ c\tau = \frac{L_{xy}}{{p_T(\psi)\over m(\psi)}\cdot F_{corr}}$, where
$F_{corr}$ is a correction factor, estimated by Monte-Carlo simulations, which links 
the $\psi$ boost factor  $\beta_T\gamma$ to that of the $B$~\cite{CDF7193}.

The prompt component of the signal is parametrised by  $c\tau=0$ (a single vertex), 
the component coming from $B$ decay is represented by an exponential, whose lifetime is  
$\tau_b$ and which is convoluted with the resolution function.

The $c\tau$ distribution  is fit in each $p_T$-bin with an unbinned 
log-likelihood function. The noise is allowed to vary within the normalisation uncertainty extracted
from the sidebands. The fraction of $\psi$ coming from $b$, $f_b(p_T)$, obtained by CDF is 
displayed as a function of $p_T$ in \cf{fig:prompt_fB}.

\begin{figure}[h]
\centering
 \includegraphics[height=6.5cm]{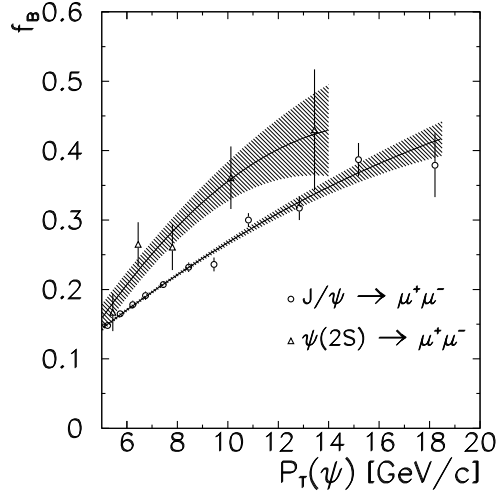}
 \caption{ Fraction of $\psi$ from $B$ decay as a function of $p_T$ 
(Reprinted figure from Ref.~\protect\refcite{CDF7997a} with permission of American Physical Society. Copyright (1997)).}
\label{fig:prompt_fB}
\end{figure}

The $\psi$ production cross section from $B$ decays is thus extracted by 
multiplying\footnote{In order to reduce statistical fluctuations, $f_b$ is fit by a parabola
weighted by the observed shape of the cross section\cite{CDF7193}.} $f^{fit}_b(p_T)$
by the inclusive $\psi$ cross section. Multiplying the latter by $(1-f^{fit}_b(p_T))$, 
one obviously gets the prompt-production cross section (cf. \cf{fig:prompt_result}).

\begin{figure}[h]
\centering
 \mbox{{\includegraphics[width=0.5\textwidth]{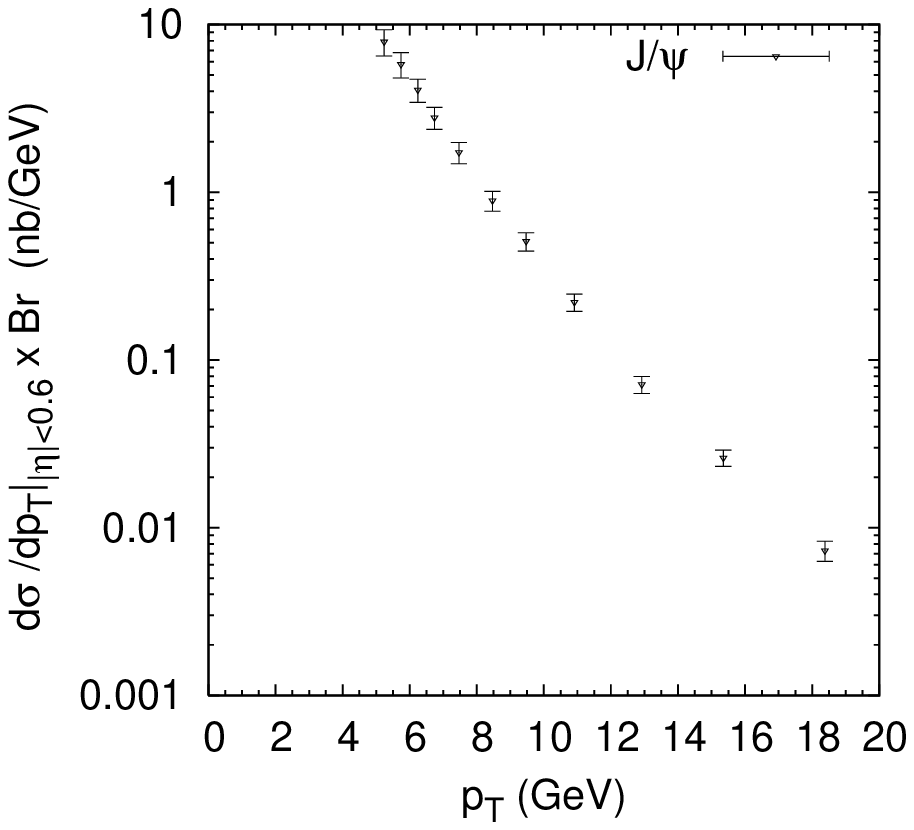}}
      {\includegraphics[width=0.5\textwidth]{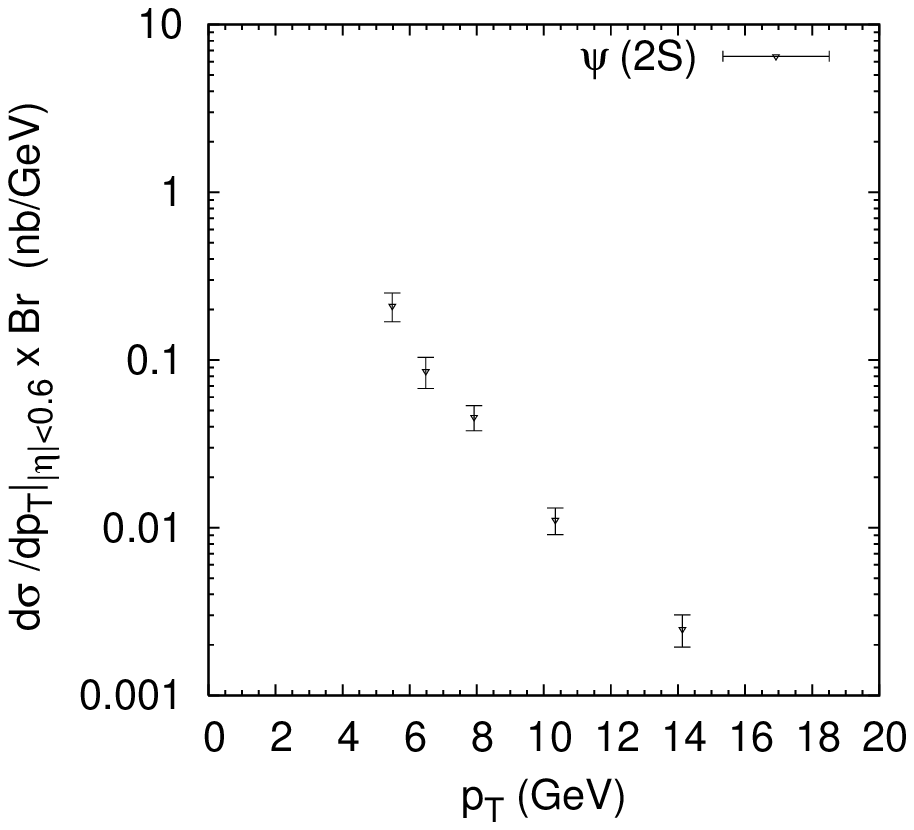}}}
 \caption{$\frac{d\sigma}{dp_T}{\cal B}$ from the prompt component of $\psi$ (data from Ref.~\protect\refcite{CDF7997a}).}
\label{fig:prompt_result}
\end{figure}

We remind the reader than for $\psi(2S)$ the prompt production identifies with the direct one.

\subsubsection{Disentangling the direct production of $J/\psi$}\label{subsec:psidir}

The problem here is to subtract the $J/\psi$ coming from $\chi_c$ decay, assuming 
that this is the only source of prompt $J/\psi$ besides the direct production after
subtraction of $ F(b \!\!\!/)_{\psi(2S)}^{J/\psi}$, the {\it prompt} fraction of $J/\psi$ that
come from $\psi(2S)$. The latter is evaluated by CDF from the $\psi(2S)$ cross section from the previous
section and from Monte-Carlo simulation of the decays $\psi(2S)\to~J/\psi X$ where $X=\pi\pi$, $\eta$
and $\pi^0$. The delicate point here is the detection of the photon emitted during the 
radiative decay of the $\chi_c$. 

The sample they use is 34367 $J/\psi$ from which $32642\pm 185$ 
is the number of real $J/\psi$ when the estimated background is removed. 

The requirements to select the photon were as follow:

\begin{itemize}
\item an energy deposition of at least 1 GeV in a cell of the central electromagnetic calorimeter;
\item a signal in the fiducial volume of the strip chambers (CES);
\item the absence of charged particles pointing to the photon-candidate cell (the no-track cut).
\end{itemize}

The direction of the photon is determined from the location of the signal in the strip chambers and from 
the event interaction point. All combinations of the $J/\psi$ with all photon candidates 
that have passed these tests are made and the invariant-mass difference defined as  
$\Delta M = M(\mu^+\mu^-\gamma)-M(\mu^+\mu^-)$
can then be evaluated. As expected, the distribution $\Delta M$ shows of a clear peak 
from $\chi_c$ decays is visible near $\Delta M= 400$ MeV. Yet, distinct 
signals for $\chi_{c,1}$ and $\chi_{c,2}$ are not resolved as the two states are separated by 45.6~MeV 
and as the mass resolution of the detector is predicted to be respectively 50 and 55~MeV.

Eventually, the $\Delta M$ distribution obtained from the data is fit with a 
gaussian and with the background fixed by the procedure explained above but with a free normalisation. 
The parameter of the gaussian then leads to the number of signal events:  $1230 \pm 72$ $\chi_c$.

The analysis of the direct $J/\psi$ signal is done within four $p_T$-bins:  
$ 4 < p_T^{J/\psi}<6$,  $ 6 < p_T^{J/\psi} < 8$ ,  $8 < p_T^{J/\psi}< 10$ and  $  p_T^{J/\psi}> 10$ GeV. 
For $p_T^{J/\psi} > 4.0$ GeV and $|\eta^{J/\psi}|<0.6$,  CDF finds that the fraction of $J/\psi$ from $\chi_c$ is then
\begin{equation}
F_{\chi}^{J/\psi} =27.4 \% \pm 1.6\%(stat.) \pm 5.2\% (syst.)
\end{equation}

The last step now is the disentanglement of the prompt $\chi_c$ production, that is the determination of 
$F(b \!\!\!/)_\chi^{J/\psi} $. Let us here draw the reader's attention to the fact that by
selecting prompt $J/\psi$ (cf. section~\ref{subsec:psipr}), $J/\psi$ produced by {\it non-prompt} $\chi_c$
have also been eliminated; it is thus necessary to remove only the prompt production by $\chi_c$ decay, and
nothing else. This necessitates the knowledge of $F(b \!\!\!/)_\chi^{J/\psi}$, the {\it prompt} fraction of $J/\psi$ that come from $\chi$.

The latter is calculated as follows:
\eqs{
F(b \!\!\!/)_\chi^{J/\psi}=F_\chi^{J/\psi}\frac{1-F^\chi_b}{1-F^{J/\psi}_b},
}
where $N_b^\chi$, $N_b^{J/\psi}$ are the number of reconstructed $\chi_c$ and $J/\psi$ from $b$'s, 
 $F_b^\chi$, $F_b^{J/\psi}$ are the corresponding fractions.

$F_b^{J/\psi}$ (or $f_b$) is known as seen in the section~\ref{subsec:psipr}; $F_b^\chi$
 is obtained in same way and is $17.8 \% \pm 0.45\%$ for $p_T>4.0$ GeV. Consequently,
\begin{equation}
F(b \!\!\!/)_\chi^{J/\psi} =29.7 \% \pm 1.7\%(stat.) \pm 5.7\% (syst.)
\end{equation}
and its evolution as a function of $p_T$ is shown in \cf{fig:direct_jpsi} (left) . It is also 
found that the fraction of directly produced  $J/\psi$ is
\begin{equation}\label{eq:jpsi_dir_prod_rate}
 F_{direct}^{J/\psi}=64 \% \pm 6 \%, 
\end{equation} 
and is almost constant from 5 to 18 GeV in $p_T$ (see \cf{fig:direct_jpsi} (left)). 
We therefore conclude from the analysis of CDF that the direct production is the 
principal contribution to $J/\psi$.

\begin{figure}[h]
\centering
 \mbox{{\includegraphics[width=0.5\textwidth]{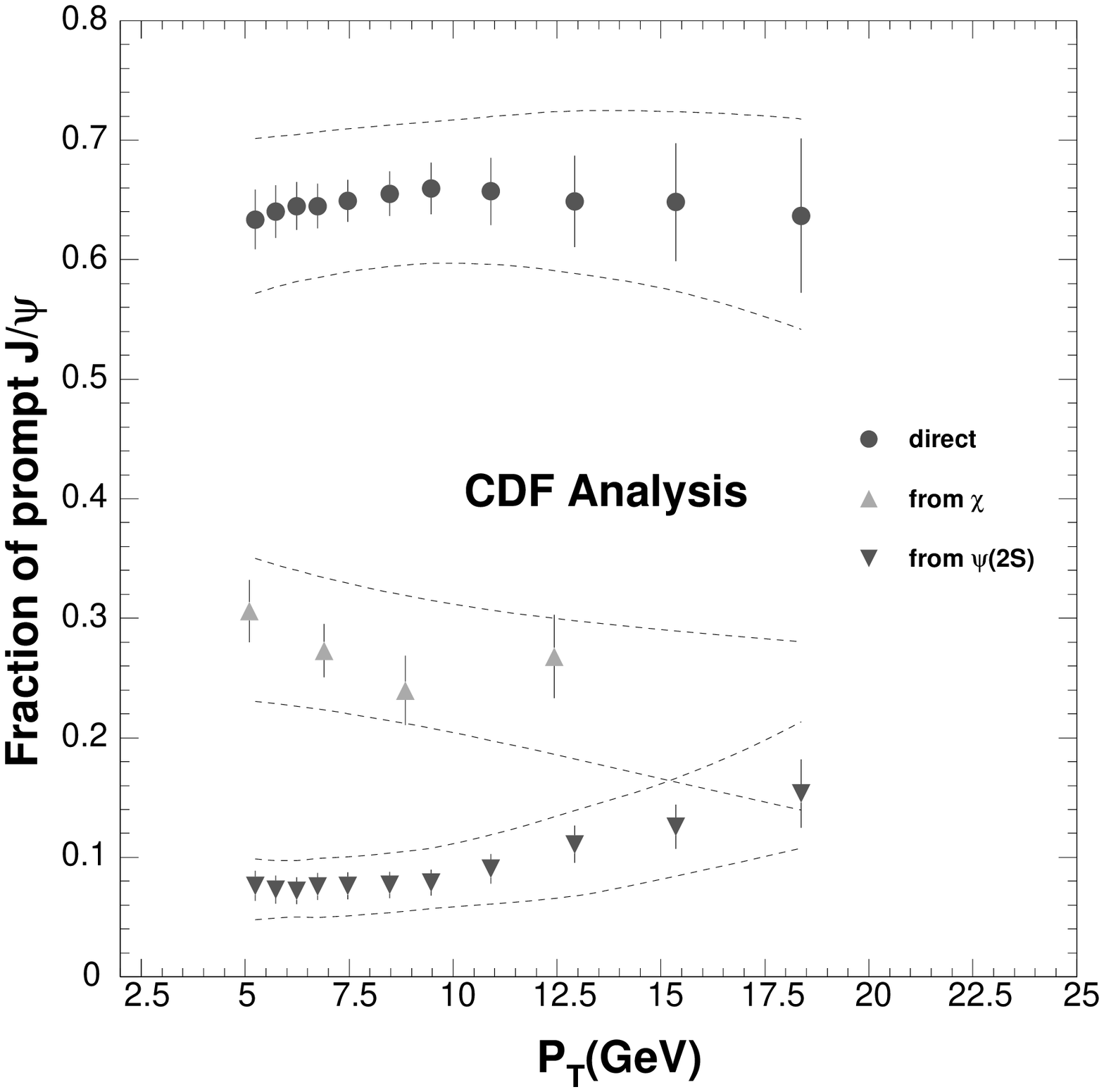}}
      {\includegraphics[width=0.5\textwidth]{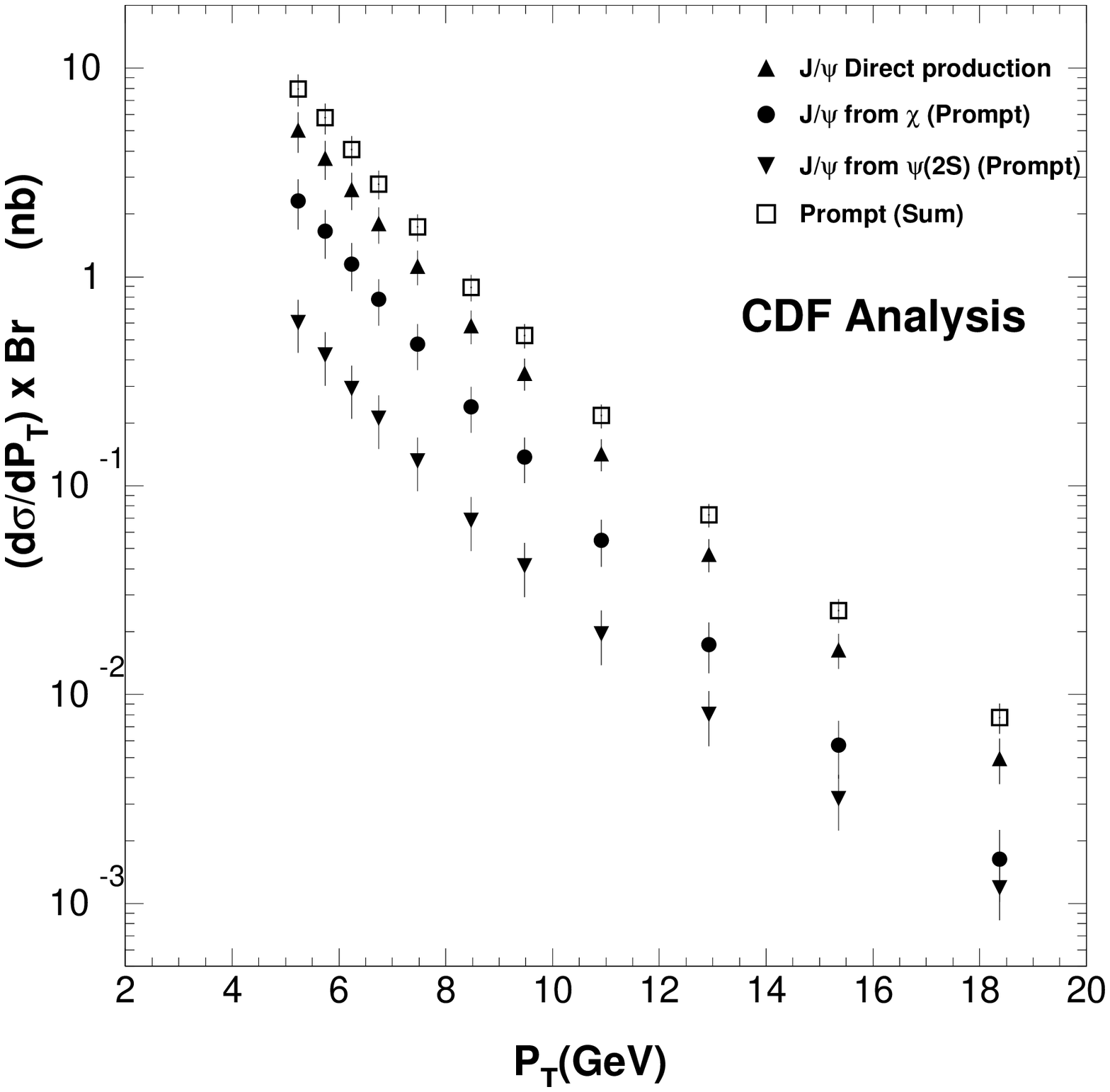}}}
 \caption{(left) Fractions of $J/\psi$ with the contribution of
$b$'s removed. The error bars correspond to statistical 
uncertainty. The dashed 
lines show the upper and lower bounds corresponding to the statistical and systematic
uncertainties combined (from~\protect\refcite{new1}). (right) Differential cross section for prompt  production of 
$J/\psi\to \mu^- \mu^+$ as a function of $p_T$ (Reprinted figure from Ref.~\protect\refcite{CDF7997b} with permission of American Physical Society. Copyright (1997)).}
\label{fig:direct_jpsi}
\end{figure}

In order to get the cross section of direct $J/\psi$ production, it is sufficient now
to extract the contribution of $\psi'$ obtained by Monte-Carlo simulation and of $\chi_c$ obtained
by multiplying the cross section of prompt production by the factor $F(b \!\!\!/)_\chi^{J/\psi}$, 
which is a function of $p_T^{J/\psi}$. The different cross sections are displayed 
in~\cf{fig:direct_jpsi} (right).

\subsubsection{Prompt $J/\psi$ production at $\sqrt{s}=1.96~{\rm TeV}$}

The first results of the run II for prompt $J/\psi$ production at $\sqrt{s}=1.96$ TeV
have recently been published in Ref.~\refcite{Acosta:2004yw}.  They correspond to an
integrated luminosity of 39.7 pb$^{-1}$. The inclusive $J/\psi$ cross section
was measured for $P_T$ from 0 to 20 GeV and the prompt signal was extracted
from $P_T=1.25$~GeV. The rapidity domain is still $-0.6 < y < 0.6$.

We do not give here the details of the experimental analysis which is thoroughly exposed
in Ref.~\refcite{Acosta:2004yw}. The prompt-signal extraction follows the same lines as
done for the analysis previously exposed. The prompt $J/\psi$ cross section obtained
is plotted in~\cf{fig:dsdpt_jpsi_prompt-run2}.

\begin{figure}[H]
\centering\includegraphics[height=7cm]{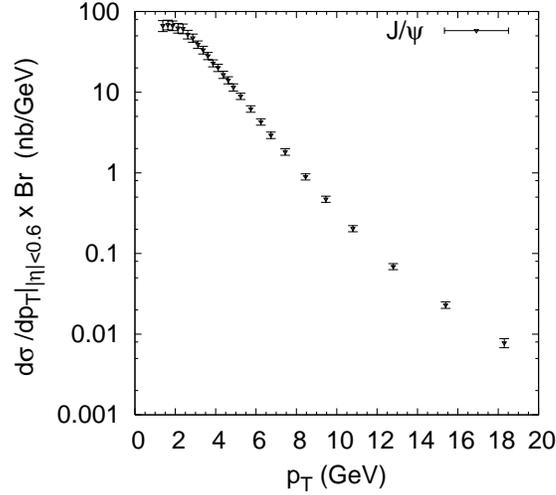}
 \caption{Prompt $J/\psi$ cross section as measured by CDF at $\sqrt{s}=1.96$ TeV (statistical and 
systematical errors have been combined quadratically for this plot); the data are 
from Ref.~\protect\refcite{Acosta:2004yw}.}
\label{fig:dsdpt_jpsi_prompt-run2}
\end{figure}

\subsection{CDF measurement of the $\Upsilon$ production cross sections}\label{sec:CDF_upsi_prod}

In  this section, the results by CDF on
 $\Upsilon$ production in $p\bar p$ at $\sqrt{s}=1.8$ TeV are exposed. These results were exposed 
in two Letters (Refs.~\refcite{CDF7595,Acosta:2001gv}), and we shall mainly focus on the second one, 
which considered data collected in 1993-95 and corresponding to an integrated luminosity 
of $77 \pm 3$ pb$^{-1}$. The number of candidates are $4430\pm 95$ for $\Upsilon(1S)$,
$1114\pm 65$ for $\Upsilon(2S)$ and $584\pm 53$ for $\Upsilon(3S)$.

The cross section for $\Upsilon(1S)$ is shown in \cf{fig:upsi-dsdpt} (left), 
for $\Upsilon(2S)$ in \cf{fig:upsi-dsdpt} (middle) and for $\Upsilon(3S)$ 
in \cf{fig:upsi-dsdpt} (right).

\begin{figure}[H]
\centering
 \mbox{{\includegraphics[width=0.3\textwidth]{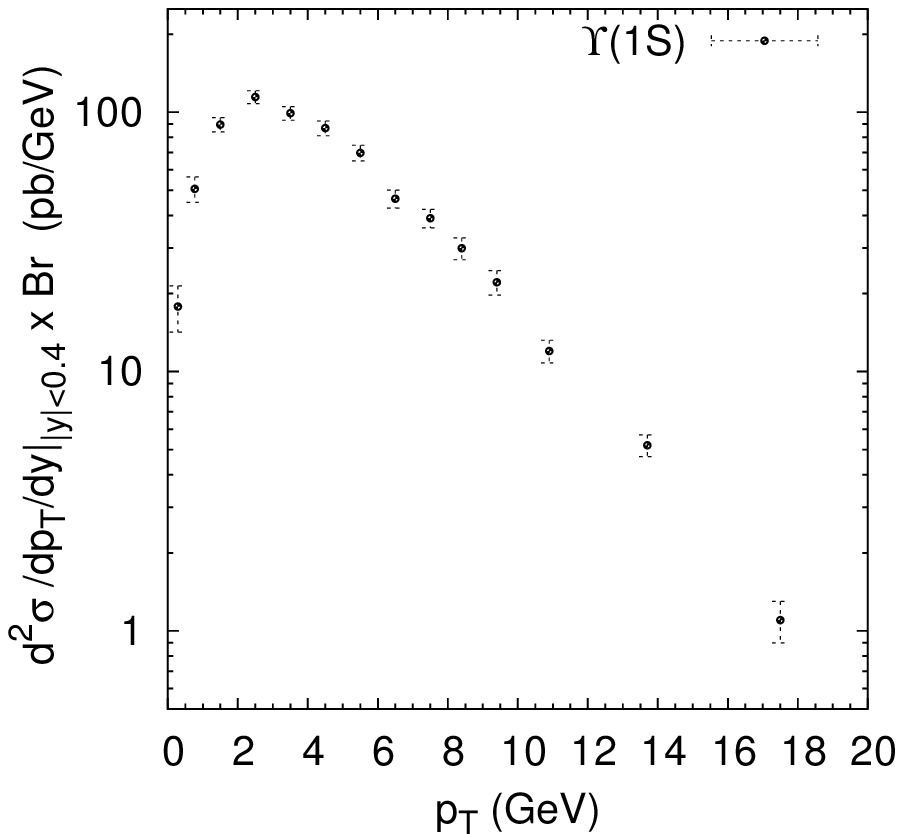}}
{\includegraphics[width=0.3\textwidth]{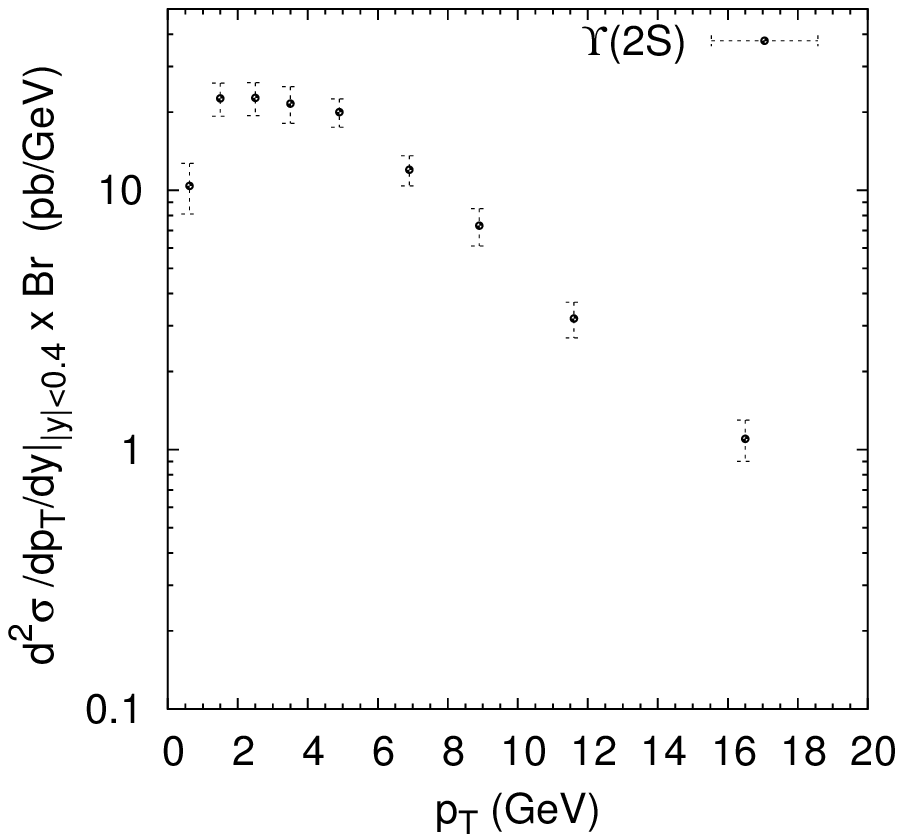}}
      {\includegraphics[width=0.3\textwidth]{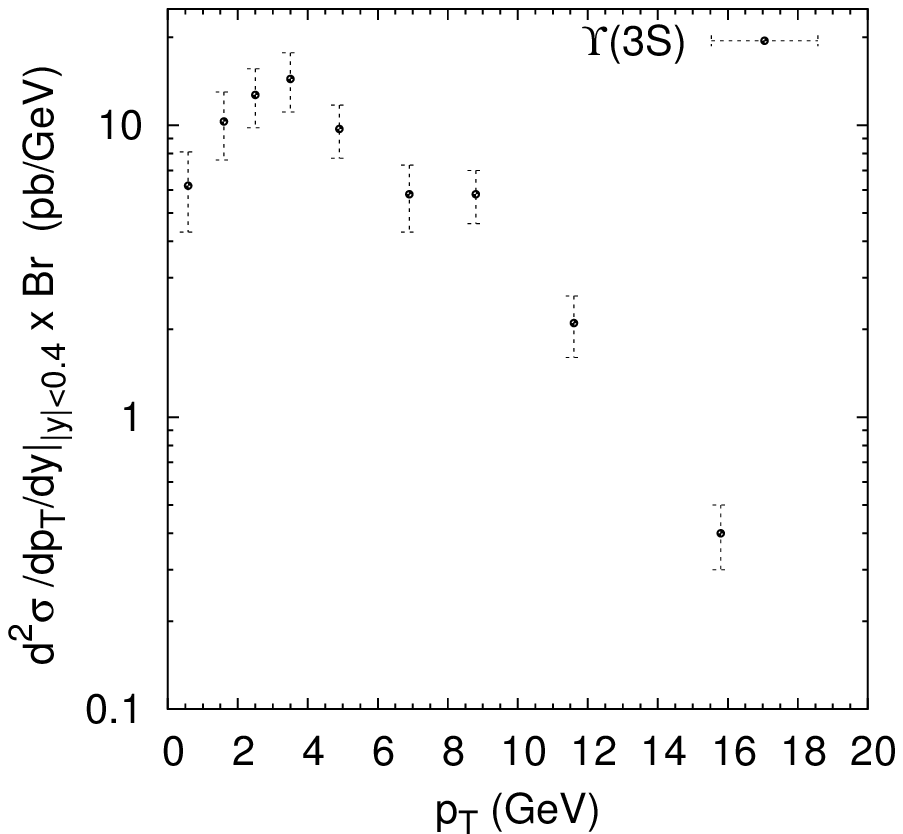}}}
\caption{left: Differential cross section of  
$\Upsilon(1S)\to \mu^- \mu^+$ as a function of $p_T$ for 
$|y|<0.4$. middle: Idem for $\Upsilon(2S) \to \mu^- \mu^+$. right: Idem for 
$\Upsilon(3S) \to \mu^- \mu^+$ (data from Ref.~\protect\refcite{Acosta:2001gv}).}
\label{fig:upsi-dsdpt}
\end{figure}

\subsubsection{Disentangling the direct production of $\Upsilon(1S)$}\label{subsec:upsidir}

The analysis\cite{Affolder:1999wm} presented here is for the most part the same as 
described in section~\ref{subsec:psidir}.
It is based on 90 pb$^{-1}$ of data collected during the 1994-1995 run. The measurement 
has been constrained to the range $p_T>8.0$ GeV because the energy of the photon emitted during the decay 
of $\chi_b$ decreases at low $p_T$ and ends up to be too small for the photon to be detected
properly. In the same spirit, analysis relative to $\Upsilon(2S)$ has not been carried out, once
again because of the lower energy of the radiative decay. Concerning $\Upsilon(3S)$, except for the
unobserved $\chi_b(3P)$ (which is nevertheless supposed to be below the $B\bar B$ threshold),
no states are supposed to be possible parents. 

A sample of 2186 $\Upsilon(1S)$ candidates is obtained from which $1462\pm 55$ 
is the estimated number of $\Upsilon(1S)$ after subtraction of the background. 
In the sample considered, photons likely to come from $\chi_b$ decay are selected as 
for the $J/\psi$ case, except for the energy deposition in the central electromagnetic 
calorimeter, which is lowered to 0.7 GeV.

For $p_T^{\Upsilon(1S)} > 8.0$ GeV and $|\eta^{\Upsilon(1S)}|<0.7$,  
the fractions of $\Upsilon(1S)$ from $\chi_b(1P)$ and  from  $\chi_b(2P)$ are measured by CDF to be
\eqs{
F_{\chi_b(1P)}^{\Upsilon(1S)} =&27.1 \% \pm 6.9\%(stat.) \pm 4.4\% (syst.),\\
F_{\chi_b(2P)}^{\Upsilon(1S)} =&10.5 \% \pm 4.4\%(stat.) \pm 1.4\% (syst.).
}

The feed-down from the $S$-waves $\Upsilon(2S)$ and $\Upsilon(3S)$ is obtained by Monte-Carlo 
simulations of these decays normalised to the production cross sections
discussed in section~\ref{sec:CDF_upsi_prod}. It is found that for $p_T^{\Upsilon(1S)} > 8.0$ GeV
the fraction of $\Upsilon(1S)$ from $\Upsilon(2S)$ and $\Upsilon(3S)$ are
respectively
\eqs{
F_{\Upsilon(2S)}^{\Upsilon(1S)} =10.1 \%^{+7.7}_{-4.8}\%,
F_{\Upsilon(3S)}^{\Upsilon(1S)} =0.8 \%^{+0.6}_{-0.4}\%.
}

Concerning the unobserved $\chi_b(3P)$, a maximal additional contribution 
is taken into account by supposing than all the $\Upsilon(3S)$ are due to 
$\chi_b(3P)$ and from theoretical expectation for the  decay of this state, 
a relative rate of $\Upsilon(1S)$ from $\chi_b(3P)$ can obtained. This rate
is less than $6\%$.

Eventually the fraction of directly produced $\Upsilon(1S)$ is found to be 
\eqs{
F_{direct}^{\Upsilon(1S)} &=50.9 \% \pm 8.2\%(stat.) \pm 9.0\% (syst.).
}


\subsection{Polarisation study}
\label{sec:polarisation}

As the considered bound states, $\psi$ and $\Upsilon$ are massive spin-1 particles, 
they have three polarisations. In addition to measurements of their 
momentum after their production, the experimental set-up of CDF is sufficiently refined
to provide us with a measurement of their spin alignment through an analysis of 
the angular distribution of the dimuon pairs from the decay.

The CDF collaboration has carried out two analyses, one for the $\psi$ 
states\cite{Affolder:2000nn} -- for which the feed-down
from $b$ decay has been subtracted, but not the indirect component in the case of the 
$J/\psi$ -- and another for $\Upsilon(nS)$~\cite{Acosta:2001gv}.

In the following, we shall proceed to a brief outline of the analysis 
and give its main results.

\begin{figure}[H]
\centerline{\includegraphics[width=5cm]{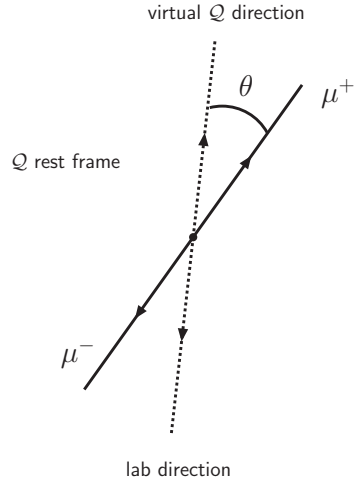}}
\caption{ Definition of the angle $\theta$ used in the polarisation analysis of a quarkonium
${\cal Q}$.}
\label{fig:pol_def_theta}
\end{figure}

The polarisation state of the quarkonium can be deduced from the angular dependence of
its decay into $\mu^{+}\mu^{-}$. Taking the spin quantisation axis along  the quarkonium 
momentum direction in the $p\bar{p}$ c.m. frame, we define $\theta$ as the angle  
between the $\mu^{+}$  direction in the quarkonium rest frame  
and the quarkonium  direction in the lab frame (see \cf{fig:pol_def_theta}). 
Then the normalised angular distribution $I(\cos \theta)$ is 
given\footnote{For a derivation, see the Appendix A of Ref.~\refcite{Cropp:2000zk}.} by
\eqs{
  I(\cos \theta) = \frac{3}{2(\alpha+3)}(1+\alpha \ \cos^2\theta)
 \label{def_alpha}
}
where the interesting quantity is 
\eqs{
\alpha=\frac{\frac{1}{2}\sigma_T-\sigma_L}{\frac{1}{2}\sigma_T+\sigma_L}.
}
$\alpha=0$ means that the mesons are unpolarised, $\alpha=+1$ corresponds to
a full transverse polarisation and $\alpha=-1$ to a longitudinal one.

As the expected behaviour is biased by muons cuts -- for instance there exists 
a severe reduction of the acceptance  as $\theta$ approaches 0 and 180 degrees, due 
to the $p_T$ cuts on the muons --, the method followed by CDF was to
compare measurements, not with a possible $(1+\alpha \ \cos^2\theta)$ distribution, 
but with distributions obtained after simulations of quarkonium decays taking account  
the geometric and kinematic acceptance of the detector as well as the reconstruction efficiency.

The parts of the detector used are the same as before, with the additional central muon upgrade (CMP) 
outside the CMU. 

\subsubsection{Study of the $\psi$'s polarisation by CDF} \label{subsec:psi_detect_pol}

We give here the results relative to the CDF analysis published in 2000~\cite{Affolder:2000nn}.
The data used correspond to an integrated luminosity of 110 pb$^{-1}$ collected
between 1992 and 1995. 

\paragraph{Disentangling prompt production for $\psi$}

The measured fraction of $J/\psi$ mesons which come from $b$-hadron decay,
$F^{J/\psi}_b$,  is measured to increase from $(13.0\pm 0.3)\%$ at $p_T = 4 $ GeV to 
$(40\pm 2)\%$ at 20 GeV and for $\psi'$ mesons, 
$F^{\psi'}_b$ is  $(21\pm2)\%$ at 5.5 GeV and $(35\pm4)\%$  at  20 GeV.

Within a 3-standard-deviation mass window around the $J/\psi$ peak, 
the data sample is of 180 000 $J/\psi$ events. In order to study the effect of $p_T$, 
the data are divided into seven $p_T$-bins from 4 to 20 GeV.
Because the number of $\psi'$  events is lower, data for $\psi'$  
are divided into three  $p_T$-bins from 5.5 to 20 GeV.

\paragraph{$J/\psi$ polarisation measurement}

The polarisation is obtained using
a $\chi^2$ fit of the data to a weighted sum of transversely 
polarised and longitudinally polarised templates. The weight obtained with the fit
provides us with the polarisation. Explanations relative to
 procedure used can be found in Refs.~\refcite{Affolder:2000nn} and~\refcite{Cropp:2000zk}. 

Note however that the polarisation is measured in each $p_T$ bin and that separate polarisation 
measurements for direct $J/\psi$ production and for production via $\chi_c$ and 
$\psi'$ decays was found to be unfeasible. Let us recall here that $\chi_c$ and 
$\psi'$ were shown to account for 36$\pm$6\% of the prompt production 
(cf.~\ce{eq:jpsi_dir_prod_rate}) and to be mostly constant in the considered $p_T$ range.

Except in the lowest $p_T$ bins, the systematic uncertainties are much smaller than 
the statistical one. The values obtained for $\alpha_{Prompt}$ and $\alpha_{from B}$ 
are given in \ct{tab:res_pol_psi1s} and plotted in \cf{fig:evol_alpha_jpsi}.

\begin{table}[H]
\tbl{Fit results for $J/\psi$ polarisation, with
statistical and systematic uncertainties (Ref.~\protect\refcite{Affolder:2000nn}).}{
\begin{tabular}{|c|c|c|c|} \hline
$p_T$ bin (GeV) & Mean $P_T$  (GeV) & $\alpha_{Prompt}$ & $\alpha_{from B}$ \\ \hline \hline
$4-5$   & 4.5  & $0.30\pm0.12\pm0.12$    & $-0.49\pm0.41\pm0.13$   \\ 
$5-6$   & 5.5  & $0.01\pm0.10\pm0.07$    & $-0.18\pm0.33\pm0.07$  \\ 
$6-8$   & 6.9  & $0.178\pm0.072\pm0.036$ & $0.10\pm0.20\pm0.04$  \\
$8-10$  & 8.8  & $0.323\pm0.094\pm0.019$ & $-0.06\pm0.20\pm0.02$  \\
$10-12$ & 10.8 & $0.26\pm0.14\pm0.02$    & $-0.19\pm0.23\pm0.02$  \\
$12-15$ & 13.2 & $0.11\pm0.17\pm0.01$    & $0.11\pm^{0.31}_{0.28}\pm0.02$  \\
$15-20$ & 16.7 & $-0.29\pm0.23\pm0.03$   & $-0.16\pm^{0.38}_{0.33}\pm0.05$  \\ \hline
\end{tabular}
\label{tab:res_pol_psi1s}
}
\end{table}

\begin{figure}[H]
  \centerline{\includegraphics[width=10cm]{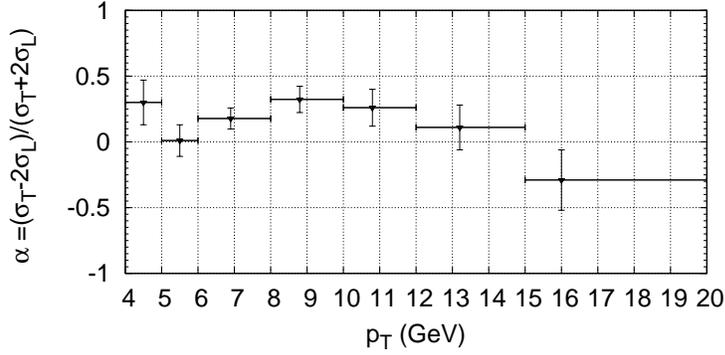}}
\caption{$\alpha_{Prompt}$ for  $J/\psi$ fit  for $|y| < 0.6$.
The error bars denote statistical and systematic uncertainties added in quadrature.}
\label{fig:evol_alpha_jpsi}
\end{figure}

\paragraph{$\psi'$ polarisation measurement}

The procedure to obtain $\alpha_{Prompt}$ and $\alpha_{from B}$ is similar. Weighted
simulations of the angular distribution are fit to the data and the weight obtained gives the 
polarisation. The $|\cos \theta |$ distributions in the two $c\tau$ sub-samples 
are fit simultaneously. As  there is no expected radiative decay from 
higher excited charmonia, we are in fact dealing with {\it direct} production.  

Anew, the systematic uncertainties\cite{Affolder:2000nn} are much smaller than 
the statistical uncertainties.  The values obtained for $\alpha_{Prompt}$ and $\alpha_{from B}$ 
are given in \ct{tab:res_pol_psi2s} and plotted in \cf{fig:evol_alpha_psi2s}.

\begin{figure}[H]
  \centerline{\includegraphics[width=10cm]{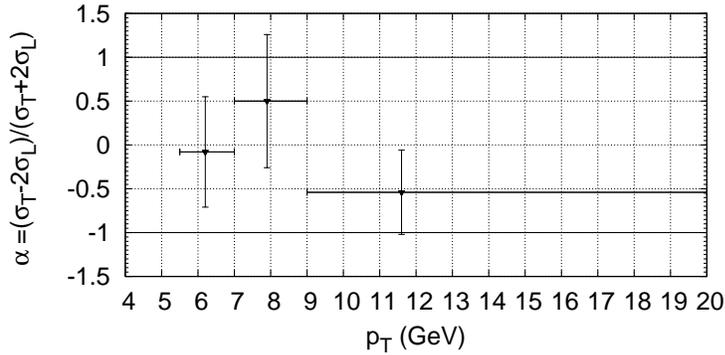}}
\caption{$\alpha_{Prompt}$ for  $\psi'$ fit  for $|y| < 0.6$.
 Error bars denote statistical and systematic uncertainties added in quadrature.}
\label{fig:evol_alpha_psi2s}
\end{figure}

\begin{table}[H]
\tbl{Fit results for $\psi(2S)$ polarisation, with
statistical and systematic uncertainties (Ref.~\protect\refcite{Affolder:2000nn}).}{
\begin{tabular}{|c|c|c|c|} 
\hline
$p_T$ bin (GeV) & Mean $p_T$  (GeV) & $\alpha_{Prompt}$ & $\alpha_{from B}$ \\ \hline\hline
$5.5-7.0$  & 6.2  & $-0.08\pm0.63\pm0.02$ & $-0.26\pm1.26\pm0.04$ \\
$7.0-9.0$  & 7.9  & $ 0.50\pm0.76\pm0.04$ & $-1.68\pm0.55\pm0.12$ \\
$9.0-20.0$ & 11.6 & $-0.54\pm0.48\pm0.04$ & $ 0.27\pm0.81\pm0.06$ \\ \hline
\end{tabular}
}
\label{tab:res_pol_psi2s}
\end{table}

\subsubsection{Study of the $\Upsilon(1S)$} \label{subsec:upsi_detect_pol}

The measurements are made in the region of $p_T$ from 0 to 20 GeV and with $|y|<0.4$ and the 
data are separated in four $p_T$-bins\cite{Acosta:2001gv}. In \ct{tab:res_pol_upsi1s} are given 
the results for $\alpha$ and the same values are plotted in \cf{fig:cdf_ups1s_pol}. 
Our conclusion is that $\Upsilon(1S)$ seems to be mostly produced unpolarised.

\begin{figure}[H]
\centerline{\includegraphics[width=10cm]{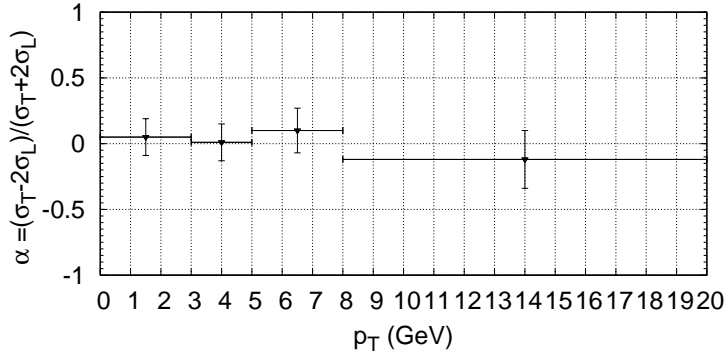}}
\caption{ $\alpha$ for  $\Upsilon(1S)$ fit  for $|y| < 0.4$.}
\label{fig:cdf_ups1s_pol}
\end{figure}

\begin{table}[H]
\tbl{Fit results for $\Upsilon(1S)$ polarisation (Ref.~\protect\refcite{Acosta:2001gv}).}{
\begin{tabular}{|c|c|} 
\hline
$p_T$ bin (GeV) & $\alpha$ \\ \hline\hline
$0.0-3.0$  & $+0.05\pm0.14$  \\
$3.0-5.0$  &  $+ 0.01\pm0.14$ \\
$5.0-8.0$ & $+0.10\pm0.17$ \\ 
$8.0-20.0$ & $-0.12\pm0.22$ \\ 
\hline
\end{tabular}
}
\label{tab:res_pol_upsi1s}
\end{table}

\subsubsection{New preliminary measurements by CDF for prompt $J/\psi$}

To complete the review of the measurements of quarkonium polarisation done by the CDF collaboration, we
give in~\cf{fig:pol_prelim} the preliminary one for prompt $J/\psi$  for RUN II with an integrated luminosity
of $188\pm 11$ pb$^{-1}$ for $J/\psi$ with $5 \leq P_T \leq  30$ GeV and 
$|y| \leq 0.6$.

\begin{figure}[h]
\centering
\includegraphics[height=6cm]{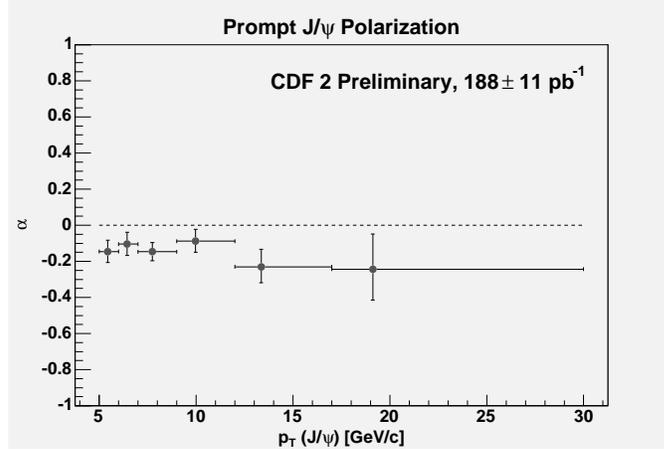}
\caption{Preliminary measurements of the prompt production of $J/\psi$ as measured by CDF for RUN II
(from Ref.~\protect\refcite{pol_prelim}).}
\label{fig:pol_prelim}
\end{figure}

\subsection{PHENIX analysis for $J/\psi$ production cross sections}

In this section, we present the first measurements of $pp\to J/\psi+ X$ at RHIC obtained
by the PHENIX experiment\cite{Adler:2003qs} at a c.m. energy of $\sqrt{s}=200$ GeV. The analysis
was carried out by detecting either dielectron or dimuon pairs. The data were taken  during
the run at the end of 2001 and at the beginning of 2002. The data 
amounted to 67 nb$^{-1}$ for $\mu^+\mu^-$  and\footnote{The difference results from different
cuts on the extrapolated vertex position.} 82 nb$^{-1}$ for $e^+e^-$.
The $B$-decay feed-down is estimated to contribute less than 4\% at $\sqrt{s}= 200$ GeV
and is not studied separately. The production is thus assumed here to be nearly totally
prompt, feed-down from $\chi_c$ is expected to exist though\cite{Nayak:2003jp,Klasen:2003zn}.

The net yield of $J/\psi$ within the region $ 2.80 \hbox { GeV} < 
M(pair) < 3.40 \hbox { GeV}$ was found to be $46.0 \pm 7.4$ for electrons, whereas
for muons, it was $65.0 \pm 9.5$ $J/\psi$  within the region 
$ 2.71 \hbox { GeV} < M(pair) < 3.67 \hbox { GeV}$.  The cross section as a function
of $p_T$ is shown for the two analyses on \cf{fig:dsdpt_PHENIX}.

\begin{figure}[H]
\centerline{\includegraphics[width=11cm]{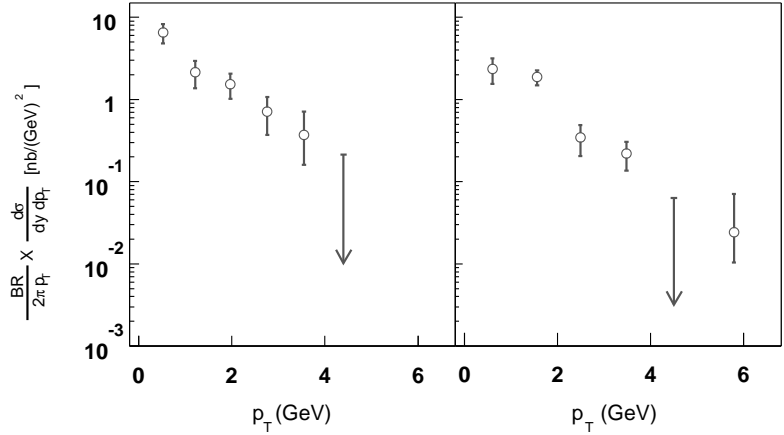}}
\caption{The differential cross section measured: left for dielectrons, right for dimuons (Ref.~\protect\refcite{Adler:2003qs}).}
\label{fig:dsdpt_PHENIX}
\end{figure}

\section{Review of contemporary models for production at the Tevatron and RHIC}

\subsection{Soft-Colour Interaction vs Colour Evaporation Model}\label{subsec:SCI}

Introduced\cite{Edin:1995gi,Edin:1996mw} as a new way to explain the observation of 
rapidity gaps in deep inelastic scattering at HERA, the Soft-Colour Interaction (SCI) model 
was also applied to quarkonium production, in particular in the context of  hadron-hadron 
collisions (Tevatron\cite{Edin:1997zb} and LHC\cite{Damet:2001gu}).

The main idea of the model is to take into consideration soft interactions occurring below a 
small momentum scale, which we shall call here $Q_0^2$,  in addition to those considered 
in the hard interaction through Feynman graphs. Unfortunately, we do not 
have satisfactory tools to deal with such soft QCD interactions. Nevertheless, the interesting point of
these interaction --emphasised by the SCI model-- lies in the fact that these  matter 
only for colour, since they cannot affect significantly the parton momenta. Therefore, one 
possibility is to implement them via Monte Carlo (MC) event generators by solely exchanging 
the colour state of two selected {\it softly interacting} partons with a probability $R$. From rapidity 
gap studies, $R$ is close to 0.5.

\subsubsection{The model in hadroproduction}

To what concerns the hard part, Edin~\etal~followed a procedure similar to the CEM:
 the prompt cross section to get a given quarkonium $\cal Q$ is obtained from the 
one to get a colour-singlet $Q\bar Q$  pair with an invariant mass between $2m_Q$ and $2m_{\bar qQ}$ 
after distributing it  between the different states of a 
family (charmonium or bottomonium). In the CEM, it is supposed --from $SU(3)$ counting-- to be 
one ninth of the total\footnote{The epithet {\it total} refers to the colour: singlet plus octet 
configurations, not to a possible integration over $P_T$. The dependence on  $P_T$ of $\sigma$ is 
implied and not written down to simplify notation.} cross section $\sigma_{Q\bar Q}$.

The latter can be computed by Monte Carlo (MC) event generators, which can be coupled to the SCI procedure
(possible exchange of colour state between some partons). The computation of $\sigma_{Q\bar Q}$ 
can be effectively done either with NLO matrix elements (which include
 gluon-fragmentation into a colour-octet pair) 
or with {\sc Pythia}\cite{pythia} --including LO matrix elements and parton showers-- 
(which also contains gluon-fragmentation into a colour-octet pair). Choosing {\sc Pythia}
has the advantage of introducing even higher contributions which can be significant.
To illustrate this, we have reproduced a comparison (see~\cf{fig:CEM_compar}) of two CEM 
calculations: one with NLO matrix-element implementation and another with {\sc Pythia} Monte Carlo 
including parton showers. $m_c$ was set\footnote{these are the values to be taken to 
reproduce\cite{Mariotto:2001sv} fixed-target measurements.} to 1.5 GeV, $\rho_{J/\psi}$
was taken to be 0.5, and  $\rho_{\psi'}$ 0.066. These two choices disagree clearly with
simple spin statistics, but are necessary to reproduce the normalisation of the data. 

\begin{figure}[h]
\centering
\includegraphics[height=6cm]{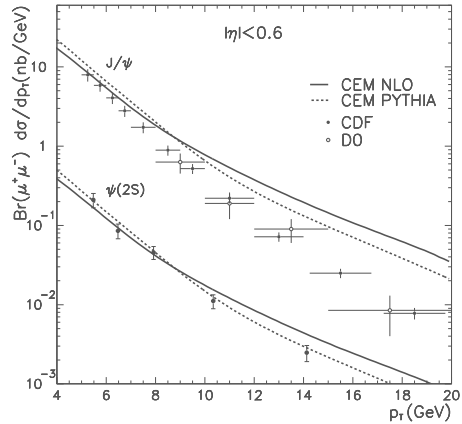}
\caption{Comparison between $J/\psi$ and $\psi'$ {\it prompt} production cross sections within CEM
with NLO matrix elements and with {\sc Pythia}. The data are from 
CDF~\protect\cite{CDF7595} and D$\emptyset$~\protect\cite{Abachi:1996jq} 
(Reprinted figure from Ref.~\protect\refcite{Mariotto:2001sv} with kind permission of Springer Science and Business Media. Copyright (2002)).}
\label{fig:CEM_compar}
\end{figure}

\subsubsection{The results of the model}

The cross section to produce the colour-singlet heavy-quark pair (see~\ce{eq:CEM1}) is computed here 
with {\sc Pythia} which is coupled to the SCI model: once the selected partons --with a probability $R$--
 have exchanged their colour, only the events producing a colour-singlet heavy-quark pair are retained. 
This directly gives the singlet-state cross section, 
whereas in the CEM $\frac{1}{9}\sigma_{Q\bar Q}$ is used. It is also integrated 
in the mass region between $2m_Q$ and $2m_{\bar qQ}$. The cross section to produce 
one given quarkonium $\cal Q$ is obtained using:
\eqs{\label{eq:sharing_SCI}
\sigma_{\cal Q}=\frac{\Gamma_{\cal Q}}{\sum_i \Gamma_{i}} \sigma_{onium},
}
with $\Gamma_{i}= \frac{2 J_i+1}{n_i}$ where $J_i$ is the total angular momentum and $n_i$ the main 
quantum number. $\rho_{\cal Q}$ is thus effectively replaced by $\frac{\Gamma_{\cal Q}}{\sum_i \Gamma_{i}}$
and is not free to vary anymore.

The effect of SCI can be either to turn a colour-singlet pair into a colour-octet one or the contrary. 
The latter case is important, since it opens the possibility of gluon fragmentation into a 
quarkonium at order $\alpha^3_s$ (see \cf{fig:gluon_frag_SCI})\footnote{The same thing 
happens implicitly in the CEM since one ninth of $\sigma_{Q\bar Q}$ gives colour-singlet pair 
irrespectively of the kinematics and thus of which diagram in the hard part contributes the most.
This can also be compared to the Colour-Octet Mechanism as one can extract a fragmentation
function from the simulation\cite{Edin:1997zb}.}. 
This explains why the model reaches a reasonable agreement with data to what concerns the 
slope vs $p_T$ -- the same is equally true for the CEM. Note that the effect of SCI's
on the cross section depends on the partonic state (through the 
number of possible SCI's), and thus on the transverse momentum of the produced quarkonium. The final 
colour-singlet $p_T$ slope is slightly steeper than the initial $Q\bar Q$ one and in better
agreement with data (compare~\cf{fig:CEM_compar} and \cf{fig:ptdep_SCI_PRD} (left))

\begin{figure}[H]
\centering
\includegraphics[height=5cm]{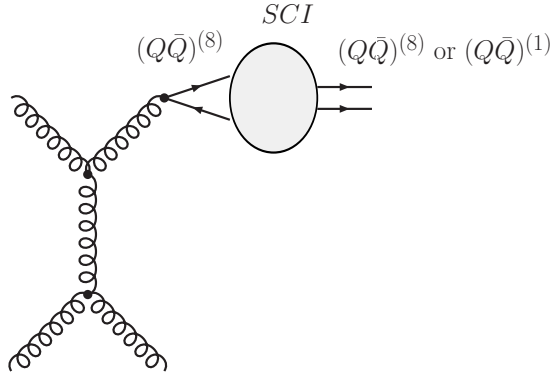}
\caption{Gluon fragmentation process (part of  NLO contributions) giving a singlet state through SCI.}
\label{fig:gluon_frag_SCI}
\end{figure}

The very interesting point of this approach is its simplicity. 
$R$ is the only free parameter and is kept at 0.5, which is the value chosen to reproduce the rate of 
rapidity gaps. However, 
the cross section depends little on it. Putting $R$ to 0.1 instead of 0.5 decreases the cross section
by 30\%. 

On the other hand, in \ce{eq:sharing_SCI}, the $\frac{1}{n}$ suppression for radially
excited states does not fundamentally follow from spin statistics; it is solely motivated by the ratio
of the leptonic decay width. 

Another drawback is the strong dependence upon the heavy-quark mass\footnote{The same dependence
exists for the CEM.}. This is mainly 
due to modified boundary values in the integral from $2m_Q$ to $2m_{\bar qQ}$, much less
to a change in the hard scale and in the mass value entering the matrix elements. Changing $m_c$ 
from 1.35 GeV to 1.6 GeV decreases the cross section by a factor of 3, and from 1.35 GeV to 
1.15 GeV increases it by a factor of 2. The other dependences 
(on $\Lambda_{QCD}$, PDF sets, ...) are not significant.

\begin{figure}[H]
\centering
\mbox{{\includegraphics[width=0.5\textwidth]{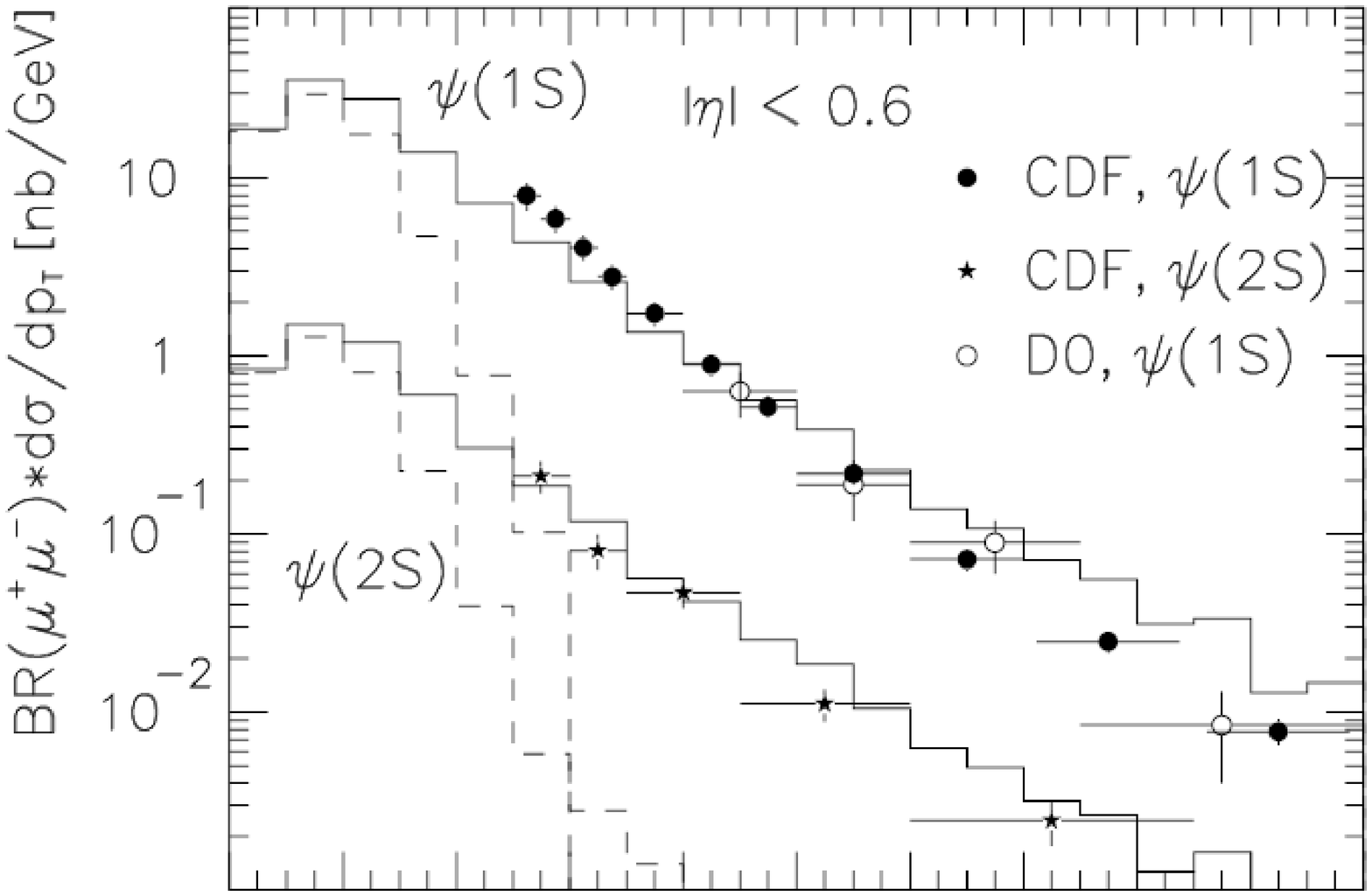}}\quad\quad
      {\includegraphics[width=0.5\textwidth]{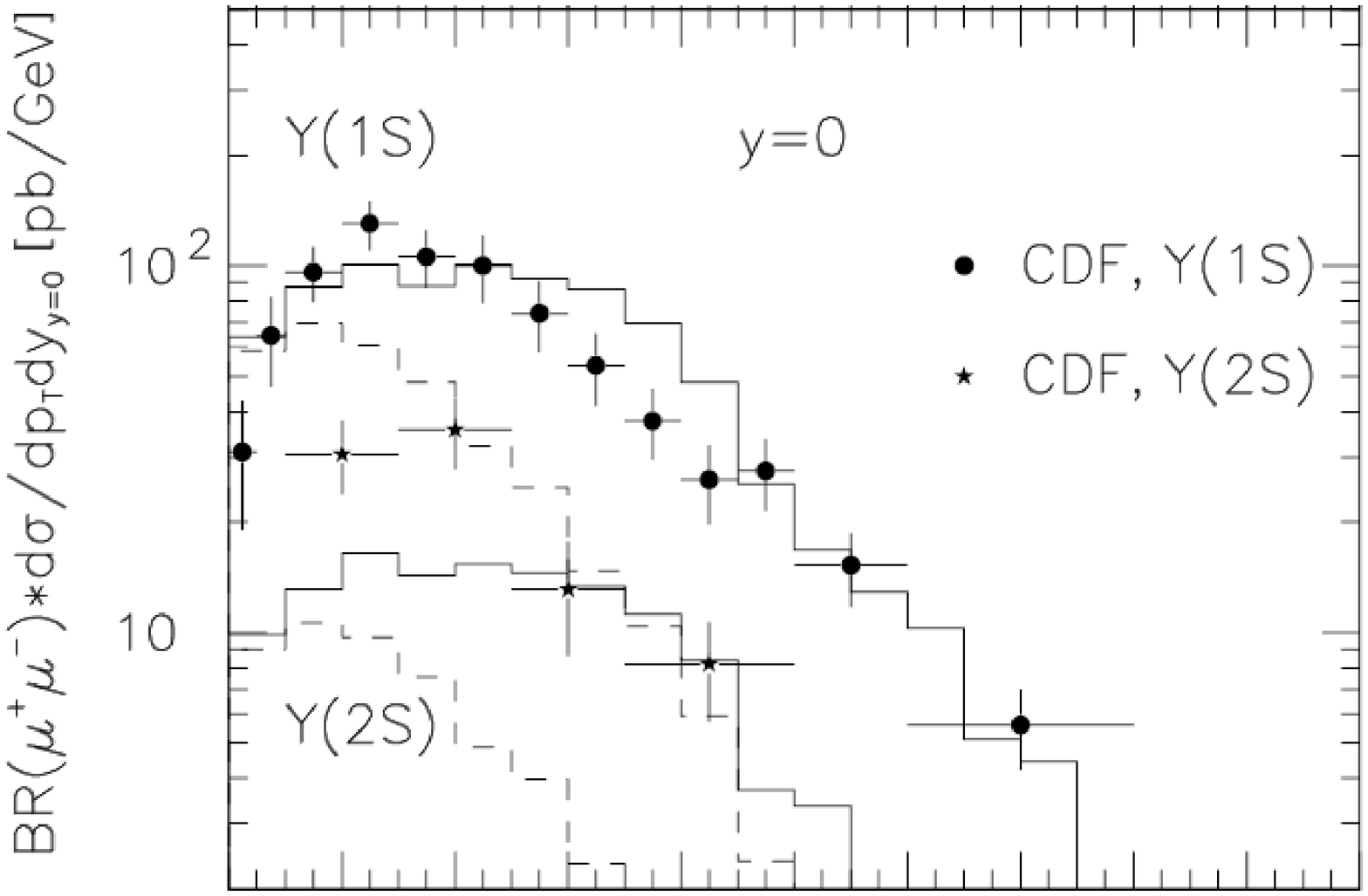}}}
\caption{SCI calculations for prompt production of $J/\psi$, $\psi'$ (left) and $\Upsilon(1S)$ and 
$\Upsilon(2S)$ (right). The dashed lines correspond to the LO matrix-element alone, the solid one  
to the LO matrix-elements  with parton showers 
(Reprinted figures from Ref.~\protect\refcite{Edin:1997zb} with permission of American Physical Society. Copyright (1997)).}
\label{fig:ptdep_SCI_PRD}
\end{figure}

\subsubsection{Improving the mapping }

As one can imagine, physically, during the transition between the $Q\bar Q$ pair --produced 
by the hard interaction with an invariant mass between  $2m_Q$ and $2m_{\bar qQ}$-- and the 
physical quarkonium $\cal Q$, this invariant mass is likely to be modified. It
is then reasonable to suppose\cite{Mariotto:2001sv} that it is smeared to a mass $m$ 
around its initial value $m_{Q\bar Q}$ with such distribution: 
\eqs{ G_{sme}(m_{Q\bar Q},m)= \exp\left(-\frac{(m_{Q\bar Q}-m)}{2 \sigma_{sme}}\right).}

Supposing that the width of the quarkonium resonances tends to zero, the probability that a 
pair with an invariant mass $m_{Q\bar Q}$ gives a quarkonium ${\cal Q}_i$ is given by:
\eqs{{\cal P}_i(m_{Q\bar Q})\simeq\frac{\Gamma_i G_{sme}(m_{Q\bar Q},m_{{\cal Q}_i})}
{\sum_j \Gamma_j G_{sme}(m_{Q\bar Q},m_{{\cal Q}_j})},}
still with $\Gamma_{i}= \frac{2 J_i+1}{n_i}$.

This smearing enables quark pairs with invariant masses above the heavy-meson 
threshold $2m_{\bar qQ}$ to produce stable quarkonia, but the reverse is true also, with 
a probability $A(m_{Q\bar Q})$ equal to 
$\frac{1}{2}\hbox{erfc}\left( \frac{2m_{\bar qQ}-m_{Q\bar Q}}{\sqrt{2\pi}\sigma_{sme}}\right)$.

The cross section to get a quarkonium ${\cal Q}_i$ is thus given Ref.~\protect\refcite{Mariotto:2001sv} by
\eqs{
\sigma_i=\int^{\sqrt{s}}_{2m_Q} dm_{Q\bar Q} \frac{d\sigma_{Q\bar Q}}{dm_{Q\bar Q}}
(1-A(m_{Q\bar Q})){\cal P}_i(m_{Q\bar Q}),
}
with $\frac{d\sigma_{Q\bar Q}}{dm_{Q\bar Q}}$ calculated either within CEM (with $\frac{1}{9}$ 
from $SU(3)$) or within SCI.

Using this procedure, the production ratio $\psi'$ over $J/\psi$ is better 
reproduced\cite{Mariotto:2001sv} than for simple spin statistics as well as the different 
components of the prompt $J/\psi$ as measured by CDF\cite{CDF7997b}. On the other hand, due to 
the proximity in mass of the $\chi_{c1}$ and $\chi_{c2}$, their production ratio is not
affected and remains $3:5$ slightly in contradiction with the CDF measurements\cite{Affolder:2001ij}
consistent with a ratio of 1. Finally, this refinement affects the cross sections only by a 
factor of 20\%, what is well within the model uncertainty.

\subsection{NRQCD: including the colour-octet mechanism}\label{subsec:NRQCD}

In 1992, Bodwin~\etal~considered new Fock-state contributions in order to
cancel exactly the IR divergence in the light-hadron decays of $\chi_{c1}$ (and $h_c$) at LO. This decay
proceeds via two gluons, one real and one off-shell, which gives the 
light-quark pair forming the hadron. When the first gluon becomes soft, the decay width diverges.
The conventional treatment\cite{barbieri:div_p_wave1,barbieri:div_p_wave2,barbieri:div_p_wave3,barbieri:div_p_wave4}, which amounts to regulating the divergence 
by an infrared cut-off identified with the binding energy of the bound state, was not satisfactory: 
it supposed a logarithmic dependence of $\Psi'(0)$ upon the binding energy. They looked at
this divergence as being a clear sign that the factorisation was violated in the CSM.

Their new Fock states for $\chi_c$ were \eg~a gluon plus a $c\bar c$ pair, 
in a $^3 S_1$ configuration and in 
a {\it colour-octet} state. The decay of this Fock state occurred through
the transition of the coloured pair into a gluon (plus the other
gluon already present in the Fock state as a spectator). 
This involved a new phenomenological parameter, $H_8$, which was related to 
the probability that the $c\bar c$ pair of the $\chi_c$ be in 
a colour-octet $S$-wave state. The key point of this procedure was that a 
logarithmic dependence on a new scale $\Lambda$ -- a typical momentum scale
for the light quark -- appeared naturally within the effective field theory Non-Relativistic
Quantum Chromodynamic (NRQCD)\cite{Caswell:1985ui,Bodwin:1994jh}.

This effective theory is based on a systematic expansion in both
$\alpha_s$ and $v$, which is the quark velocity within the bound state.  For charmonium, 
$v_c^2\simeq 0.23$ and for bottomonium $v_b^2\simeq 0.08$. One of the main novel features
of this theory is the introduction of dynamical gluons in the Fock-state decomposition of the physical
quarkonium states. In the case of $S$-wave orthoquarkonia ($^3S_1$), we schematically have, in the Coulomb
gauge:
\eqs{\label{eq:NRQCD_decomp1}
|{\cal Q}_Q\rangle=&{\cal O}(1) | Q\bar Q [\ ^3S_1^{(1)}]\rangle
+{\cal O}(v) | Q\bar Q [\ ^3P_J^{(8)}g]\rangle
+{\cal O}(v^2) | Q\bar Q [\ ^1S_0^{(8)}g]\rangle\\
&+{\cal O}(v^2) | Q\bar Q [\ ^3S_1^{(1,8)}gg]\rangle
+{\cal O}(v^2) | Q\bar Q [\ ^3D_J^{(1,8)}gg]\rangle
+\dots
}
whereas for  $P$-wave orthoquarkonia ($^3P_J$), the decomposition is as follows
\eqs{\label{eq:NRQCD_decomp2}
|{\cal Q}_{Q_J}\rangle=&{\cal O}(1) | Q\bar Q [\ ^3P_J^{(1)}]\rangle
+{\cal O}(v) | Q\bar Q [\ ^3S_1^{(8)}g]\rangle
+\dots
}
In these two formulae, the superscripts (1) and (8) indicate the colour state of the $Q \bar Q$ pair.
The ${\cal O}(v^n)$ factors indicate the order in the velocity expansion at which the corresponding
Fock state participates to the creation or the annihilation of quarkonia. This follows from the 
{\it velocity scaling rules} of NRQCD (see \eg~Ref.~\refcite{Bodwin:1994jh}).

In this formalism, it is thus possible to demonstrate, in the limit of large quark mass, the 
factorisation between the short-distance -- and perturbative -- contributions and 
the hadronisation of the $Q \bar Q$, described by non-perturbative
matrix elements defined within NRQCD. For instance, the differential cross section for the production of
a quarkonium $\cal Q$ associated with some other hadrons $X$ reads
\eqs{\label{eq:facto_nrqcd}
d \sigma ({\cal Q}+X)=\sum d \hat\sigma (Q\bar Q[^{2S+1}L_J^{(1,8)}]+X)
\langle{\cal O}^{\cal Q}[^{2S+1}L_J^{(1,8)}] \rangle,
}
where the sum stands for $S$, $L$, $J$ and the colour.

The long-distance matrix element (LDME) $\langle{\cal O}^{\cal Q}[^{2S+1}L_J^{(1,8)}] \rangle$ 
takes account of the transition between the $Q \bar Q$ pair and the final physical state ${\cal Q}$.
Its power scaling rule comes both from the suppression in $v$ of the Fock-state component
$[^{2S+1}L_J^{(1,8)}]$ in the wave function of $\cal Q$ and
from the scaling of the NRQCD interaction  responsible for the transition.

Usually, one defines ${\cal O}^{\cal Q}[^{2S+1}L_J^{(1,8)}]$ as 
the production operator that creates and annihilates a point-like $Q \bar Q$ pair in the specified state. 
This has the following general expression 
\eqs{\label{eq:def_op_prod}
{\cal O}^{\cal Q}[^{2S+1}L_J^{(1,8)}]&=\chi^\dagger K \psi\left(\sum_X\sum_{J_z}|{\cal Q}+X\rangle
\langle{\cal Q}+X|\right)  \psi^\dagger K \chi\\
&=\chi^\dagger K \psi (a^\dagger_{\cal Q} a_{\cal Q}) \psi^\dagger K \chi,}
where\footnote{The second line of~\ce{eq:def_op_prod} is nothing but a short way of 
expressing this operator.} $\psi$ and $\chi$ are Pauli spinors and the matrix $K$ 
is  a product of colour, spin and covariant derivative factors.
These factors can be obtained from the NRQCD lagrangian\cite{Bodwin:1994jh}. For instance,
${\cal O}^{\cal Q} [\ ^3S_1^{(1)}]=\chi^\dagger \vect \sigma\psi 
  (a^\dagger_{\cal Q} a_{\cal Q})\psi^\dagger \vect \sigma\chi$. However some transitions
require the presence of chromomagnetic ($\Delta L=0$ and $\Delta S=\pm1$) and chromoelectric
($\Delta L=\pm1$ and $\Delta S=0$) terms for which the expressions of $K$ are more complicated.

On the other hand, the general scaling rules relative to the LDME's\cite{Kramer:2001hh} give:
\eqs{
\langle{\cal O}^{\cal Q}[^{2S+1}L_J^{(1,8)}] \rangle=v^{3+2L+2E+4M},
}
where $E$ and $M$ are the minimum number of chromoelectric and chromomagnetic transitions for the 
$Q\bar Q$ to go from the state $[^{2S+1}L_J^{(1,8)}]$ to the dominant quarkonium ${\cal Q}$ Fock
state.

The idea of combining fragmentation as the main source of production with allowed
transitions between a $\chi_c$ to a $^3S_1$ in a colour-octet state, was applied 
by Braaten and Yuan\cite{Braaten:1994kd}.
Indeed, similar formulae to the one written for fragmentation within the CSM can be written for 
fragmentation functions in NRQCD\cite{Braaten:1994vv}:
\eqs{
D_{g\to {\cal Q}}(z,\mu)=\sum d_{[^{2S+1}L_J^{(1,8)}]}(z,\mu) 
\langle{\cal O}^{\cal Q}[^{2S+1}L_J^{(1,8)}] \rangle, 
}
where $d_{[\cdot]}(z,\mu)$ accounts for short-distance contributions 
and does not depend on which  $\cal Q$ is involved.
But since the theoretical predictions for prompt $J/\psi$ production did not disagree dramatically 
with data and since there was no possible $\chi_c$ decay to $\psi'$, a possible 
enhancement of the $\chi_c$ cross section by colour-octet mechanism (COM) was not seen at that 
time as a key-point both for $J/\psi$ and $\psi'$ production.

However, in the case of $J/\psi$ and $\psi'$ production, COM could still matter, but in a different 
manner: fragmentation of a gluon
into a $^3P_J^{(8)}$ is possible with solely one gluon emission and fragmentation into a $^3S_1^{(8)}$
requires no further gluon emission 
(at least in the hard process --described by $d_{[\cdot]}(z,\mu)$-- and not in the soft process
associated with $\cal O$). Concerning
the latter process, as two chromoelectric transitions are required 
for the transition $|c\bar cgg\rangle$ to
$|c\bar c\rangle$, the associated LDME $\langle{\cal O}^{\psi}[^{3}S_1^{(8)}] \rangle$ was 
expected to scale as $m_c^3 v^7$. In fact, $d_{[^{3}S_1^{(8)}]}$, the contribution 
to the fragmentation function of the short-distance process $g\to\!\!\ ^{3}S_1^{(8)}$ was already known
since the paper of Braaten and Yuan\cite{Braaten:1994kd} and could be used here, 
as the hard part $d_{[\cdot]}(z,\mu)$ 
of the fragmentation process is independent of the quarkonium.

\begin{figure}[h]
\centering
\includegraphics[width=8.4cm]{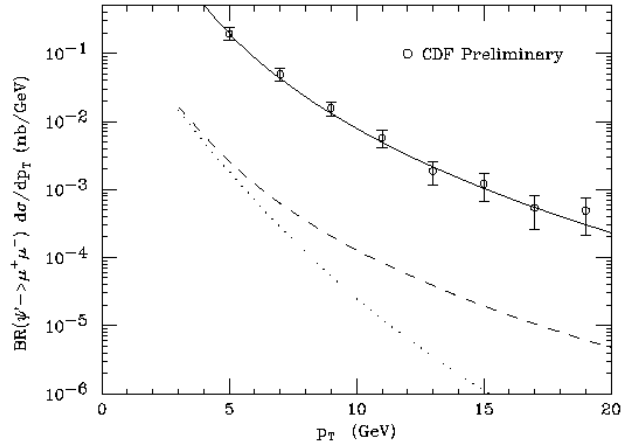}
\caption{Differential cross section versus $p_T$ of the CSM (dotted: LO; dashed: fragmentation and LO) 
production and of the COM fragmentation (solid curve) to be compared with CDF 
preliminary  measurements of the direct production of $\psi'$~\protect\cite{:1994dx} 
(Reprinted figure from Ref.~\protect\refcite{Braaten:1994vv} with permission of American Physical Society. Copyright (1995)).}
\label{fig:dsdpt_COM_vs_CDF}
\end{figure}

In a key paper, Braaten and Fleming\cite{Braaten:1994vv} combined everything together 
to calculate, for the Tevatron, the fragmentation rate of a 
gluon into an octet $^3S_1$ that subsequently evolves into a $\psi'$. They obtained,
with\footnote{This corresponds to a suppression of 25 compared to the ``colour-singlet matrix
element'' $\langle{\cal O}^{\psi(2S)}[^{3}S_1^{(1)}]\rangle$, which scales as $m_c^3 v^3$. This is 
thus in reasonable agreement with a $v^4$ suppression.}  
$\langle{\cal O}^{\psi(2S)}[^{3}S_1^{(8)}] \rangle=4.2 \times 10^{-3}$ GeV$^3$, 
a perfect agreement with the CDF preliminary data\cite{:1994dx}: compare
these predictions in~\cf{fig:dsdpt_COM_vs_CDF} with the CSM fragmentation ones 
in~\cf{fig:dsdpt_braaten-94} (right).

Following these studies, a complete survey on the colour-octet mechanism was made in two papers
by Cho and Leibovich\cite{Cho:1995vh,Cho:1995ce}. The main achievements of these papers
were the calculation of colour-octet $P$-state contributions to $\psi$, the predictions for prompt and 
direct colour-octet $J/\psi$ production with  also the $\chi_c$ feed-down calculation, the first
predictions for $\Upsilon$ and the complete set of $2\to 3$ parton processes like $ij\to c \bar c +k$,
all this in agreement with the data.

\subsubsection{Determination of the LDME's} 

Here we expose the results relative to the determination of the Long Distance Matrix Elements
of NRQCD necessary to describe the production of $^3S_1$ quarkonium. We can distinct between 
two classes of matrix elements : the colour singlet ones, which are fixed as we shall 
see, and the colour octet ones which are fit to reproduce the observed cross sections as a 
function of $p_T$.

In fact, as already mentioned, NRQCD predicts that there is an infinite 
number of Fock-state contributions to the  production of a quarkonium ${\cal Q}$ and 
thus an infinite number of LDME. Practically, one is driven to truncate the series; this
is quite natural in fact since most of the contributions should be suppressed by factor
of at least ${\cal O}(\frac{v^2}{c^2})$, where $v$ is the quark velocity in the bound state.

For definiteness, the latest studies accommodating the production rate of $^3S_1$ states retain
only the colour singlet state with the same quantum numbers as the bound state and colour 
octet $P$-wave states and singlet $S$-wave. In this context, the CSM 
can be thought as a further approximation to the NRQCD formalism, where we keep solely the leading
terms in $v$.

\paragraph{Colour-singlet LDME}

In this formalism, factorisation tells us that each contribution is the product of 
a perturbative part and a non-perturbative matrix element, giving, roughly speaking, the probability
that the quark pair perturbatively produced will evolve into the considered physical bound state. If one 
transposes this to the CSM, this means the wave function at the origin corresponds to
this non-perturbative element. This seems reasonable since the wave function squared is also
a probability.

Yet, one has to be cautious if one links production processes with decay processes. In 
NRQCD, two different matrix elements are defined for the ``colour singlet'' production and decay, and
they are likely to be different and independent.

The only path left to recover the CSM is the use of a further approximation, the 
vacuum saturation approximation. The latter tells us how the matrix element for the 
decay is linked to the one for the production. This enables us to relate the wave function
at the origin appearing in $\Gamma_{\ell\ell}$ (Eq. (3.1) of~Ref.~\refcite{Lansberg:2005aw}) to the colour-singlet 
NRQCD matrix element
for production. This gives:
\eqs{\label{eq:LDME_CSM}
\langle {\cal O}^{^3S_1}[^3S^{(1)}_1]\rangle=18 \left|\psi(0) \right|^2 +{\cal O}(v^4).
}

The conclusion that could be drawn within NRQCD is that the extraction of non-perturbative
input for production from the one for decay is polluted by factors of ${\cal O}(v^4)$, this is
also true for extraction from potential models.

In \ct{tab:LDME_psi_singlet}, we give the colour-singlet LDME for the $J/\psi$ and the $\psi'$. The 
result for the different potentials are deduced from the solutions of~Ref.~\refcite{Kopeliovich:2003cn}. 
These
LDME's --up to a factor 18-- are those that are to be used in CSM calculations. The values differ from
the one used in Refs.~\refcite{CSM_hadron1,CSM_hadron2,CSM_hadron3} because of modifications in the measured values 
of $\Gamma_{\ell\ell}$, 
NLO QCD corrections to $\Gamma_{\ell\ell}$ and also in the potential used to obtain the wave function at the origin.

\begin{table}[H]
\tbl{Colour-singlet LDME for the $J/\psi$ and the $\psi'$ determined from the leptonic decay width 
and from various potentials. The error of ${\cal O}(v^4)$ as shown in~\ce{eq:LDME_CSM} should be implied. 
Values are given in units GeV${}^3$.}{
\footnotesize
$\begin{array}{|c|c|c|c|c|}
\hline
\cal Q & \mbox{Leptonic decay} & \mbox{BT Potential Ref.~\protect\refcite{Buchmuller:1980bm,Buchmuller:1980su}}& \mbox{POW potential Ref.~\protect\refcite{Martin:1980jx}}& \mbox{Cornell Potential Ref.~\protect\refcite{cornell_potential1,cornell_potential2}} \\
\hline\hline
J/\psi  & 1.1 \pm 0.1 & 1.16 & 1.4  & 2.06\\
\psi'   & 0.7 \pm 0.1 & 0.76 & 0.76 & 1.29 \\
\hline
\end{array}
$}
\label{tab:LDME_psi_singlet}
\end{table}

In \ct{tab:LDME_upsi_singlet}, we expose the results of~Ref.~\refcite{Braaten:2000cm} concerning
the colour-singlet LDME for $\Upsilon(nS)$, \ie\ $\langle{\cal{O}}^{\Upsilon(nS)}[{}^3S^{(1)}_1]\rangle$.
The error quoted for the value from potential models expresses the variation of the latter 
when passing from one to another.

\begin{table}[H]
\tbl{Colour-singlet LDME for the $\Upsilon (nS)$ determined from the leptonic decay width 
and from potentials models. Values are given in units GeV${}^3$ (Ref.~\protect\refcite{Braaten:2000cm}).}
{
$\footnotesize
\begin{array}{|c|c|c|}
\hline
\cal Q & \mbox{Leptonic decay} & \mbox{Potential Models} \\
\hline\hline
\Upsilon (1S) & 10.9\pm 1.6 & 10.8\pm 5.5 \\
\Upsilon (2S) & 4.5 \pm 0.7 & 5.0 \pm 1.8 \\
\Upsilon (3S) & 4.3 \pm 0.9 & 3.7 \pm 1.5 \\ \hline 
\end{array}
$}
\label{tab:LDME_upsi_singlet}
\end{table}

\paragraph{Colour-octet LDME's}

As said above, three intermediate colour-octet states are currently considered in the description
of $^3S_1$ production. These are ${}^1\!S^{(8)}_0$, $^3\!P^{(8)}_0$ and ${}^3S^{(8)}_1$. The corresponding
LDME's giving the probability of transition between these states and the physical 
colour-singlet $^3S_1$ state are not known and are to be fit to the data.

Unfortunately, the perturbative amplitudes to produce a ${}^1\!S^{(8)}_0$ or $^3\!P^{(8)}_0$
have the same $p_T$ slope and their coefficient cannot be determined apart. Therefore, one defines $k$
as the ratio between these two amplitudes. From it, one defines the following combination
\eqs{ M_{k}^{\cal Q}({}^1S^{(8)}_0,{}^3P^{(8)}_0)\equiv \langle {\cal{O}}^{\cal Q} 
[{}^1\!S^{(8)}_0]\rangle  + k \frac{\langle {\cal{O}}^{\cal Q} [{}^3\!P^{(8)}_0]\rangle}{m_c^2},}
 which is fit to the data. In the following, we expose the results obtained by different analyses 
using various PDF set and parameter values. 
In the following tables, the first error quoted is statistical,
 the second error, when present, reflects the variation of the fit LDME when the renormalisation 
and factorisation scales is set to $\mu=1/2\,\sqrt{p_T^2+4 m_c^2}$  and to $2\,\sqrt{p_T^2+4 m_c^2}$.
The agreement with the data being actually good, there is no real interest to plot the cross sections given
by the fits.

\begin{table}[H]
\tbl{ Fit values of $J/\psi$ production LDME's from $\frac{d\sigma}{dp_T}$
at the Tevatron. Values are given in units $10^{-2}$~GeV${}^3$ (Numbers from Ref.~\protect\refcite{Kramer:2001hh}).}
{$ \footnotesize
\begin{array}{|c||c|c||c|c|c|}
\hline
 \mbox{Reference} & \multicolumn{2}{c|}{\mbox{PDF}} & \langle
 {\cal{O}}^{J/\psi}[{}^3S^{(8)}_1] \rangle &
 M_{k}^{J/\psi}({}^1S^{(8)}_0,{}^3P^{(8)}_0) & k \\ \hline\hline
 \mbox{Cho-Leibovich Ref.~\protect\refcite{Cho:1995ce}} &
 \multicolumn{2}{c|}{\mbox{MRS(D0) Ref.~\protect\refcite{Martin:1993zi}}} & 0.66 \pm
 0.21 & 6.6 \pm 1.5 & 3 \\ \hline &
 \multicolumn{2}{c|}{\mbox{CTEQ4L Ref.~\protect\refcite{Lai:1996mg}}} & 1.06 \pm
 0.14^{+1.05}_{-0.59} & 4.38 \pm 1.15^{+1.52}_{-0.74} & \\
 \mbox{Beneke-Kr\"amer Ref.~\protect\refcite{Beneke:1996yw}} &
 \multicolumn{2}{c|}{\mbox{GRV-LO(94) Ref.~\protect\refcite{Gluck:1995uf}}} & 1.12 \pm
 0.14^{+0.99}_{-0.56} & 3.90 \pm 1.14^{+1.46}_{-1.07} & 3.5 \\ &
 \multicolumn{2}{c|}{\mbox{MRS(R2) Ref.~\protect\refcite{Martin:1996as}}} & 1.40 \pm
 0.22^{+1.35}_{-0.79} & 10.9 \pm 2.07^{+2.79}_{-1.26} & \\ \hline &
 \multicolumn{2}{c|}{\mbox{MRST-LO(98) Ref.~\protect\refcite{Martin:1998sq}}} & 0.44
 \pm 0.07 & 8.7 \pm 0.9 & \\
 \raisebox{2ex}[-2ex]{
 \mbox{Braaten-Kniehl-Lee Ref.~\protect\refcite{Braaten:1999qk}}} &
 \multicolumn{2}{c|}{\mbox{CTEQ5L Ref.~\protect\refcite{Lai:1999wy}}} & 0.39 \pm 0.07
 & 6.6 \pm 0.7 & \raisebox{2ex}[-2ex]{3.4} \\ \hline
 \mbox{Kr\"amer Ref.~\protect\refcite{Kramer:2001hh}} &
 \multicolumn{2}{c|}{\mbox{CTEQ5L Ref.~\protect\refcite{Lai:1999wy}}} & 1.19 \pm 0.14
 & 4.54 \pm 1.11 & 3.5 \\[0.5mm] \hline
\end{array}
$
}
\label{table2}
\end{table}

\begin{table}[H]
\tbl{Same as Table~\ref{table2} for $\psi(2S)$ production.  Values
 are given in units $10^{-2}$~GeV${}^3$ (Numbers from Ref.~\protect\refcite{Kramer:2001hh})}{
$ \footnotesize
\begin{array}{|c||c|c||c|c|c|}
\hline
 \mbox{Reference} & \multicolumn{2}{c|}{\mbox{PDF}} & \langle
 {\cal{O}}^{\psi(2S)}[{}^3S^{(8)}_1] \rangle &
 M_{k}^{\psi(2S)}({}^1S^{(8)}_0,{}^3P^{(8)}_0) & k   \\ \hline\hline
 \mbox{Cho-Leibovich Ref.~\protect\refcite{Cho:1995ce}} &
 \multicolumn{2}{c|}{\mbox{MRS(D0) Ref.~\protect\refcite{Martin:1993zi}}} & 0.46 \pm
 0.1 & 1.77 \pm 0.57 & 3 \\ \hline &
 \multicolumn{2}{c|}{\mbox{CTEQ4L Ref.~\protect\refcite{Lai:1996mg}}} & 0.44 \pm
 0.08^{+0.43}_{-0.24} & 1.80 \pm 0.56^{+0.62}_{-0.30} & \\
 \mbox{Beneke-Kr\"amer Ref.~\protect\refcite{Beneke:1996yw}} &
 \multicolumn{2}{c|}{\mbox{GRV-LO(94) Ref.~\protect\refcite{Gluck:1995uf}}} & 0.46 \pm
 0.08^{+0.41}_{-0.23} & 1.60 \pm 0.51^{+0.60}_{-0.44} & 3.5 \\ &
 \multicolumn{2}{c|}{\mbox{MRS(R2) Ref.~\protect\refcite{Martin:1996as}}} & 0.56 \pm
 0.11^{+0.54}_{-0.32} & 4.36 \pm 0.96^{+1.11}_{-0.50} & \\ \hline &
 \multicolumn{2}{c|}{\mbox{MRST-LO(98) Ref.~\protect\refcite{Martin:1998sq}}} & 0.42
 \pm 0.1 & 1.3 \pm 0.5 & \\
 \raisebox{2ex}[-2ex]{
 \mbox{Braaten-Kniehl-Lee Ref.~\protect\refcite{Braaten:1999qk}}} &
 \multicolumn{2}{c|}{\mbox{CTEQ5L Ref.~\protect\refcite{Lai:1999wy}}} & 0.37 \pm 0.09
 & 0.78 \pm 0.36 & \raisebox{2ex}[-2ex]{3.4}  \\ \hline 
\mbox{Kr\"amer Ref.~\protect\refcite{Kramer:2001hh}} &
 \multicolumn{2}{c|}{\mbox{CTEQ5L Ref.~\protect\refcite{Lai:1999wy}}} & 0.50 \pm 0.06
 & 1.89 \pm 1.11 & 3.5 \\ \hline
\end{array}
$
}
\label{table3}
\end{table}

In the following tables (Tables 12-16), the results of Cho and Leibovich are for $p_T > 3.5~{\rm GeV} $ on the data 
of\cite{CDF7595} and the ones of Braaten~\etal~are for $p_T > 8~{\rm GeV} $ 
and $m_b=4.77\,$GeV was chosen.

\begin{table}[H]
\tbl{Same as Table~\ref{table2} for $\Upsilon(1S)$ production.  Values
 are given in units $10^{-2}$~GeV${}^3$.}{
$ \footnotesize
\begin{array}{|c||c|c||c|c|c|}
\hline
 \mbox{Reference} & \multicolumn{2}{c|}{\mbox{PDF}} & \langle
 {\cal{O}}^{\Upsilon(1S)}[{}^3S^{(8)}_1] \rangle &
 M_{k}^{\Upsilon(1S)}({}^1S^{(8)}_0,{}^3P^{(8)}_0) & k   \\ \hline\hline
 \mbox{Cho-Leibovich Ref.~\protect\refcite{Cho:1995ce}} &
 \multicolumn{2}{c|}{\mbox{MRS(D0) Ref.~\protect\refcite{Martin:1993zi}}} & 0.6 \pm
 0.2 & 0.4 \pm 0.5 & 5\\ \hline &
 \multicolumn{2}{c|}{\mbox{MRST-LO(98) Ref.~\protect\refcite{Martin:1998sq}}} & 
 0.4 \pm 0.7^{-1.0}_{+0.7}&20.2 \pm 7.8^{+11.9}_{-8.5} & \\
 \raisebox{2ex}[-2ex]{
 \mbox{\parbox{3cm}{Braaten Fleming Leibovich Ref.~\protect\refcite{Braaten:2000cm}}}} &
 \multicolumn{2}{c|}{\mbox{CTEQ5L Ref.~\protect\refcite{Lai:1999wy}}} &  2.0 \pm 4.1^{-0.6}_{+0.5}
 & 13.6 \pm 6.8^{+10.8}_{-7.5} & \raisebox{2ex}[-2ex]{5}  \\ \hline
\end{array}
$
}
\label{table4}
\end{table}

\begin{table}[H]
\tbl{Same as Table~\ref{table2} for $\Upsilon(2S)$ production.  Values
 are given in units $10^{-2}$~GeV${}^3$.}{
$ \footnotesize
\begin{array}{|c||c|c||c|c|c|}
\hline
 \mbox{Reference} & \multicolumn{2}{c|}{\mbox{PDF}} & \langle
 {\cal{O}}^{\Upsilon(2S)}[{}^3S^{(8)}_1] \rangle &
 M_{k}^{\Upsilon(2S)}({}^1S^{(8)}_0,{}^3P^{(8)}_0) & k   \\ \hline\hline
 \mbox{Cho-Leibovich Ref.~\protect\refcite{Cho:1995ce}} &
 \multicolumn{2}{c|}{\mbox{MRS(D0) Ref.~\protect\refcite{Martin:1993zi}}} & 0.4 \pm
 0.1 & 0.5 \pm 0.4 & 5\\ \hline &
 \multicolumn{2}{c|}{\mbox{MRST-LO(98) Ref.~\protect\refcite{Martin:1998sq}}} & 
 17.4 \pm 6.4^{+7.0}_{-5.1} & -9.5 \pm 11.1^{-2.8}_{+2.1} & \\
 \raisebox{2ex}[-2ex]{
 \mbox{\parbox{3cm}{Braaten Fleming Leibovich Ref.~\protect\refcite{Braaten:2000cm}}}} &
 \multicolumn{2}{c|}{\mbox{CTEQ5L Ref.~\protect\refcite{Lai:1999wy}}} & 16.4 \pm 5.7^{+7.1}_{-5.1}
 & -10.8 \pm 9.7^{-3.4}_{+2.0}& \raisebox{2ex}[-2ex]{5}  \\  \hline
\end{array}
$
}
\label{table5}
\end{table}

\begin{table}[H]
\tbl{Same as Table~\ref{table2} for $\Upsilon(3S)$ production.  Values
 are given in units $10^{-2}$~GeV${}^3$.}{
$ \footnotesize
\begin{array}{|c||c|c||c|c|c|}
\hline
 \mbox{Reference} & \multicolumn{2}{c|}{\mbox{PDF}} & \langle
 {\cal{O}}^{\Upsilon(3S)}[{}^3S^{(8)}_1] \rangle &
 M_{k}^{\Upsilon(3S)}({}^1S^{(8)}_0,{}^3P^{(8)}_0) & k   \\ \hline\hline &
 \multicolumn{2}{c|}{\mbox{MRST-LO(98) Ref.~\protect\refcite{Martin:1998sq}}} & 
 3.7 \pm 2.1^{+1.7}_{-1.3} & 7.5 \pm 4.9^{+3.4}_{-2.5} & \\ \raisebox{2ex}[-2ex]{
 \mbox{\parbox{3cm}{Braaten Fleming Leibovich Ref.~\protect\refcite{Braaten:2000cm}}}}
 &
 \multicolumn{2}{c|}{\mbox{CTEQ5L Ref.~\protect\refcite{Lai:1999wy}}} & 3.6 \pm 1.9^{+1.8}_{-1.3}
 & 5.4 \pm 4.3 ^{+3.1}_{-2.2}& \raisebox{2ex}[-2ex]{5}  \\\hline
\end{array}
$
}
\label{table6}
\end{table}

\subsubsection{Polarisation predictions}

A straightforward and unavoidable consequence of the NRQCD solution to the $\psi'$ anomaly was
early raised by Cho and Wise\cite{Cho:1994ih}: the $\psi'$, produced by a fragmenting (and real) gluon
through a colour octet state, is 100\% transversally polarised. They in turn suggested a test of 
this prediction,~\ie~the measurement of the lepton angular distribution in $\psi'\to \ell^+ \ell^-$, 
which should behave as
$\frac{d\Gamma}{d \cos \theta} (\psi'\to \ell^+ \ell^-) \propto (1+\alpha \cos^2 \theta)$,
with $\alpha$=1 for 100\% transversally polarised particles since the spin symmetry of NRQCD prevents
soft-gluon emissions to flip the spin of $Q\bar Q$ states.
 
In parallel to the extraction studies of LDME, the evolution of this polarisation variable $\alpha$
as a function of $p_T$ was thus predicted by different groups and compared to measurement of the CDF
collaboration (see section \ref{sec:polarisation}).

If we restrict ourselves to the high-$p_T$ region, where the fragmenting gluon is transversally polarised, 
the polarisation can only be affected\footnote{We mean $\alpha\neq+1$.} by $v$ corrections (linked to 
the breaking of the NRQCD spin symmetry) or $\alpha_s$ corrections different than the ones already 
included in the Altarelli-Parisi
evolution of the fragmentation function $D$. Indeed, emissions of 
hard\footnote{with momenta higher than $\Lambda_{NRQCD}$.} gluons are likely to flip the spin 
of the $Q\bar Q$ pair. These corrections have been considered by Beneke~\etal\cite{Beneke:1996yw}.

\begin{figure}[h]
\centering
\mbox{{\includegraphics[width=0.3\textwidth]{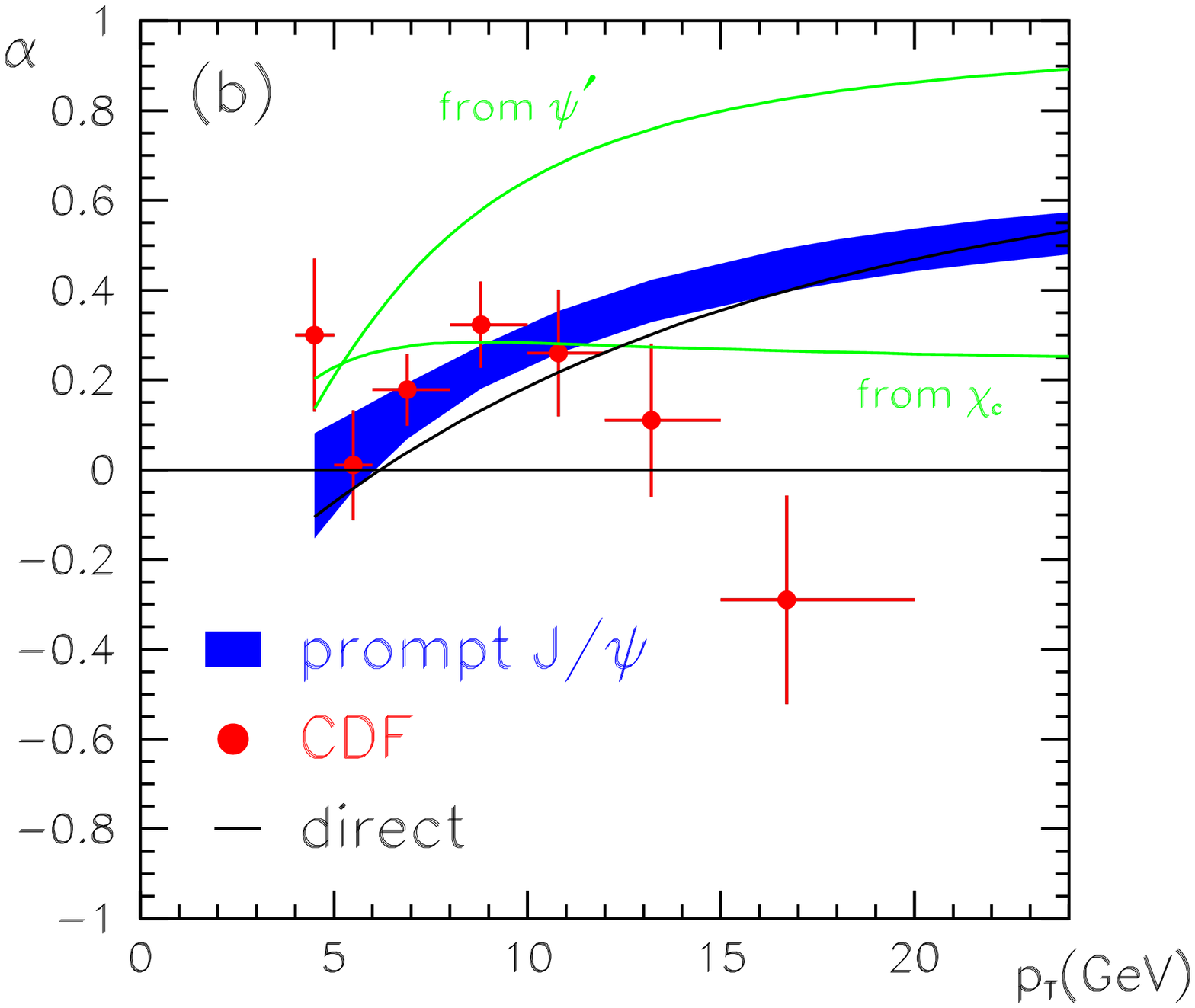}}\quad
      {\includegraphics[width=0.3\textwidth]{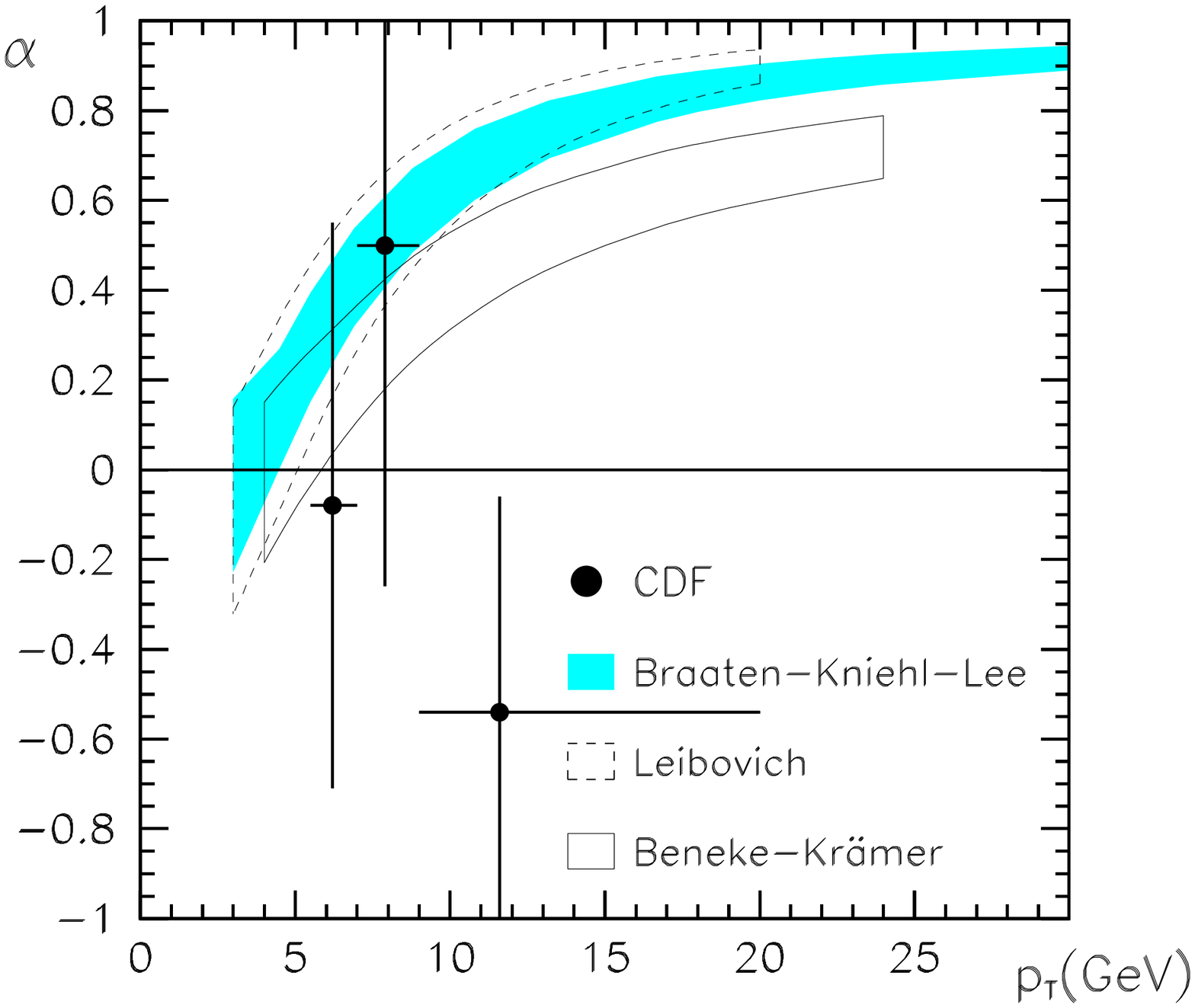}}\quad
      {\includegraphics[width=0.28\textwidth]{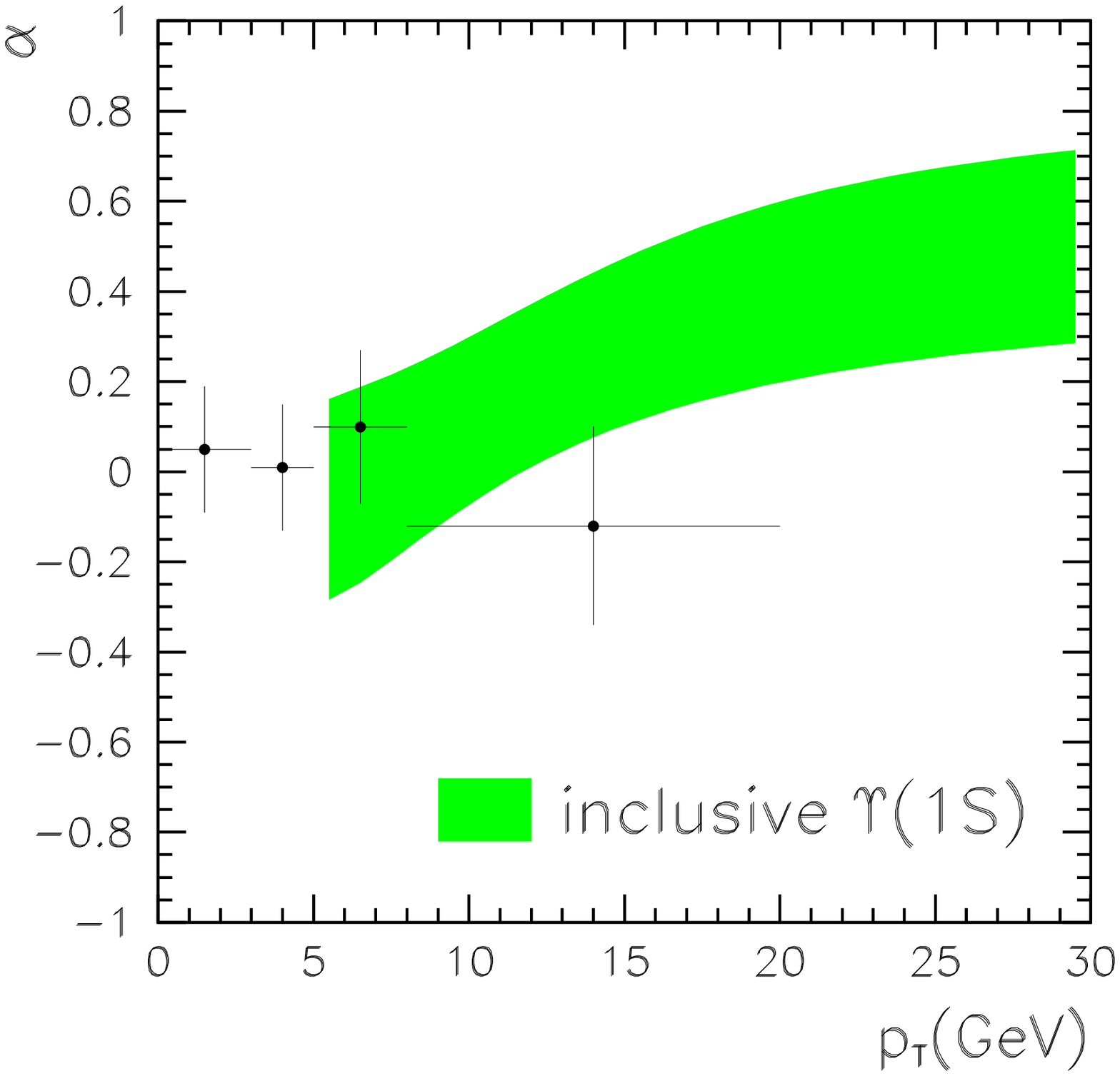}}
}
\caption{NRQCD calculation of $\alpha(p_T)$ for (left) $J/\psi$ (Braaten~\etal: Ref.~\protect\refcite{Braaten:1999qk}), 
(middle) $\psi'$ (Leibovich: Ref.~\protect\refcite{Leibovich:1996pa}; Braaten~\etal: Ref.~\protect\refcite{Braaten:1999qk}; Beneke~\etal: Ref.~\protect\refcite{Beneke:1996yw}) and (right) $\Upsilon(1S)$ (Braaten-Lee: Ref.~\protect\refcite{Braaten:2000gw}).(left reprinted figure 
from Ref.~\protect\refcite{Braaten:1999qk} with permission of American Physical Society. Copyright (2000);
from Ref.~\protect\refcite{Braaten:1999qk}, middle and right from Ref.~\protect\refcite{yr})}
\label{fig:pol_NRQCD}
\end{figure}

In \cf{fig:pol_NRQCD}, we show the various polarisation calculations from NRQCD for prompt $J/\psi$, 
direct $\psi'$ and $\Upsilon(1S)$ with feed-down from higher bottomonium taken into account. The least
that we may say is that NRQCD through gluon fragmentation is not able to describe the present 
data on polarisation, especially if the trend to have $\alpha \leq 0$ at high $p_T$ is confirmed by future 
measurements.

Motivated by an apparent discrepancy in the hierarchy of the LDME's for $J/\psi$ and $\psi'$, Fleming,
Rothstein and Leibovich\cite{Fleming:2000ib} proposed  different scaling rules belonging to
NRQCD$_c$. Their prediction for the cross-section was equally good and for polarisation they
predicted that $\alpha$ be close to $1/3$ at large $p_T$.

This latter proposal nevertheless raises some questions since the very utility of the scaling rules
was to provide us with the evolution of the unknown matrix elements of NRQCD when the quark velocity $v$
changes, equally when one goes from one quarkonium family to another. Indeed, a LMDE scaling
as ${\cal O}(v^4)$ may be bigger than another scaling as ${\cal O}(v^3)$ since we have no control
on the coefficient multiplying the $v$ dependence. On the other hand, a comparison of the same
LDME for a charmonium and the corresponding bottomonium is licit. Now, if the counting rules are 
modified between
charmonia and bottomonia, the enhancement of the predictive power due to this scaling rules is likely to be
reduced to saying that the unknown coefficient should not be that large and an operator scaling as
${\cal O}(v^4)$ is conceivably suppressed to one scaling as ${\cal O}(v^3)$, not more than
a supposition then.

\subsection{$k_T$ factorisation and BFKL vertex}\label{subsec:kt_fact}

If one considers the production of charmonium or bottomonium at hadron colliders such that the Tevatron,
for reasonable values of $p_T$ and of the rapidity $y$, it can be initiated by partons with momentum 
as low as a few percent of that of the colliding hadrons. In other words, we are dealing with 
processes in the low Bjorken $x$ region. In that region we usually deal with the BFKL equation, 
which arises from the resummation of $\alpha_s \log(1/x)$ factors in the partonic distributions. 
This process of resummation involves what we can call Balitsky-Fadin-Kuraev-Lipatov (BFKL) 
effective vertices and the latter can be used in other processes than the evolution of parton 
distributions.

On the other hand, the $k_T$ factorisation 
approach\cite{Catani:1990xk,Catani:1990eg,Collins:1991ty,Camici:1996st,Camici:1997ta,Ryskin:1999yq}, 
which generalises the collinear approach to nonvanishing transverse momenta for the initial
partons, can be coupled to the NLLA vertex to describe production processes, such as those of 
heavy quark or even quarkonium.

This combination of the $k_T$ factorisation for the initial partons and the NLLA BFKL effective
vertex\cite{Fadin:1996nw} for the hard part can be thought as the natural 
framework to deal with low $x$ processes since its approximations  are especially valid 
in this kinematical region. As an example, very large contributions from NLO in the collinear factorisation
are already included in the LO contributions of this approach. A typical case is the 
fragmentation processes at large $p_T$ in quarkonium production.

Since this approach is thought to be valid for low $x$ processes, but still at a 
partonic scale above $\Lambda_{\rm QCD}$, this can be also referred to as the semi-hard approach.
One can find an useful review about the approach, its applications and its open questions in
the two following papers of the small-x collaboration\cite{Andersson:2002cf,Andersen:2003xj}.

\subsubsection{Differences with the collinear approach}

Practically, compared to the collinear approach, we can highlight two main differences. 
First, instead of appealing to collinear parton distributions, we 
are going to employ the unintegrated PDF ${\cal F} (x,{\bf q}_T)$ for which there exist 
different parameterisations, exactly as for the collinear case. The discussion of the 
difference between these is beyond the scope of this review.

These are however related to the usual PDF's by:
\eqs{
  x g(x,\mu^2)=\int^{\mu^2}_0 \frac{d{\bf q}_T^2}{{\bf q}_T^2} {\cal F} (x,{\bf q}_T).
}

Secondly, the hard part of the process is computed thanks to effective vertices derived 
in~Ref.~\refcite{Fadin:1996nw} or following the usual Feynman rules of pQCD using
an extended set of diagram\cite{Baranov:2002cf} due to the off-shellness
of the $t$-channel gluons and then using what is called the Collins-Ellis ``trick''\cite{Collins:1991ty} 
which consists in the following replacement: $\overline{\epsilon^\mu(q_i)\epsilon^{\star\nu}(q_i)}=
\frac{q_{iT}^\mu q_{iT}^\nu}{|q_{iT}|^2}$. Let us present here the first approach and give the 
expression for the $Q\bar Q$ production vertex ${\cal V}$\cite{Hagler:2000dd}:
\begin{equation}  
\label{BFKL_NLLA_vertex} 
{\cal V}^{ab}=-g_1 g_2 u(k_1)\left(  
T^{a}T^{b}\,{\cal B}(q_1,q_2,k_{1},k_{2})-T^{b}T^{a}{\cal B}^{T}(q_1,q_2,k_{2},k_{1})%
\right) \bar v(k_2),    
\end{equation} 
where $a$, $b$ are the gluon colour indices, $T^{c}=\frac{1}{2} \lambda^c$, 
$k_1$, $k_2$ the (on-shell) heavy-quark and antiquark momenta, $q_1$, $q_2$, the
(off-shell) gluon momentum, and $g_1$, $g_2$ the strong coupling constant which
is evaluated at two different scales, $q_1^2$ and $q_2^2$ respectively. The  
expression for ${\cal B}(q_1,q_2,k_{1},k_{2})$ (and ${\cal B}^T$) is a sum of two terms
\eqs{
\label{eq:BFKL_NNLA_vertex_QQ}  
&{\cal B}(q_1,q_2,k_{1},k_{2})=\gamma ^{-}\frac{\ks q_{1\perp }-\ks k_{1\perp }-m}{%
(q_{1}-k_{1})^{2}-m^{2}}\gamma ^{+}-\frac{\gamma _{\beta }\Gamma  
^{+\,-\,\beta }(q_{2},q_{1})}{(k_{1}+k_{2})^{2}}\,,  \\
&\left({\cal B}^{T}(q_1,q_2,k_{2},k_{1})=\gamma ^{+}\frac{\ks q_{1\perp }-\ks k_{2\perp }+m}{%
(q_{1}-k_{2})^{2}-m^{2}}\gamma ^{-}-\frac{\gamma _{\beta }\Gamma  
^{+\,-\,\beta }(q_{2},q_{1})}{(k_{1}+k_{2})^{2}}\right).
}
where $\Gamma ^{+\,-\,\beta }(q_{2},q_{1})=2(q_{1}-q_{2})^{\beta  
}-2q_{1}^{+}n^{-\beta }+2q_{2}^{-}n^{+\beta }-2q_1^2\,\frac{n^{-\beta }}{%
q_{1}^{-}+q_{2}^{-}}+2q_2^2\,\frac{n^{+\beta }}{q_{1}^{+}+q_{2}^{+}}$.

The first  corresponds to the contribution of
B in \cf{fig:BFKL_NNLA_vertex_QQ}; the second, with an $\frac{1}{s}$ propagator, is 
linked to a transition between
two (off-shell) $t$-channel gluons (Reggeons) and a (off-shell) gluon which 
subsequently splits into the heavy quark (C in \cf{fig:BFKL_NNLA_vertex_QQ}). 

\begin{figure}[h]
\centering
\includegraphics[height=6cm]{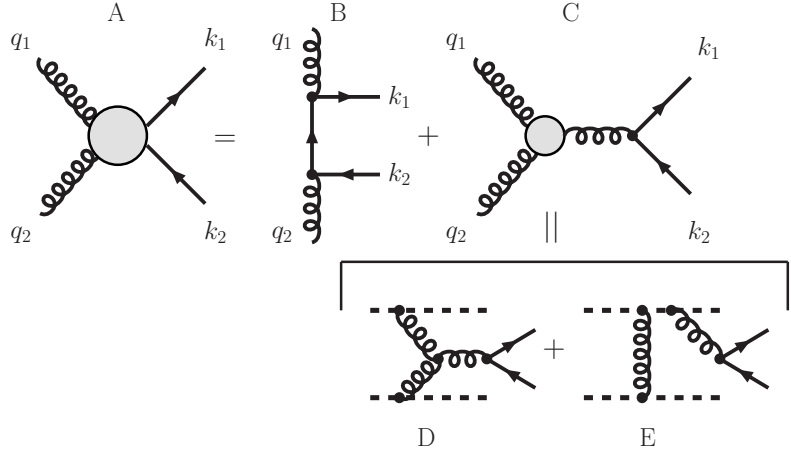}
\caption{Different components in the $Q\bar Q$ production BFKL vertex.}
\label{fig:BFKL_NNLA_vertex_QQ}
\end{figure}

This term is not only derived from a triple gluon vertex (D of \cf{fig:BFKL_NNLA_vertex_QQ}). 
The complication is expected since the gluons are off-shell. Indeed
if we want to deal with on-shell particles to impose current conservation, 
we have to go back to the particles --here the initial hadrons--
which have emitted these $t$-channel gluons. Those can directly emit the third
gluon in addition to the $t$-channel gluon (E of \cf{fig:BFKL_NNLA_vertex_QQ}),
similarly to a Brem\ss trahlung contribution.
This kind of emission gives birth to the two terms: $-2q_1^2\,\frac{n^{-\beta }}{  
q_{1}^{-}+q_{2}^{-}}+2q_2^2\,\frac{n^{+\beta }}{q_{1}^{+}+q_{2}^{+}}$. The factors
$q_1^2$ and $q_2^2$ account for the fact that there is only one $t$-channel gluon
in this type of process, the denominator comes from the propagator of the particle which 
has emitted the $s$-channel gluon.

The vertex of~\ce{eq:BFKL_NNLA_vertex_QQ} has been used successfully for 
the description of open beauty production\cite{Hagler:2000dd} and was also
shown to give a large contribution to the production of $\chi_{c1}$ in the Colour-Singlet 
approach\cite{Hagler:2000de}.

In the case of $\psi$ and $\Upsilon$ production in the Colour-Singlet approach, 
we need to consider the vertex $g^\star g^\star \to Q\bar Q g$ rather than
 $g^\star g^\star \to Q\bar Q$ to conserve $C$-parity. This would correspond
to NNLLA corrections and these are not known yet. However, the reason of this 
complication ($C$-parity) is also from where one finds the solution since 
it sets the contributions
of the unknown diagrams to zero when projected on a colour-singlet $^3S_1$ 
$Q \bar Q$ state. The expressions are finally the same with the addition
of one gluon emission on the quark lines (see~Ref.~\refcite{Hagler:2000eu}).

The cross section is then obtained after the integration on the 
transverse momentum of the gluons\footnote{The fractions of momentum
carried by the gluons $x_{1,2}=\frac{q^+_{1,2}}{P^+_{1,2}}$ are automatically
integrated during this integration.} and on the final state momenta ($P$ is
the quarkonium momentum, $k$ that of the final state gluon).
As in the usual CSM, the heavy-quark pair is projected on a colour-singlet $^3S_1$ state and their relative
momentum is set to zero ($k_1=k_2$ and $P=k_1+k_2$). This gives\footnote{To make
connection with the ``trick'' of Collins and Ellis\cite{Collins:1991ty} as presented 
in\cite{Baranov:2002cf}, note the presence in the 
denominators of $q^2_{1T}$ and $q^2_{2T}$ which would come from the following 
replacement of the $t$-channel gluon polarisation vectors: $
\overline{\epsilon^\mu(q_i)\epsilon^{\star\nu}(q_i)}=\frac{q_{iT}^\mu q_{iT}^\nu}{|q_{iT}|^2}$.
The vectors in the numerators $q_{iT}^\mu q_{iT}^\nu$ would appear \eg~in the three first
terms of $\Gamma^{+\,-\,\beta }$.}
\eqs{
\sigma_{P_1P_2\to {\cal Q}gX}=\frac{1}{16 (2\pi)^4}\frac{1}{8^2}\int 
\frac{d^3P}{P^+}\frac{d^3k}{k^+} d^2\vect q_{1T}d^2\vect q_{2T}
&\delta^2(\vect q_{1T}-\vect q_{2T}-\vect k_{1T}-\vect k_{2T})\times\\&\times
\frac{{\cal F}(x_1,\vect q_{1T})}{(q^2_{1T})^2} \frac{{\cal F}(x_1,\vect q_{2T})}{(q^2_{2T})^2}
\left|{\cal M}\right|^2,
}
where -- $a$ and $b$ being the colour indices of the $t$-channel gluons, 
$i$ and $j$ those of the heavy quarks 
($\left\langle 3i,\bar{3}j|1\right\rangle=\frac{\delta^{ij}}{\sqrt{3}}$) and  ${\cal R}(0)$  the
radial part of the Schr\"odinger wave function at the origin (in position space)  -- 
\eqs{{\cal M}^{ab}=\sum_{i,j}\sum_{L_{z},S_{z}}\frac{1}{\sqrt{m}}&\left\langle 
0,0;1,S_{z}|1,J_{z}\right\rangle
\left\langle 3i,\bar{3}j|1\right\rangle
\frac{{\cal R}(0)}{\sqrt{4\pi}}\times\\&\times
{\rm Tr}\left(\frac{\ks \epsilon(S_z)(\ks k_1+m)}{\sqrt{M}} 
{\cal V}_{Q_i\bar Q_j g}^{ab}(q_1,q_2,k_{1},k_{2}=k_{1},k)\right).
}

Different projections can be used if one studies other quarkonia than the 
$^3S_1$. This formalism can be also combined with the COM by projecting
the heavy-quark pair on a colour-octet state and introducing a LDME to give
the probability for the non-perturbative transition into the physical
quarkonium, exactly as in the collinear approach presented in 
section~\ref{subsec:NRQCD}. Of course, as we shall see in the results,
these LDME's will have modified values compared to the ones of the collinear fits.

\subsubsection{Results for $J/\psi$, $\psi'$ and $\Upsilon(1S)$}

Let us first present the result of H\"agler \etal\cite{Hagler:2000eu} who have used the KMS unintegrated 
PDF\cite{Kwiecinski:1997ee}. Since the hard part of the process is modified
compared to the collinear case, it is not surprising that the ratio $k$ of the
cross section of ${}^1\!S^{(8)}_0$ and ${}^3\!P^{(8)}_0$ is different, the slopes are
still similar, though; an independent fit would not again distinguish the two LDME's associated with these
processes. The ratio $k$ lies between 6 at low $P_T$ and 4.5 and high 
$P_T$. They did the fit with these two contributions
alone; their results are given in~\ct{tab:table_ktfact_1}. The first combination (only $^1S^{(8)}_0$)
is plotted in \cf{fig:sigma_hagler_1}.

\begin{table}[H]
\tbl{Same as~\ct{table2} for $J/\psi$ production in the $k_T$ factorisation
approach and the NLLA BFKL vertex. Values  are given in units $10^{-2}$~GeV${}^3$.}{
$ \footnotesize
\begin{array}{|c||c|c|c|c|c|c|}
\hline
 \mbox{} & {\mbox{PDF}} & \langle
 {\cal{O}}^{J/\psi}[{}^3S^{(8)}_1] \rangle &
 M_{k}^{J/\psi}({}^1S^{(8)}_0,{}^3P^{(8)}_0) & k  & \\ \hline\hline 
\mbox{H\"agler \etal~1} &  {\mbox{KMS Ref.~\protect\refcite{Kwiecinski:1997ee}}} & 
 0.032 \pm 0.012 & 7.0 \pm 0.5 &  5 &\mbox{only} ^1S^8_0 \\ 
\hline 
\mbox{H\"agler \etal~2} &
{\mbox{KMS Ref.~\protect\refcite{Kwiecinski:1997ee}}} & 0.05\pm 0.012 & 6 \pm 0.5 & 5 & \mbox{only} ^3P^8_J\\ \hline
\end{array}
$
}
\label{tab:table_ktfact_1}
\end{table}

\begin{figure}[h]
\centering
\includegraphics[height=6cm]{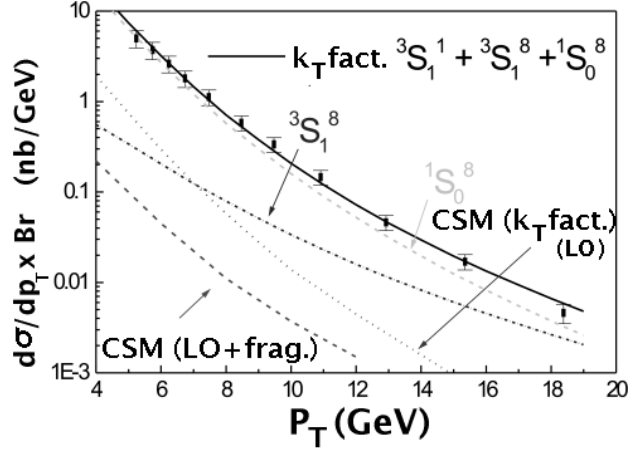}
\caption{$J/\psi$ production cross sections at $\sqrt{s}=1.8$ TeV for the CSM in the 
collinear approach (LO + fragmentation), 
in the $k_T$ factorisation approach for the Colour-Singlet contribution and
the Colour-Octet ($^3S_1^{(8)}$ and $^1S_0^{(8)}$) compared to the direct measurement of
CDF~\protect\cite{CDF7997b}
(Reprinted figure from Ref.~\protect\refcite{Hagler:2000eu} with permission of American Physical Society. Copyright (2001)).}
\label{fig:sigma_hagler_1}
\end{figure}

To what concerns the colour-singlet contribution, as can be seen on \cf{fig:sigma_hagler_1}, 
it is more than one order of magnitude
larger than the LO CSM contribution, and also larger than fragmentation-CSM contribution. 
The same trend is confirmed by the $\psi'$ case\cite{Yuan:2000cp}. This can be partially
explained (up to a factor 2.5) by a genuine different choice for the scale at which 
$\alpha_s$ should be evaluated (cf. the factors $g_1$ and $g_2$ in \ce{BFKL_NLLA_vertex}).

On the other hand, the colour-octet LDME $\langle {\cal{O}}^{J/\psi}[{}^3S^{(8)}_1] \rangle$ is 
thirty times smaller than in the collinear fits (compare~\ct{table2} and~\ct{tab:table_ktfact_1})
whereas $M_{k}^{J/\psi}({}^1S^{(8)}_0,{}^3P^{(8)}_0)$ is similar. In Ref.~\refcite{Hagler:2000eu},
it is however emphasised
that $\langle {\cal{O}}^{J/\psi}[{}^3S^{(8)}_1] \rangle =0$ would give a much worse fit,
whereas, according to Baranov\cite{Baranov:2002cf}, $\langle {\cal{O}}^{J/\psi}[{}^3S^{(8)}_1] \rangle$
can be set\footnote{In fact, the values obtained in this work are not from fit. The theoretical
uncertainties highlighted there were too large to make a fit meaningful\cite{baranov_private}.} 
to 0 with the unintegrated PDF of Ref.~\refcite{Blumlein:1995eu}. The latter analysis of Baranov is
dominated by the Colour-Singlet contribution and due to the longitudinal polarisation of the initial 
off-shell gluons\cite{baranov_private},
the polarisation parameter $\alpha$ should be negative and close to -1 as soon as $P_T$ reaches 6~GeV
both for $J/\psi$ and $\psi'$. The same results for $J/\psi$ and $\psi'$ as presented in Fig.~3 
of Ref.~\refcite{Yuan:2000qe} are irrelevant since they do not take into account the weights of the 
different contributions in the particle yield. The polarisation of
the quarkonia from isolated colour-octet channels is not measurable and 
it is obvious from the cross-section plots (Fig.~1 and Fig.~2 of Ref.~\refcite{Yuan:2000qe})
that none of the colour-octet channels can reproduce the cross section alone, they should be
combined. 

To conclude this section, let us mention the combination of the $k_T$ factorisation approach 
with the colour-octet gluon fragmentation by Saleev~\etal\cite{Saleev:2003ys}. 
The hard part which is considered is $g^*g^* \to g$ where the $g^*$ are reggeised gluons 
(again distributed according to an unintegrated PDF) and where the gluon, $g$, 
subsequently fragments into a quarkonium via the colour-octet state $^3S_1^{(8)}$, 
exactly like in the collinear approach.
The LDME values used were the one from the collinear fit\cite{Braaten:1999qk} (see \ct{table2}) : 
$\langle {\cal{O}}^{J/\psi}[{}^3S^{(8)}_1] \rangle= 0.44 \times 10^{-2}$ 
GeV$^3$ and $\langle {\cal{O}}^{\psi'}[{}^3S^{(8)}_1] \rangle= 0.42 \times 10^{-2}$ GeV$^3$. The
agreement is reasonable contrasting with the conclusion from the LO $k_T$ factorisation
analysis of H\"agler \etal\cite{Hagler:2000eu} which requires a strong suppression of the ${}^3S^{(8)}_1$ channel.

\subsection{Durham group: enhanced NNLO contributions}

As already said, $^3S_1$ quarkonia produced by gluon fusion 
are necessarily accompanied by a third gluon in the final state. Indeed, it is required for 
$C$-parity conservation and in the case of
semi-inclusive reaction, as the ones we have been considering so far, this gluon cannot 
 come from the initial states (see \cf{fig:durham_diag} (left)).

As we have seen, the classical description --through the CSM-- of this kind of production 
(especially at LO) in QCD severely underestimates the production rates as measured at the Tevatron and even
at RHIC. In their work\cite{Khoze:2004eu}, V.A. Khoze~\etal~considered the special case 
where this third gluon, 
attached to the heavy-quark loop, couples to another parton of the colliding hadrons and produces
the gluon needed in the final state. Indeed, it is likely that the large number of possible graphs 
--due to the large number of available gluons at large $s$-- may
compensate the $\alpha_s$ suppression. The parton multiplicity $n$ behaves like
$\log s$ and this process (see \cf{fig:durham_diag} (right)) can be considered as the LO amplitude in the
BFKL approach whereas it is NNLO in pQCD.

\begin{figure}[h]
\centering
\includegraphics[height=6cm]{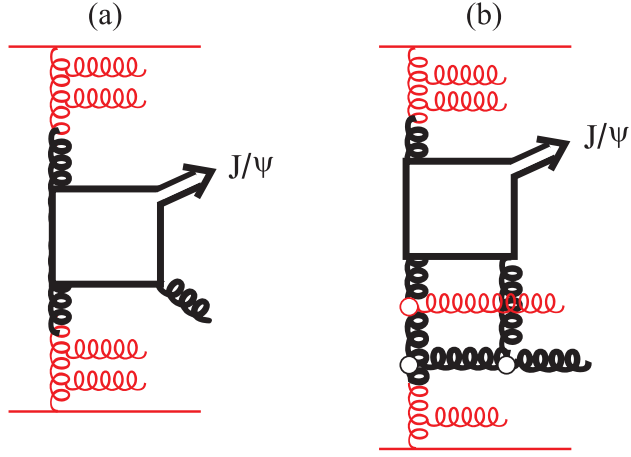}
\caption{(left) Usual LO pQCD in the CSM; (right)
NNLO pQCD or LO BFKL contributions. In both cases, the $gg \to {\cal Q}g$ sub-process is shown in bold and 
the two quarks entering the $\cal Q$ are on-shell.}
\label{fig:durham_diag}
\end{figure}

\subsubsection{Integrated cross section}

Since the two $t$-channel gluon off the quark loop are now in a colour-octet symmetric state,
the real part of the amplitude is expected to dominated by its imaginary part 
(in \cf{fig:durham_diag} (right) the two quarks and the gluon is the $s$-channel are then put 
on-shell as well as the remaining quark entering the $\cal Q$). 
One then chooses to work  in the region where the rapidity between the $\cal Q$ and the final 
gluon $p$ is large (\ie~when $\hat s \gg M_{\cal Q}$) since it should dominate and one
gets\cite{Khoze:2004eu} the following expressions for the imaginary part of the amplitude for the two
possible Feynman graphs (the two other ones with the loop momentum reverted give a factor two):

\eqs{
\Im m A^a=& \frac{3}{8} d^{abc} g (4 \pi \alpha_s)^{5/2}\times \\\times &
\int d\ell_T^2 
\frac{\tr(\ks \ep(p_1) (\frac{\kks P}{2}+m_Q) \ks\ep(P) \ks p_2 (\frac{-\kks P}{2}-\ks\ell+m_Q) \ks p_2
 (\frac{\kks P}{2}-\ks p_1+m_Q))}{ 2 \pi \hat s ((\frac{P}{2}-p_1)^2-m_Q^2)(\ell^2-\lambda_g^2)((q+\ell)^2-\lambda_g^2)}}

\eqs{
\Im m A^b=& -\frac{3}{8} d^{abc} g (4 \pi \alpha_s)^{5/2}\times \\\times &
\int d\ell_T^2 
\frac{\tr(\ks \ep(p_1) (\frac{\kks P}{2}-\ell+m_Q)  \ks p_2 (\frac{-\kks P}{2}+m_Q) \ks\ep(P)\ks p_2
 (\frac{\kks P}{2}-\ell-\ks p_1+m_Q))}{ 2 \pi \hat s ((\frac{P}{2}-\ell-p_1)^2-m_Q^2)(\ell^2-\lambda_g^2)((q+\ell)^2-\lambda_g^2)},
}
$g$ is a quantity related, up to some known factors, to $|\psi(0)|^2$ through the leptonic decay 
width ($g^2=\frac{3 \Gamma_{\ell\ell}M_{\cal Q}}{64 \pi \alpha_{QED}^2}$) such that 
the $^3S_1$ vertex reads $\frac{g}{2} (\kks P+M_{\cal Q})\ep^*(P)$. $\lambda_g$ is an effective gluon mass to avoid logarithmic
infrared singularity, motivated by possible confinement effects.

The partonic differential cross section thus reads:
\eqs{\frac{d\hat \sigma}{dP^2_T}=\frac{|A|^2}{16 \pi \hat s^2},}
with $A=2(A^a+A^b)$. The hadronic cross section is obtained with the help of
\eqs{
\frac{d \sigma}{dy dP^2_T}=\int \frac{dx_2}{x_2}x_1g(x_1)x_2g(x_2)\frac{d\hat \sigma(\hat s,P^2_T)}{dP^2_T},
}
$\hat s=s x_1 x_2$ with $\sqrt{s}$ the collision energy in the hadronic frame,
$y$ is the rapidity of the $\cal Q$ also in the latter frame, $g(x_i)$ are the
gluon PDF.

Due to the low-$x$ behaviour of the PDF, the main contribution to the integral comes from the 
lowest value of
$x_2$, that is $x_2\simeq \frac{\sqrt{M^2_{\cal Q}+P^2_T}}{\sqrt{s}} e^{-y}$ (and thus
$x_1\simeq \frac{\sqrt{M^2_{\cal Q}+P^2_T}}{\sqrt{s}} e^{y}$).
For reasons exposed in Ref.~\refcite{Khoze:2004eu}, the $x_2$-integration\footnote{~\ie~setting  $x_2$ to 
$\frac{\sqrt{M^2_{\cal Q}+P^2_T}}{\sqrt{s}} e^{-y}$.}
region is to be extended over the whole kinematically available rapidity interval $\Delta y$, this gives
\eqs{
\frac{d \sigma}{dy dP^2_T}=x_1(y)g(x_1(y))x_2(y)g(x_2(y))\Delta y \frac{d\hat \sigma(\hat s(x_1,x_2),P^2_T)}{dP^2_T},
}
with $x_{1,2}(y)=\frac{\sqrt{M^2_{\cal Q}+P^2_T}}{\sqrt{s}} e^{\pm y}$ and 
\eqs{
\Delta y=\ln(\frac{x^2_{max}s}{M^2_{\cal Q}+P^2_T}), 
}
with $x_{max}$ set to 0.3 to exclude contributions when the third gluon couples to partons
with $x>0.3$ -- this would have normally been suppressed by the PDF in conventional calculations. 
See Ref.~\refcite{Khoze:2004eu} for further
discussion about uncertainties linked to those approximations. Integrating
over $P_T$, for $\sqrt{s}=1.96$ TeV with the LO MRST2001 gluon PDF\cite{Martin:2002dr} at the 
scale $\mu=0.5 \sqrt{M^2_{\cal Q}+P^2_T}$, for $|y|<0.6$ and $\lambda_g=0.8$ GeV, 
the integrated cross section is
\eqs{\sigma \simeq 2.7 \mu b.}

This seems in agreement with the latest measurement by CDF\cite{Acosta:2004yw} at $\sqrt{s}=1.96$ 
TeV in the whole $P_T$ range but for the total cross section only; the extraction of the prompt 
signal was only done for $P_T >1.25$ GeV.

As exposed in Ref.~\refcite{Khoze:2004eu}, this calculation is affected by the following uncertainties:

\begin{enumerate}
\item Choice of the effective gluon mass, $\lambda_g$:  for $\lambda_g= 0.5, 0.8, 1.0$ GeV, $\sigma$
is 2.0, 2.7, 4.0 $\mu b$;
\item Choice of the factorisation $\mu_F$ scale (at which the PDF are evaluated) and 
renormalisation $\mu_R$ scale (at which $\alpha_s$ is evaluated): defining 
$\mu_0=\sqrt{M^2_{\cal Q}+P^2_T}$, setting $\mu_F=\mu_R$ 
to $0.5\, \mu_0$, $\mu_0$, $2\, \mu_0$, $\sigma$ is 2.7, 2.3, 1.5 $\mu b$;
\item Choice of the cut-off $x_{max}$: its variation introduces NLL corrections in the BFKL approach.
\end{enumerate}

Beside those, we have the usual uncertainties linked to the PDF (especially at low $x$) and a 
possible $K$-factor or equally higher-order pQCD corrections.

Using the same parameters and setting $y=0$, one can in turn compute the cross sections for $\psi(2S)$, 
but also for $\Upsilon(nS)$  at $\sqrt{s}=1.96$ TeV and at $\sqrt{s}=14$ TeV (see Table 18).

\begin{table}[H]
\centering
\tbl{Direct cross section calculations Ref.~\protect\refcite{Khoze:2004eu}.}{
\begin{tabular}{|c||c|c|c|c|c|}
\hline 
$\left.d\sigma/dy \right|_{y=0}$ & $J/\psi$ ($\mu b$) & $\psi(2S)$ ($\mu b$) & $\Upsilon(1S)$ 
($nb$)& $\Upsilon(2S)$ ($nb$) & $\Upsilon(3S)$ ($nb$) \\
\hline \hline 
$\sqrt{s}=1.96$ TeV & 2.2 & 0.6 & 40 & 12 & 9\\
$\sqrt{s}=14$   TeV & 8.1 & 2.5 & 310& 100 & 80\\
\hline
\end{tabular}
}
\label{table:durham}
\end{table}

\subsubsection{$P_T$ differential cross section}

Unfortunately, one cannot rely on the amplitude written above the compute the $P_T$ 
differential cross section 
(see Ref.~\refcite{Khoze:2004eu}). As a makeshift, they use a simple parameterisation 
for the partonic cross section based on dimensional counting:
\eqs{
\frac{d \hat \sigma}{dP^2_T} \propto g^2 \alpha^5_s 
\frac{\ln\left(\frac{x^2_{max}s}{M^2_{\cal Q}+P^2_T}\right)}
{(M^2_{\cal Q}+P^2_T)^3}.
}

Taking again $x_{max}=0.3$ and normalising the distribution by equating its integral over
$P^2_T$ to the previous one, one obtains a reasonable agreement with the data from CDF. 
Again, the comparison is somewhat awkward since their prediction (for $\sqrt{s}=1.96$ TeV) 
is only for direct production (what they call ``prompt'') whereas the data at $\sqrt{s}=1.96$ TeV are 
still only for prompt (and for $P_T > 1.25$ GeV only).

\subsubsection{Other results}

Since some NNLO processes seem to have enhanced contributions, a second class 
of diagrams was considered where the two $t$-channel gluons ``belong to two different pomerons'', or two 
different partons showers. This kind of contributions can be related to the single diffractive
 cross section (see Ref.~\refcite{Khoze:2004eu}). It was however found that this class of diagrams 
contributes less that the one considered above, though it may compete with it at large $\sqrt{s}$.

In the same fashion, they consider associative production ${\cal Q}+ c\bar c$, for which they expect
$\hat \sigma (\psi c \bar c)$ to be close to 2 $nb$, which gives for the hadronic cross 
section\cite{Khoze:2004eu}:
\eqs{\left.\frac{d\sigma (\psi c \bar c)}{dy}\right|_{y=0}=x_1 g(x_1) x_2 g(x_2) 
\hat \sigma (\psi c \bar c)\approx 0.05 \mu b.} In this case, it is about only 1 \% of the other contributions.

\subsection{CES: Comover Enhancement Scenario}

\subsubsection{General statements}

In their model\cite{Hoyer:1998ha,Marchal:2000wd}, P. Hoyer and S. Peign\'e 
postulate the existence of a perturbative
rescattering of the heavy-quark pair off a comoving colour field, which
arises from gluon radiations preceding the heavy-quark pair production.
The model is first developed for low $P_T \lesssim m$ production 
\cite{Hoyer:1998ha}, and generalised
in a second work\cite{Marchal:2000wd}, to large $P_T \gg m$, where 
quarkonium production is dominated by fragmentation.
In general (\ie~at low and large $P_T$), the model assumes a rich colour 
environment in the fragmentation region of any (coloured) parton, modelled 
as a comoving colour field. In the case of large-$P_T$ production, the 
comoving field is assumed to be produced by the fragmenting parton DGLAP 
radiation. The strength and precise shape of the comoving field are 
not required to be the same at small and large $P_T$. 

If, as they suppose, the presence of the comoving field is responsible for an enhancement of the 
production cross section, as it is absent in photon-hadron collisions
(no colour radiation from the photon and no fragmentation
in the $P_T$ region where the data are taken)
there would not be any increase in photo-production. Indeed, the NLO
CSM cross section\cite{Kramer:1995nb} fits well the data and no modification is needed. 

Taking benefit of this assumed perturbative character of the scattering and assuming 
simple properties of the comoving colour field -- namely a classical isotropic\footnote{ in the CMS
of the heavy-quark pair.} colour field --, they are able to carry out the
calculation of the rescattering, even when two rescatterings of the heavy-quark pair 
are required, as in the gluon fragmentation case\cite{Marchal:2000wd}, 
to produce a colour-singlet $^3S_1$ state. In the latter case, which is relevant 
at large transverse momentum,  the $\psi'$ (as well as directly produced $J/\psi$) 
is predicted to be produced unpolarised. 

Another assumption, motivated by the consideration of the relevant time scales, 
is that the heavy quarks propagate nearly on-shell
both before and after the rescattering. In the latter case, the assumption is comparable to 
the static approximation of the CSM. Furthermore, the rescattered quark pair is projected
on a colour-singlet state which has the same spin state as the 
considered $\cal Q$, similarly to the CSM.

Note that since the strength of the field is unknown, 
only cross-section ratios and polarisations can be predicted 
in the framework of this model, not absolute normalisations.
Let us review the high-$P_T$ case\cite{Marchal:2000wd} which interests us most.

\subsubsection{Results of the model}

Considering two perturbative scatterings as illustrated in \cf{fig:CES_highpt} and working
in a first order approximation for quantities like $\frac{\vect \ell_i}{m}$ and
$\frac{\vect q}{m}$ for on-shell quarks before and after the rescattering, they show\cite{Marchal:2000wd} 
that the rescattering amplitude to form a colour-singlet $^3S_1$ quarkonium from a gluon can 
be cast in the following
simple form\footnote{the details of the calculation can be found in Ref.~\refcite{Marchal:2000wd}.}:

 \begin{equation}
{\cal M}(^3S_1,S_z) = \frac{- {\cal R}(0)  g_s^3}{2 \sqrt{6\pi m^3}} \,
d_{ab_1b_2} \vect G_1 \cdot \vect \ep(\lambda_g) \, \vect G_2 \cdot \vect \ep(S_z)^*,
     \label{eq:highptsampl}
\end{equation}
where $\lambda_g$ is the polarisation of the fragmenting gluon, $\vect G_i = \vect \Gamma_f(\vect \ell_i) \times \vect \ell_i$ involves the colour field $\Gamma_f$ produced by the fragmentation, $\ell_1$ and
$\ell_2$ are --as depicted in \cf{fig:CES_highpt}-- the momenta of the two rescattering gluons. As usual, 
${\cal R}(0)$ is the radial wave function at origin. Apart from the presence of rescatterings, their 
approach follows the same lines as the CSM with a projection on static colour-singlet states with
the proper spin state.

Very interesting pieces of information can be extracted from this formula: since \ce{eq:highptsampl} is 
independent of $\Gamma_f^0$, solely transverse gluons emerging from the comoving gluon field 
contribute to $^3S_1$ quarkonium production at high $P_T$. Gauge invariance is also 
preserved: ${\cal M}(\Gamma_f^{\mu}(\ell_i) \to \ell_i^{\mu}) = 0$ since
the quarks are on-shell. Besides this, the most interesting part
is the {\it factorised} dependence on the fragmenting gluon and quarkonium polarisations.
They appear in two different scalar products 
with the quantity $\vect G$ involving the comoving field $\Gamma_f$. The polarisation of
the quarkonium depends solely on the  properties of the comoving field through $\vect G_2$.

This outcome of this perturbative calculation is totally at variance with the 
statement of Cho and Wise\cite{Cho:1994ih}, \ie~a $^3S_1$ produced by
a high $P_T$ (thus real and transverse) gluon is to be transversally polarised. More
precisely, if the comoving field is supposed to be isotropically distributed in the comoving
$Q\bar Q$ rest frame, then they predict an unpolarised production of $^3S_1$ 
quarkonia in the high $P_T$ region.

\begin{figure}[H]
\centering
\includegraphics[width=7cm]{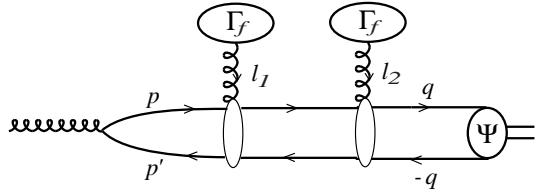}
\caption{Rescattering scenario for a fragmenting gluon into a colour-singlet $^3S_1$ quarkonium.}
\label{fig:CES_highpt}
\end{figure}

To what concerns $P$-waves, they predict that ``rather independently of the form of
$\Gamma_f$'', $\chi_1$ should be ``longitudinally polarised if the rescattering
scenario dominates $\chi_1$  production at high $P_T$''. Finally, we should emphasise
that the approach at high $P_T$ applies equally for charmonium and bottomonium as long
as gluon fragmentation matters. This sole consideration of gluon 
fragmentation can be seen as a drawback: indeed $c$-quark fragmentation seems to be 
bigger than gluon one since it is one order of $\alpha_s$ less (see \cf{fig:dsdpt_divers} (left)).

In conclusion, this simple production model seems quite interesting 
(and motivated by phenomenology \cite{Hoyer:1998ha}), even though the properties of
this comoving field are poorly known (to obtain the results mentioned here, it was supposed
to be classical and isotropically distributed in the $Q\bar Q$ rest frame). The existence of 
such perturbative reinteraction should be also searched for in other processes involving quarkonia.
A qualitative agreement between the predictions of the model and future data on 
polarisation and cross-section ratios would be quite intriguing and might 
indicate the importance of comovers in quarkonium production. 

\subsection{Non-static and off-shell contributions to the CSM}

As we have seen, none of the models reviewed above provides an entirely conclusive
solution to the heavy-quarkonium production problem, especially without the introduction
of new mechanisms and associated free parameters. Considering that the CSM is still 
the most natural way to describe heavy-quarkonium production, it is legitimate to wonder 
whether some of the hypotheses
of this model were actually justified. One of these is in fact
common to the CSM, the COM and the CES  and can be referred to as the static approximation, in the
sense that one considers the hard part of the process solely for configurations
where the heavy quarks are on-shell and at rest in the frame of the quarkonium to be 
created.

This approximation is usually justified by the fact that the wave function, parameterising
the amplitude of probability for the binding, should be peaked at the origin and expanding
its product with the hard amplitude, one has only to keep the first non-vanishing term
in the Taylor series. In the latter, the hard part is evaluated at this static and on-shell
configuration.

This is legitimate as long as the hard part is ``well-behaved'', \eg~does not present singularities
in the non-static region. In order to test the validity of this supposition, one has to reconsider the 
basis of quarkonium ($\cal Q$) production in field theory. In the following, we review a consistent and 
systematic scheme to go beyond this static approximation proposed by J.~P.~Lansberg, J.~R.~Cudell 
and Y.~L.~Kalinovsky. In the latter scheme, when off-shell 
configurations are taken into account, there exist new contributions at the leading-order 
for $^3S_1$ production by gluon fusion. These can compete with the ones considered in the CSM.

This approach requires the use of 3-point vertices depending on 
the relative momentum of the constituent quarks of the $\cal Q$
which can be paralleled to Bethe-Salpeter vertex functions. The arbitrary 
normalisation of these
is fixed by the calculation of the $\cal Q$ leptonic width. 
Gauge invariance, which could be spoilt by non-local contributions, is preserved by the 
introduction of 4-point vertices.

\subsubsection{Description of the bound-state nature of the quarkonia}

The transition 
$q\bar q\rightarrow {\mathcal Q}$ can be described by the following
 3-point function\cite{Lansberg:2005pc,Lansberg:2005aw}:
\begin{eqnarray}\label{vf}
\Gamma^{(3)}_{\mu}(p,P) = \Gamma(p,P) \gamma_\mu,
\end{eqnarray}
with $P\equiv p_{1}-p_{2}$ the total momentum of the bound state, and $p\equiv(p_{1}+p_{2})/2$ 
the relative momentum of the bound quarks. This choice amounts to describing the vector 
meson as a massive photon with a non-local coupling and is justified by other studies (see 
\eg~Ref.~\refcite{Burden:1996nh}).

In order to go beyond the static limit, the quarks
are not supposed to be on-shell. To allow connections with wave functions,
$\Gamma(p,P)$ is taken as a function of the square of the
relative c.m. 3-momentum $\vec p$ of the quarks, which can be written in a 
Lorentz invariant form as $\vec p^{\,\, 2}=
-p^2+\frac{(p.P)^2}{M^2}$ and  possible cuts in the vertex function $\Gamma(p,P)$ are 
neglected. Two opposite scenarios have been considered: 
\eqs{
\Gamma(p,P)=\frac{N}{(1+\frac{\vec p^{\,2}}{\Lambda^2})^2} \text{ and } \Gamma(p,P)=N e^{-\frac{\vec p^{\,2}}{\Lambda^2}} 
,}
both with a normalisation $N$ and a size parameter $\Lambda$, which
can be obtained from relativistic quark models\cite{Lambda1,Lambda2,Lambda3,Lambda4}. The 
normalisation $N$ 
can be fixed from the leptonic-decay width and the procedure to do so 
is thoroughly explained in\cite{Lansberg:2005aw,Lansberg:2005ed}.

\subsubsection{LO production diagrams}

As usual, it is natural to think of gluon-fusion process to produce quarkonia
at high-energy, especially if the value of $P_T$ is not excessively large. To conserve
$C$ parity, in the case of $^3S_1$ production, a third gluon is required; this leads to 
the consideration of $gg \to {\cal Q}g$ processes.   

As explained in Ref.~\refcite{Lansberg:2005pc}, the use of the Landau equations to determine
the discontinuities of the amplitude leads us to the distinction between 
two families of diagrams, the first is the one 
which gives the usual LO CSM contributions\cite{CSM_hadron1,CSM_hadron2,CSM_hadron3}
when the relative momentum of the heavy quarks entering the ${\cal Q}$ is set to zero,
the other family is absent in the on-shell limit and was never considered 
before Ref.~\refcite{Lansberg:2005pc} in the case of ${\cal Q}$ inclusive production. These are shown
in~\cf{fig:diag_LO_QCD}.

\begin{figure}[H]
\centerline{\mbox{
{\includegraphics[height=3.5cm]{LO_CSM_diag}}
\quad\quad\quad
{\includegraphics[height=3.5cm]{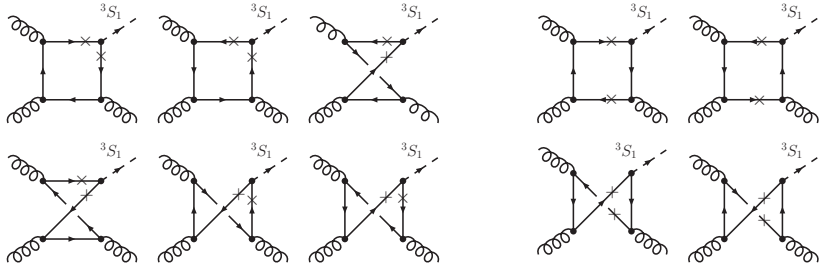}}
}}
\caption{The first family (left) has 6 diagrams
and the second family (right) has 4 diagrams 
contributing to the discontinuity of $gg \to \!\!\ ^3S_1 g$ at LO in QCD.}
\label{fig:diag_LO_QCD}
\end{figure} 

\subsubsection{Gauge invariance}

Now, the diagrams of~\cf{fig:diag_LO_QCD}~(right) are not gauge invariant.
Indeed, the vertex function $\Gamma(p,P)$ takes different values in diagrams where either the on-shell 
quark or the antiquark  touches the $\cal Q$, so that current conservation is shattered compared
to the case where the $\cal Q$ is replaced by a photon.

\begin{figure}[H]
\centerline{\mbox{{\includegraphics[height=4cm]{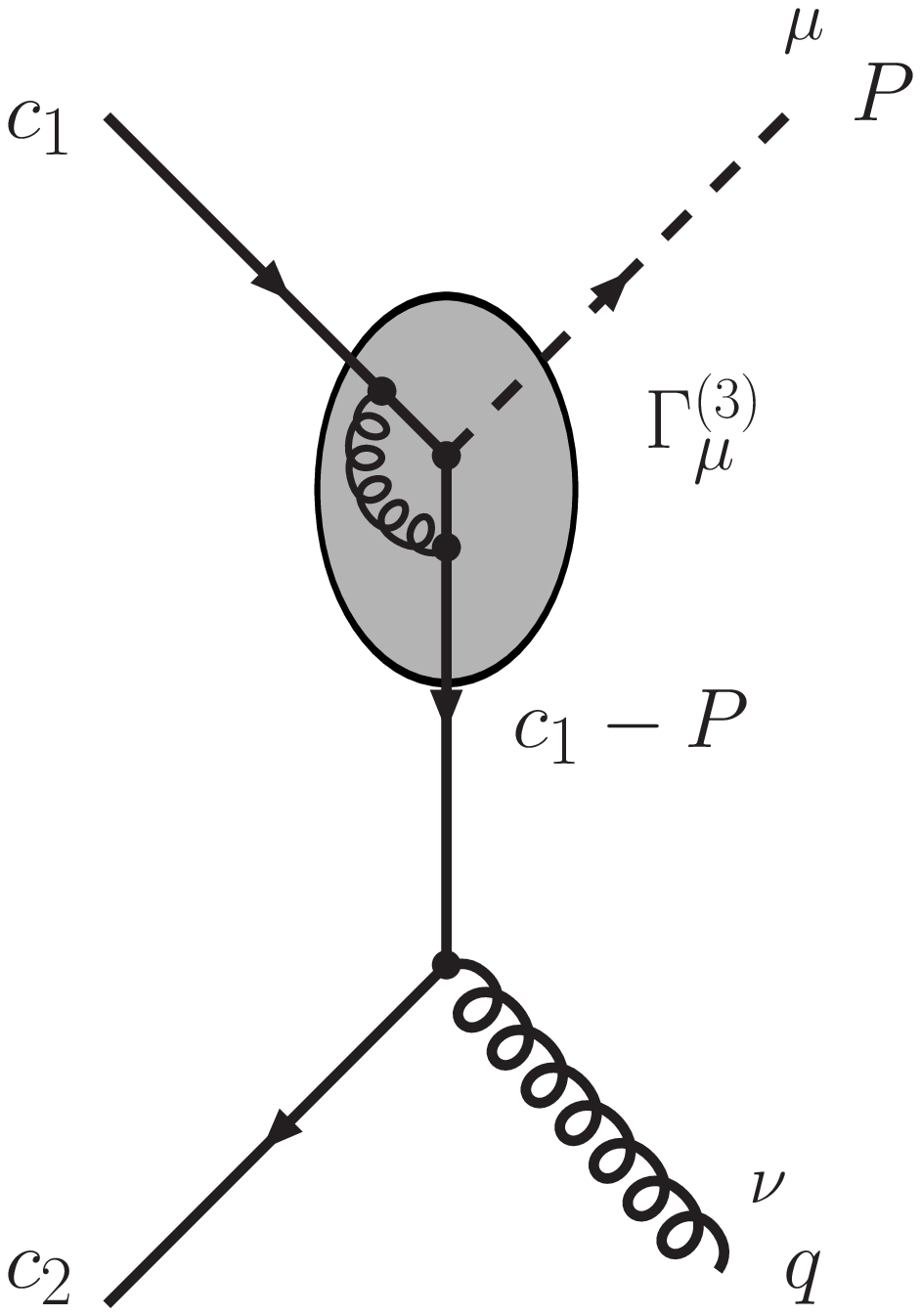}\quad\quad
\includegraphics[height=4cm]{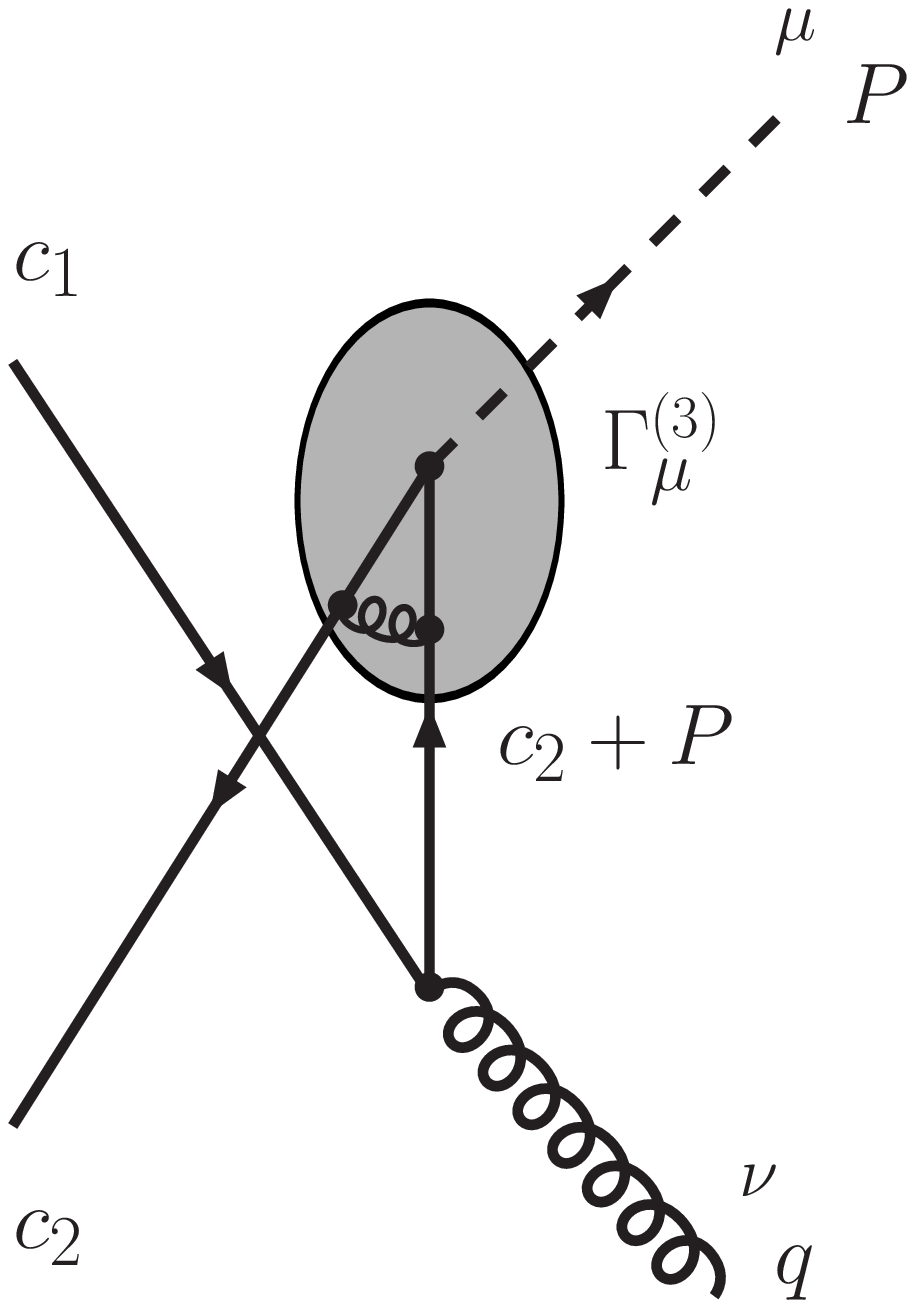}}\quad\quad
{\includegraphics[height=4cm]{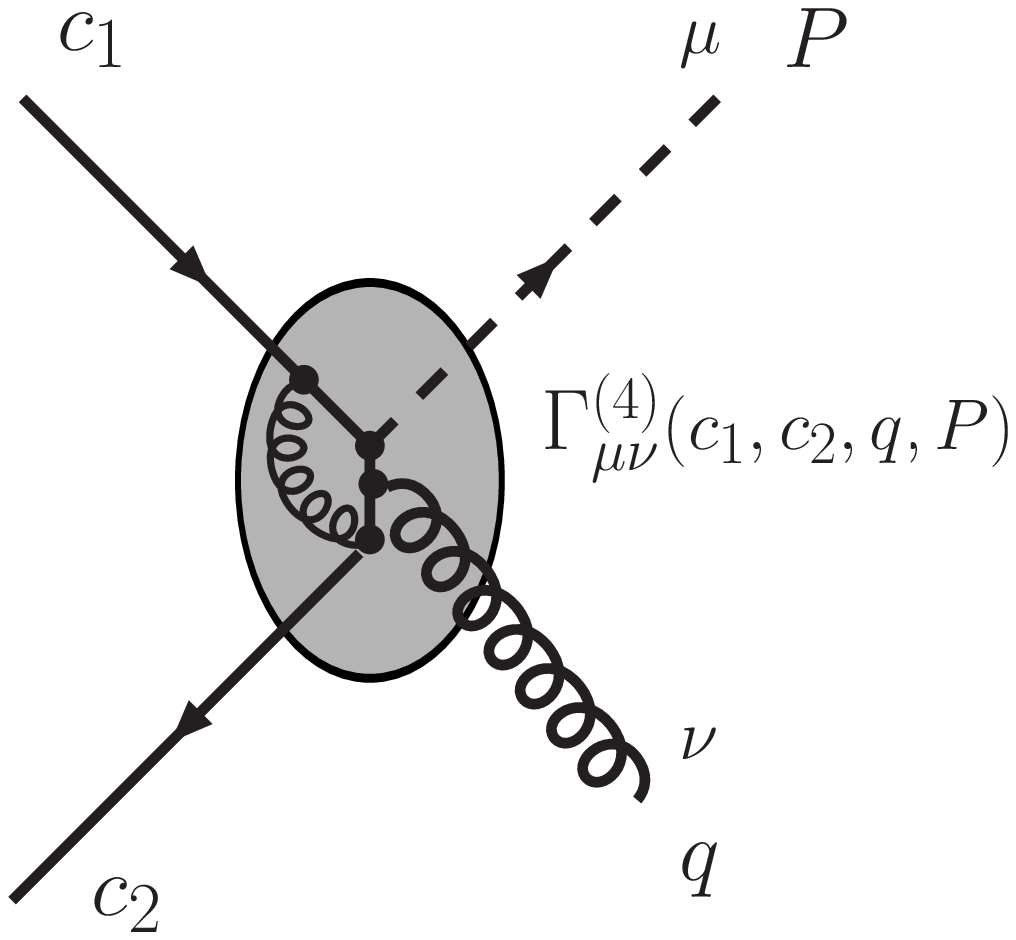}}}}
\caption{Illustration of the necessity of a 4-point vertex.}
\label{fig:illus_GI_break}
\end{figure} 

Physically, this can be understood by the lack of some diagrams:
if one considers a local vertex, then the gluon can only couple to the
quarks that enter it. For a non-local vertex, it
is possible for the gluon to connect to the quark or gluon lines inside the vertex,
as shown in \cf{fig:illus_GI_break}.
These contributions are related to a 4-point $q\bar q {\cal Q}g$ 
vertex, $\Gamma^{(4)}_{\mu\nu}(c_1,c_2,q,P)$, whose form is, in
general, unknown, although it must obey some general 
constraints\cite{Lansberg:2005pc,Lansberg:2005aw,Drell:1971vx}:

\begin{itemize}
\item
it must restore gauge invariance: its addition to the amplitude must
lead to current conservation at the gluon vertex;
\item
it must obey crossing symmetry (or invariance by $C$ conjugation) which 
can be written
\begin{equation}
\Gamma^{(4)}_{\mu\nu}(c_1,c_2,q,P,m)=-\gamma_0\Gamma_{\mu\nu}^{(4)}(-c_2,-c_1,q,P,-m)^\dagger\gamma_0;
\end{equation}
\item
it must not introduce new singularities absent from the propagators or 
from $\Gamma(p,P)$, hence it can only have denominators proportional to $(c_1-P)^2-m^2$ or
$(c_2+P)^2-m^2$;
\item
it must vanish in the case of a local vertex $\Gamma_\mu^{(3)}\propto\gamma_\mu$, 
hence we multiply it by $\Gamma(2c_1-P,P)-\Gamma(2c_2+P,P)$.
\end{itemize}

These conditions are all fulfilled by the following simple choice\cite{Lansberg:2005pc}:
\begin{eqnarray}
\Gamma^{(4)}_{\mu\nu}(c_1,c_2,P,q)&=&-i g_s T^a_{ki} \left[ \Gamma(2c_1-P,P)
-\Gamma(2c_2+P,P)\right]\nonumber \\
&\times&\left[\frac{c_{1\nu}}{(c_2+P)^2-m^2}
+\frac{c_{2\nu}}{(c_1-P)^2-m^2}\right]\gamma_\mu
\label{Gamma4}
\end{eqnarray}
where the indices of the colour matrix $T$
are defined in \cf{fig:gamma_4}, and $g_s$ is the strong coupling constant.

\begin{figure}[H]
\centering
\includegraphics[width=3cm]{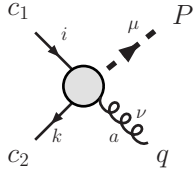}
\caption{The gauge-invariance restoring vertex, $\Gamma^{(4)}$.}
\label{fig:gamma_4}
\end{figure} 

It should be also emphasised that this choice of vertex is not unique. A first study
of the effect of different choices can be found in Ref.~\refcite{Lansberg:2005aw,Lansberg:2005gs},
where we can see that agreement with data can be reached. 
The introduction of such new vertex can be paralleled with the COM since the quark 
pair $(c_1,c_2)$ that makes the meson is now in
a colour-octet state. Such configurations are required here to restore gauge invariance.

This new $\Gamma^{(4)}$ vertex introduces two new diagrams in the calculation
of the amplitude which, now, becomes a gauge-invariant quantity. The detail of the calculation
of the polarised cross section can be found in Refs.~\refcite{Lansberg:2005pc,Lansberg:2005aw}. 

\subsubsection{Results for $J/\psi$ and $\psi'$}

We show in \cf{fig:LCK_results} (left) the results obtained in Ref.~\refcite{Lansberg:2005pc} 
for $\sqrt{s}=1800$ GeV,  $|\eta|<0.6$, $m=1.87$ GeV and $\Lambda=1.8$ GeV. 
The curves for $\sigma_{TOT}$, $\sigma_T$ and $\sigma_L$
are calculated within this approach with the new cut only (see Ref.~\refcite{Lansberg:2005pc}
for details), the LO CSM is recalculated from the expression of Refs.~\refcite{CSM_hadron1,CSM_hadron2,CSM_hadron3}.

In fact, the normalisation of the results using
the decay width removes most dependence on the choice of parameters: instead 
of a factor 100 of difference expected from $\left(\frac{N_{\Lambda=1.0}}{N_{\Lambda=2.2}}\right)^2$ 
there is less than a factor 2 at $P_T=4$ GeV and a factor 3 at $P_T=20$ GeV.
Interestingly, the dependence on $\Lambda$ is
negligible once values of the order of 1.4 GeV are taken.

\begin{figure}[ht]
\centerline{
\mbox{
{\includegraphics[width=0.39\textwidth,clip=true]{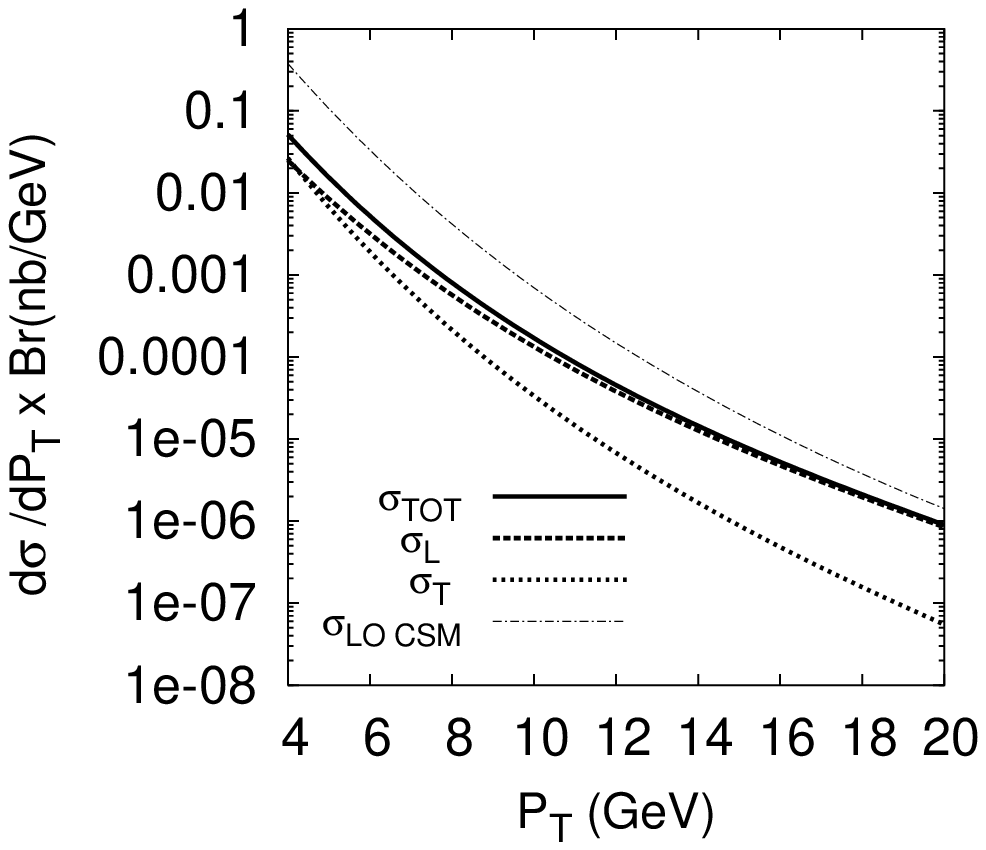}}\quad
{\includegraphics[width=0.5\textwidth,clip=true]{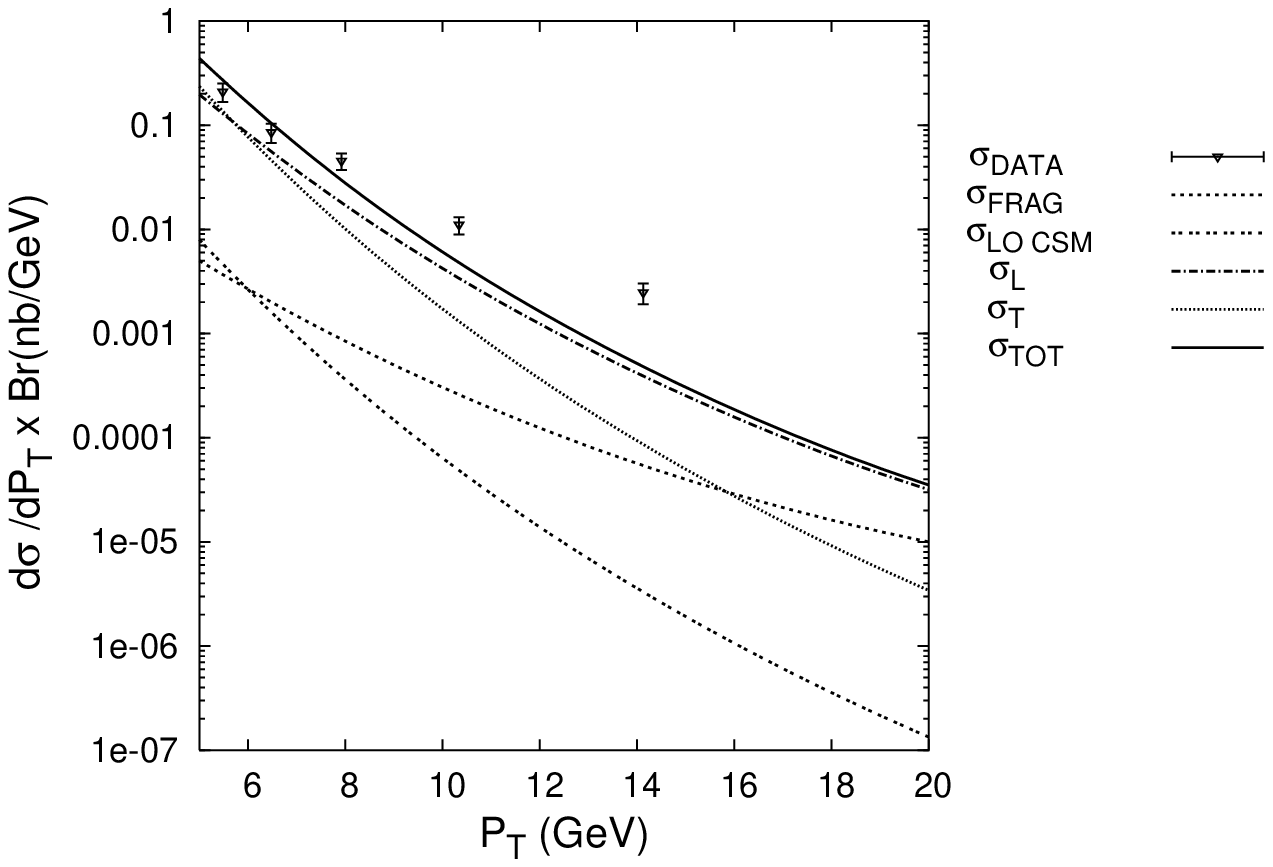}}
}
}
\caption{(left) Polarised ($\sigma_T$ and $\sigma_L$) and total ($\sigma_{TOT}$) cross sections obtained 
with a Gaussian vertex function, $m=1.87$ GeV, $\Lambda=1.8$~GeV 
and the MRST gluon distribution\protect\cite{Martin:2002dr}, to be compared with LO CSM.
(right); Polarised ($\sigma_T$ and $\sigma_L$) and total ($\sigma_{TOT}$) cross sections for $\psi'$
obtained  with a Gaussian vertex function, $a_{node}=1.334$ GeV, $m=1.87$ GeV, $\Lambda=1.8$~GeV 
and the CTEQ gluon distribution \protect\cite{Pumplin:2002vw}, to be compared with LO CSM, 
CSM fragmentation\protect\cite{Cacciari:1994dr,Braaten:1994xb} and the data from CDF\protect\refcite{CDF7997a}.
}
\label{fig:LCK_results}
\end{figure} 

We see that the contribution of the new cut cannot be neglected at large $P_T$.
It is interesting to see that it is much flatter in $P_T$ than the LO CSM, 
and its polarisation is mostly longitudinal. 
This could have been expected as scalar products of $\ep_L$ with momenta in the loop 
will give an extra $\sqrt{\hat s}$ contribution, or equivalently an extra $P_T$ 
power in the amplitude, compared to scalar products involving $\ep_T$.

Although the use of the leptonic width is a good means to fix  most of the dependence on $N$ 
in the $J/\psi$ case, it is not so for radially excited states, such as the $\psi'$.
Indeed, in this case, the vertex function must have a node. We expect that $\Gamma_{2S}(p,P)$ 
 should be well parametrised by
\eqs{\left(1-\frac{|\vect p|}{a_{node}}\right)
\frac{N'}{(1+\frac{|\vect p|^2}{\Lambda^2})^2} \text{ or  }
N'\left(1-\frac{|\vect p|}{a_{node}}\right) e^{\frac{-|\vect p|^2}{\Lambda^2}},}
where $a_{node}$ is the node position.
 
However, the position of the node is not very well-known, and it is 
not straightforward to relate our vertex with off-shell quarks to an on-shell non-relativistic
wave function. 
Let us suppose here that the node has the same position in the vertex function than
in the wave function: now, the integrand changes sign at $a_{node}$ and the positive
contribution to the integral can be compensated by the negative one for a precise
values of $a_{node}$; the latter turns out to be close to the estimated
value of the node in the wave function. In other words, our normalisation procedure, 
translated from that of the CSM, gives a large value for $N$ compared to the 
$J/\psi$ case, whereas the calculation of the $\psi'$ production amplitude is not affected 
much by the presence of a node in $\Gamma$.

\cf{fig:LCK_results}~(right) shows that for $a_{node}=1.334$ GeV, one obtains a good fit to
CDF data at moderate $P_T$ (note that the slopes are quite similar. This is at odds with 
what is commonly assumed since fragmentation processes --with a typical $1/P_T^4$ behaviour-- 
can also describe the data). The $\psi'$ is predicted to be mostly longitudinal.

In conclusion, the procedure followed by Lansberg~\etal~enables to go beyond the static 
approximation of the CSM by taking into account configurations where the quarks which 
form the quarkonium can have a nonzero relative momentum and can be off-shell.
As one can see for radially excited states like $\psi'$, very important effects can be
omitted otherwise. Another important point raised is that new cut contributions
can appear in this non-static extension. These have been computed and were shown to be 
at least non-negligible at high $P_T$. Indeed, there exist ambiguities in the way to 
preserve gauge-invariance and these might leave room for a better description of the data.

\section{Conclusions and perspectives}

In this review, we have tried to provide a wide overview about the problem
of the hadroproduction of $J/\psi$, $\psi'$, $\Upsilon(1S)$, $\Upsilon(2S)$ 
and $\Upsilon(3S)$. In order to clarify the discussion, we have limited
ourselves to the study of direct production, which does not involve 
decays of higher excited states (like $P$-waves) nor of beauty hadron  
in the charmonium case.

On the experimental side, we benefit nowadays from studies
from two hadron colliders, the Tevatron and RHIC. We have reviewed in detail
the procedure used and the results obtained by the CDF collaboration. The situation
is ideal for  $\psi'$ as they are able to extract the direct cross section as well 
as the polarisation measurement for the sole direct sample; for $J/\psi$ and
$\Upsilon(1S)$, they have extracted the direct cross section and a polarisation
measurement, including excited states contributions though; for $\Upsilon(2S)$, 
$\Upsilon(3S)$, solely the total cross section was measured. To what concerns the
Run II, they are able to go to smaller $P_T$ but the direct signal 
as well as the polarisation measurement is still under study. Preliminary data
nevertheless tend to show a confirmation that the (prompt) $J/\psi$ are in fact
produced unpolarised if not longitudinal. This is in striking contradiction
with all the NRQCD predictions. To what concerns 
RHIC collider, the PHENIX collaboration measured for the first time
the $J/\psi$ cross section at $\sqrt{s}=200$ GeV where the non-prompt
signal is negligible contrary to $\chi_c$ feed-down.

On the theoretical side, many models and calculations have been proposed since
the outbreak of the $\psi'$ anomaly in the mid-nineties. We have
reviewed six of them. In order to present them on the same footage, 
we have limited ourselves to their basic points and their most important results.

We have first started with the Soft-Colour Interaction approach which deals with Monte Carlo
simulation. It introduces the possibility of colour quantum-number exchanges between
partons produced during the collision. This effectively opens a gluon fragmentation channel
into $^3S_1$ at LO. The sole new parameter brought in is in fact kept at the same value as 
in the description of rapidity gaps but the approach is likely to suffer from  some of the
 drawbacks as the Colour-Evaporation Model.

Secondly, we have reviewed the NRQCD factorisation approach, which is usually embedded, 
in hadroproduction, in the Colour-Octet Mechanism. We have presented the initial motivations of
the approach, namely to cure IR divergences of the Colour-Singlet Model. Since it introduces
unknown non-perturbative parameters called Long-Distance Matrix Elements (LDME), we
have reproduced the values obtained by different groups, as well as the values obtained for
the Colour-Singlet matrix elements. As said before, this approach is in contradiction
with data to what concerns the polarisation. Some modifications of the scaling rules, which 
normally provide a hierarchy between the LMDE's, were proposed to reduce the discrepancy
but the confirmation of the new CDF preliminary measurement would certainly blacken the setting. 
Finally, according to a recent analysis centered on fixed-target experiments by 
F.~Maltoni \etal\cite{Maltoni:2006yp}~the universality of these LDME's is far from being observed.

Thirdly, we have reviewed the application of the $k_T$ factorisation approach to hadroproduction. We 
have emphasised
the two main differences with the collinear approaches (CSM or COM), namely the use of
unintegrated PDF's and of the effective BFKL vertices. As we have said, the LO contributions include
naturally fragmentation channels; this lets us think that the approach is perhaps more suitable than
the collinear ones. In the case of COM, we have given the values of the LDME's obtained. These are
smaller than the corresponding ones fit from the collinear cross sections and thus in better agreement
with what is expected from $ep$ data from HERA\cite{yr} and perhaps also from
fixed-target experiments\cite{Maltoni:2006yp}.

Fourthly, we have briefly reviewed a calculation proposed 
by physicists from Durham of some specific NNLO pQCD contributions of the CSM. 
These can be viewed as LO BFKL contributions and thus are expected to be enhanced compared to other
NNLO pQCD ones. Unfortunately, the method used cannot predict the $P_T$ slope of the cross section.

The fifth model reviewed deals with reinteractions with comovers. Two pictures were analysed: one
for small $P_T$ events where the comovers are created by Brem\ss trahlung of the colliding
gluons and another for high-$P_T$ events where they are produced by DGLAP radiation of the fragmenting
parton.  The authors of the model suspect such re-scatterings to be source of a possible 
enhancement of the cross section,
and if the picture holds, they predict that in the case of gluon fragmentation, the quarkonia 
are produced unpolarised. 

The last approach reviewed is a consistent scheme to go beyond the static and on-shell approximation
used, for instance, in the CSM and the CEM. The loss of gauge invariance caused by the 
introduction of such non-local effects
was shown to be curable by the introduction of 4-point vertices. In the case of $\psi'$, the 
presence of a node
in the vertex function, which is typical of radially excited states and whose effects are necessarily
neglected in the static approximation, was shown to have non-negligible effects and is therefore
expected to be analysed in other processes.

Finally, on a more formal basis, for all models, we are lacking a factorisation theorem suitable
both for charmonium and bottomonium cases. To what concerns NRQCD, even though some
modification of it are required\cite{Nayak:2005rt}, it has been shown recently 
that the (modified) picture still holds at NNLO\cite{Nayak:2005rw} at least 
for fragmentation processes which are supposed to dominate in the large $P_T$ regime.

In view of similar discrepancies between theory and data at $B$-factories, we strongly believe
that the forthcoming efforts both on the experimental side and theoretical side will
be fruitful and will certainly shed light on the interplay between the perturbative QCD and 
the bound-state physics for which quarkonium physics is archetypal.

\section*{Acknowledgements}
\addcontentsline{toc}{section}%
{\protect\numberline{}{Acknowledgements}}

This work has been initiated during a stay at Pittsburgh U. thanks to an FNRS grant. 
I would like to thank A.K.~Leibovich for his hospitality, our discussions and his help
for the redaction of the section about NRQCD. I am also grateful to J.R. Cudell, G.~Ingelman,
V.~Papadimitriou, S.~Peign\'e and L.~Szymanowski for their comments about the manuscript, 
to E.~Braaten, J. Cugnon, G. Grunberg and Yu.L. Kalinovsky for discussions and suggestions 
as well as to S.~Baranov for correspondences. Finally, I thank them  all for their encouragements.

\end{document}